\documentclass[11pt,a4paper]{article}
\usepackage{amssymb}
\usepackage{amscd}
\usepackage{amsmath}
\usepackage{graphicx}

\begin{document}


\newcounter{mo}
\newcommand{\mo}[1]
{\stepcounter{mo}$^{\bf MO\themo}$%
\footnotetext{\hspace{-3.7mm}$^{\blacksquare\!\blacksquare}$
{\bf MO\themo:~}#1}}

\newcounter{bk}
\newcommand{\bk}[1]
{\stepcounter{bk}$^{\bf BK\thebk}$%
\footnotetext{\hspace{-3.7mm}$^{\blacksquare\!\blacksquare}$
{\bf BK\thebk:~}#1}}


\newcommand{\Si}{\Sigma}
\newcommand{\tr}{{\rm tr}}
\newcommand{\ad}{{\rm ad}}
\newcommand{\Ad}{{\rm Ad}}
\newcommand{\ti}[1]{\tilde{#1}}
\newcommand{\om}{\omega}
\newcommand{\Om}{\Omega}
\newcommand{\de}{\delta}
\newcommand{\al}{\alpha}
\newcommand{\te}{\theta}
\newcommand{\vth}{\vartheta}
\newcommand{\be}{\beta}
\newcommand{\la}{\lambda}
\newcommand{\La}{\Lambda}
\newcommand{\D}{\Delta}
\newcommand{\ve}{\varepsilon}
\newcommand{\ep}{\epsilon}
\newcommand{\vf}{\varphi}
\newcommand{\vfh}{\varphi^\hbar}
\newcommand{\vfe}{\varphi^\eta}
\newcommand{\fh}{\phi^\hbar}
\newcommand{\fe}{\phi^\eta}
\newcommand{\G}{\Gamma}
\newcommand{\ka}{\kappa}
\newcommand{\ip}{\hat{\upsilon}}
\newcommand{\Ip}{\hat{\Upsilon}}
\newcommand{\ga}{\gamma}
\newcommand{\ze}{\zeta}
\newcommand{\si}{\sigma}

\def\hS{{\hat{S}}}

\newcommand{\li}{\lim_{n\rightarrow \infty}}
\def\mapright#1{\smash{
\mathop{\longrightarrow}\limits^{#1}}}

\newcommand{\mat}[4]{\left(\begin{array}{cc}{#1}&{#2}\\{#3}&{#4}
\end{array}\right)}
\newcommand{\thmat}[9]{\left(
\begin{array}{ccc}{#1}&{#2}&{#3}\\{#4}&{#5}&{#6}\\
{#7}&{#8}&{#9}
\end{array}\right)}
\newcommand{\beq}[1]{\begin{equation}\label{#1}}
\newcommand{\eq}{\end{equation}}
\newcommand{\beqn}[1]{\begin{small} \begin{eqnarray}\label{#1}}
\newcommand{\eqn}{\end{eqnarray} \end{small}}
\newcommand{\p}{\partial}
\def\sq2{\sqrt{2}}
\newcommand{\di}{{\rm diag}}
\newcommand{\oh}{\frac{1}{2}}
\newcommand{\su}{{\bf su_2}}
\newcommand{\uo}{{\bf u_1}}
\newcommand{\SL}{{\rm SL}(2,{\mathbb C})}
\newcommand{\GLN}{{\rm GL}(N,{\mathbb C})}
\def\sln{{\rm sl}(N, {\mathbb C})}
\def\sl2{{\rm sl}(2, {\mathbb C})}
\def\SLN{{\rm SL}(N, {\mathbb C})}
\def\SLT{{\rm SL}(2, {\mathbb C})}
\def\PSLN{{\rm PSL}(N, {\mathbb C})}
\newcommand{\PGLN}{{\rm PGL}(N,{\mathbb C})}
\newcommand{\gln}{{\rm gl}(N, {\mathbb C})}
\newcommand{\PSL}{{\rm PSL}( 2,{\mathbb Z})}
\newcommand{\SLZ}{{\rm SL}(2, {\mathbb Z})}
\def\f1#1{\frac{1}{#1}}
\def\lb{\lfloor}
\def\rb{\rfloor}
\def\sn{{\rm sn}}
\def\cn{{\rm cn}}
\def\dn{{\rm dn}}
\newcommand{\rar}{\rightarrow}
\newcommand{\upar}{\uparrow}
\newcommand{\sm}{\setminus}
\newcommand{\ms}{\mapsto}
\newcommand{\bp}{\bar{\partial}}
\newcommand{\bz}{\bar{z}}
\newcommand{\bw}{\bar{w}}
\newcommand{\bA}{\bar{A}}
\newcommand{\bG}{\bar{G}}
\newcommand{\bL}{\bar{L}}
\newcommand{\btau}{\bar{\tau}}

\newcommand{\tie}{\tilde{e}}
\newcommand{\tial}{\tilde{\alpha}}

\newcommand{\Sh}{\hat{S}}
\newcommand{\vtb}{\theta_{2}}
\newcommand{\vtc}{\theta_{3}}
\newcommand{\vtd}{\theta_{4}}

\def\mC{{\mathbb C}}
\def\mZ{{\mathbb Z}}
\def\mR{{\mathbb R}}
\def\mN{{\mathbb N}}

\def\frak{\mathfrak}
\def\gb{{\frak b}}
\def\gg{{\frak g}}
\def\gp{{\frak p}}
\def\gn{{\frak n}}
\def\gJ{{\frak J}}
\def\gS{{\frak S}}
\def\gL{{\frak L}}
\def\gM{{\frak M}}
\def\gG{{\frak G}}
\def\gE{{\frak E}}
\def\gF{{\frak F}}
\def\gk{{\frak k}}
\def\gK{{\frak K}}
\def\gl{{\frak l}}
\def\gh{{\frak h}}
\def\gH{{\frak H}}
\def\gs{{\frak s}}
\def\gt{{\frak t}}
\def\gT{{\frak T}}
\def\gR{{\frak R}}

\def\baal{\bar{\al}}
\def\babe{\bar{\be}}

\def\bfa{{\bf a}}
\def\bfb{{\bf b}}
\def\bfc{{\bf c}}
\def\bfd{{\bf d}}
\def\bfe{{\bf e}}
\def\bff{{\bf f}}
\def\bfg{{\bf g}}
\def\bfm{{\bf m}}
\def\bfn{{\bf n}}
\def\bfp{{\bf p}}
\def\bfu{{\bf u}}
\def\bfv{{\bf v}}
\def\bfr{{\bf r}}
\def\bfs{{\bf s}}
\def\bft{{\bf t}}
\def\bfx{{\bf x}}
\def\bfy{{\bf y}}
\def\bfM{{\bf M}}
\def\bfR{{\bf R}}
\def\bfC{{\bf C}}
\def\bfP{{\bf P}}
\def\bfq{{\bf q}}
\def\bfS{{\bf S}}
\def\bfJ{{\bf J}}
\def\bfI{{\bf I}}
\def\bfX{{\bf X}}
\def\bfY{{\bf Y}}
\def\bfz{{\bf z}}
\def\bfnu{{\bf \nu}}
\def\bfsi{{\bf \sigma}}
\def\bfU{{\bf U}}
\def\bfso{{\bf so}}

\def\clA{\mathcal{A}}
\def\clC{\mathcal{C}}
\def\clD{\mathcal{D}}
\def\clE{\mathcal{E}}
\def\clF{\mathcal{F}}
\def\clG{\mathcal{G}}
\def\clR{\mathcal{R}}
\def\clU{\mathcal{U}}
\def\clT{\mathcal{T}}
\def\clO{\mathcal{O}}
\def\clH{\mathcal{H}}
\def\clK{\mathcal{K}}
\def\clJ{\mathcal{J}}
\def\clI{\mathcal{I}}
\def\clL{\mathcal{L}}
\def\clM{\mathcal{M}}
\def\clN{\mathcal{N}}
\def\clP{\mathcal{P}}
\def\clQ{\mathcal{Q}}
\def\clV{\mathcal{V}}
\def\clW{\mathcal{W}}
\def\clZ{\mathcal{Z}}

\def\baf{{\bf f_4}}
\def\bae{{\bf e_6}}
\def\ble{{\bf e_7}}
\def\bag2{{\bf g_2}}
\def\bas8{{\bf so(8)}}
\def\baso{{\bf so(n)}}

\def\sr2{\sqrt{2}}
\newcommand{\ran}{\rangle}
\newcommand{\lan}{\langle}
\def\f1#1{\frac{1}{#1}}
\def\lb{\lfloor}
\def\rb{\rfloor}
\newcommand{\slim}[2]{\sum\limits_{#1}^{#2}}

\newcommand{\sect}[1]{\setcounter{equation}{0}\section{#1}}
\renewcommand{\theequation}{\thesection.\arabic{equation}}
\newtheorem{predl}{Proposition}[section]
\newtheorem{defi}{Definition}[section]
\newtheorem{rem}{Remark}[section]
\newtheorem{cor}{Corollary}[section]
\newtheorem{lem}{Lemma}[section]
\newtheorem{theor}{Theorem}[section]

\vspace{0.3in}
\begin{flushright}
 ITEP-TH-44/13\\
\end{flushright}
\vspace{10mm}

\begin{center}
{\LARGE{\bf
 Classification of Isomonodromy Problems\\ \vskip2mm on Elliptic Curves}
}\\
\vspace{12mm} {\Large { A. Levin}$^{\,\natural\,\,\flat}$ { M.
Olshanetsky}$^{\,\flat\,\sharp}$
 { A. Zotov}$^{\,\diamondsuit\,\flat\,\sharp}$}\\ \vspace{7mm}
 \vspace{3mm} $^\flat$ - {\sf Institute of Theoretical and Experimental Physics,
  Moscow, 117218 Russia}\\
 \vspace{2mm}$^\natural$ - {\sf Laboratory of Algebraic Geometry, GU-HSE,
 7 Vavilova Str., Moscow, 117312 Russia}\\
 \vspace{2mm} $^\sharp$ - {\sf Moscow Institute of Physics and Technology,
  Dolgoprudny, 141700
 Russia}\\
 \vspace{2mm} $^\diamondsuit$ -
 {\sf Steklov Mathematical Institute, RAS, 8 Gubkina Str., Moscow, 119991 Russia
 }
 \vspace{4mm}

 E-mails:
{\em alevin57@gmail.com}; {\em olshanet@itep.ru};  {\em zotov@mi.ras.ru}; {\em zotov@itep.ru}\\
 \vspace{5mm}
\end{center}

\vspace{0mm}

\begin{abstract}
We consider the isomonodromy problems for flat $G$-bundles over
punctured elliptic curves $\Sigma_\tau$ with regular singularities
of connections at marked points. The bundles are classified by their
characteristic classes. These classes are elements of the second
cohomology group $H^2(\Sigma_\tau,{\mathcal Z}(G))$, where
${\mathcal Z}(G)$ is the center of $G$. For any complex simple Lie
group $G$ and arbitrary class we define the moduli space of flat
bundles, and in this way construct the monodromy preserving
equations in the Hamiltonian form and their Lax representations. In
particular, they include the Painlev\'e VI equation, its
multicomponent generalizations and elliptic Schlesinger equations.
The general construction is described for punctured curves of
arbitrary genus. We extend the Drinfeld-Simpson (double coset)
description of the moduli space of Higgs bundles to the case of flat
connections. This local description allows us to establish the
Symplectic Hecke Correspondence for a wide class of the monodromy
preserving equations classified by characteristic classes of
underlying bundles. In particular, the Painlev\'e VI equation can be
described in terms of ${\rm SL}(2, {\mathbb C})$-bundles. Since
${\mathcal Z}({\rm SL}(2, {\mathbb C}))= {\mathbb Z}_2$, the
Painlev\'e VI has two representations related by the Hecke
transformation: 1) as the well-known elliptic form of the Painlev\'e
VI
 (for trivial bundles); 2) as  the non-autonomous
Zhukovsky-Volterra gyrostat (for non-trivial bundles).
\end{abstract}


\newpage

\tableofcontents



\section{Introduction}
\setcounter{equation}{0}

In this paper we suggest classification of the isomonodromy problems
for flat $G$-bundles over elliptic curves $\Sigma_\tau={\mathbb
C}/({\mathbb Z}+\tau{\mathbb Z})$ with logarithmic singularities of
connections at the marked points $z_a$, $a=1,...,n$. The Lie group
$G$ is simple and complex. As a preliminary let us briefly recall
that the isomonodromy problem on a genus $g$ complex curve
$\Sigma_g$ with $n$ marked points lead to the following (monodromy
preserving or isomonodromic) equations
  \beq{int01}
 \begin{array}{c}
 \displaystyle{
 \p_{t_j}L(z)-\p_z M_j(z)=[L(z),M_j(z)]\,,
 }
 \end{array}
  \eq
where $z$ is a local coordinate on $\Sigma_g$, $t_j$ are the moduli of
$\Sigma_{g,n}=\Sigma_g\setminus\{z_1,...,z_n\}$ ($j=1,...,3g-3+n$
for $g>1$) while $L(z)$ and $M(z)$ are $\gg=Lie(G)$-valued
functions. Equations (\ref{int01}) are compatibility conditions for
the set of linear problems
  \beq{int02}
 \begin{array}{c}
 \displaystyle{
 \left\{
 \begin{array}{l}
 (\p_z+L(z))\Psi=0\,,
 \\
 (\p_{t_j}+M_j(z))\Psi=0\,,
 \end{array}
 \right.\ \ \ \Psi=||\psi_1...\psi_{\hbox{\tiny{dim}} {{V}}} ||\,,
 }
 \end{array}
  \eq
where $V$ is a finite-dimensional $G$-module and
$\psi_k\in\Gamma(E_G)$ $k=1,...,\hbox{dim}V$ are sections of vector
bundle $E_G={\cal P}\times_G V$ for some principle $G$-bundle ${\cal
P}$. The lower equations arise from requirement that monodromies of
solutions of the upper one equation around marked points and
fundamental cycles of $\Sigma_g$ are independent of the moduli of
$\Sigma_{g,n}$.

The equations (\ref{int01}) are valid identically in $z$. They
provide finite-dimensional non-autono\-mous Hamiltonian equations.
In this way (Lax form or zero curvature form) one can represent the
Painlev\'e equations, Schlesinger systems and their generalizations.
Historically, the studies of the isomonodromy problems were
motivated by the invention of the Painlev\'e list of equations
\cite{Painleve1,Painleve102,Gambier}. It appeared that the second
order nonlinear ODEs from that list can be written as compatibility
condition of a pair of linear equations 
on the punctured Riemann sphere \cite{Fuchs,Garnier,Schlesinger}.
The later result was rediscovered around 1980 \cite{FN,JM1,JM2,JM3}.
There is now a large literature on the isomonodromy problem,
Painlev\'e equations and  different applications in mathematical
physics
\cite{ince,book,book1,IN,IN02,Wo,Dubrovin1,Bolibrukh,Bolibrukh02,Slav,Krich5,boalch1,Poberezh}.

The aim of the paper is to classify the equations (\ref{int01}),
(\ref{int02}) for $g=1$ case, i.e. for elliptic curves
$\Sigma_\tau$. The isomonodromy problems on general compact Riemann
surfaces and in particular on the torus were studied in
\cite{Oka,Iwa,Kawa}  by  analytic methods. Here we follow approach
developed in \cite{LO1,LO102,LO11}. It is based on the
non-autonomous version of the Hitchin systems
 \cite{Hitchin} on the punctured elliptic curve \cite{Nekr,Nekr02,Nekr03,LO2} (see also \cite{Takasaki1,Takasaki102,Takasaki103}).
 A close construction of the Painlev\'e equations was proposed in \cite{HW}.
The idea was to reduce gauge symmetries $\clG$ generated by
automorphisms of $E_G$ and restrict in this way to the moduli space
of flat connections. The corresponding moment map produced equation
for $L(z)$ of type:
  \beq{int03}
 \begin{array}{c}
 \displaystyle{
 \bar\p L(z)=\sum\limits_{a=1}^nS^a\delta^{(2)}(z-z_a)\,,
 }
 \end{array}
  \eq
where $S^a$ are the residues of $L(z)$ at the marked points.
Equation (\ref{int03}) together with boundary conditions for
solutions of (\ref{int01})
  \beq{int04}
 \begin{array}{c}
 \displaystyle{
 \Psi(z+1)=\clQ(z) \Psi(z)\,,\ \ \Psi(z+\tau)=\Lambda(z)\Psi(z)
 }
 \end{array}
  \eq
fixes $L(z)$. The matrices $\clQ(z)$ and $\La(z)$ are the transition
functions of the bundle $E_G$. They satisfy the cocycle condition
  \beq{int05}
 \clQ(z)\La^{-1}(z)\clQ^{-1}(z+\tau)\La(z+1)=1\,.
  \eq
Let us mention again that the same equations
(\ref{int03})-(\ref{int05}) describe the integrable (Hitchin)
systems on elliptic curve. The solutions (in some particular cases)
are known as the elliptic Gaudin models \cite{STSR,Nekr}. The
corresponding matrices $L(z)$ were constructed via some natural
generalization of the Krichever's anzats for the Lax pairs of
elliptic integrable systems \cite{Krich1}. The phase spaces of the
integrable systems are given by two parts: the set of $S^a$ - "spin"
part and the moduli of solutions of (\ref{int05}) - configuration
space for "many-body" degrees of freedom.  The isomonodromy version
of the integrable models are the non-autonomous generalizations of
the elliptic Calogero model, elliptic top and elliptic Schlesinger
systems \cite{LO11,CLOZ1,CLOZ102}. The phenomenon that the same pair
of matrices satisfies the Lax equation of integrable system and the
monodromy preserving equation (\ref{int01}) was observed in
\cite{LO97} and is known as the {\em Painlev\'e-Calogero
Correspondence}.

In general case solutions of (\ref{int05}) are classified by
characteristic classes of the bundles which are the elements of the
second cohomology group $H^2(\Sigma_\tau,{\cal Z}(G))$, where ${\cal
Z}(G)$ is the center of $G$. The classification of these solutions,
underlying Higgs bundles and corresponding integrable systems (for
the single marked point case) is given in our recent papers
\cite{LOSZ1,LOSZ2} (see also \cite{SZ,LOSZ3,LOZ3}). The
classification arises from the fact that the transition matrices
$\clQ(z)$ and $\La(z)$ can be chosen to be $z$-independent up to
scalar factor ($\clQ$ and $\La$). Then (\ref{int05}) is replaced
(with regard to the scalar factor) by
  \beq{int06}
 \clQ\La^{-1}\clQ^{-1}\La=\zeta\,,\ \ \zeta\in{\cal Z}(G)\,.
  \eq
The representative cases for $\SLN$-bundles\footnote{The
classification of bundles over elliptic curves in this case was
proposed by M. Atiyah \cite{Atiyah}. In our terms his result is
formulated as follows: $H^2(\Sigma_\tau,{\cal Z}(\SLN))\sim{\cal
Z}(\SLN)=\mZ/N\mZ$, i.e. the bundles are classified by $N$-th roots
of unity.} are $\zeta=1$ and $\zeta=\exp(\frac{2\pi \sqrt{-1}}N)$.
The corresponding integrable systems are the Calogero model and
elliptic top respectively. The non-trivial intermediate case appears
when $N=pl$. Then, the characteristic classes $\zeta=\exp(\frac{2\pi
\sqrt{-1}}Np)$ or $\zeta=\exp(\frac{2\pi \sqrt{-1}}Nl)$ correspond
to the so-called model of interacting elliptic tops
\cite{LOSZ4,LOSZ402}. The phase space in these cases has the same
dimension as the phase space of the Calogero model with spin
variables, but less number of particles ($n<N$) and greater number
of spin variables.

In this paper we extend the classification by characteristic classes
of the Higgs bundles and integrable systems
\cite{LOSZ1,LOSZ2}\footnote{An alternative classification of Higgs
bundles was suggested recently in \cite{Garcia}. The comparison of
these results with \cite{LOSZ1,LOSZ2} needs further elucidation. The
noncompact real Lie group case was discussed recently in
\cite{Hitchin2,Hitchin202}.} to the case of flat connections and
arbitrary number of punctures on elliptic curves.

We use two descriptions of the moduli space of flat connections. The
first one \cite{LO11} is a natural generalization of the
group-theoretical description of the moduli space of the Higgs
bundles in the framework of Hitchin systems \cite{Nekr,LO2}. It is
given as quotient space
  \beq{int068}
 FBun(\Si_g,G)=Conn(\Si_g,G)//\clG=FConn(\Si_g,G)/\clG
  \eq
 of flat connections
  \beq{int07}
 FConn(\Si_g,G)=\{d+\clA\,|\,d\clA+\oh\clA\wedge\clA=0\}
  \eq
  modulo
gauge group $\clG$. The space $Conn(\Si_g,G)=\{d+\clA\}$ of smooth
connections  on $E_G$ is equipped with the well-known (Atiyah-Bott)
symplectic structure $\om=\oh\int_{\Si_g}\lan\delta {\cal A}\wedge
\delta{\cal A}\ran$ \cite{Bott}. After reduction it provides the
symplectic (and the Poisson) structure for some finite-dimensional
Hamiltonian system. Its phase space $FBun(\Si_g,G)$ is the principle
homogeneous space $PT^*Bun(\Si_g,G)$ over $T^*Bun(\Si_g,G)$.

The second description of $FBun(\Si_g,G)$ comes from the
Drinfeld-Simpson's one \cite{BD,BD02} for the moduli of Higgs
bundles. It is formulated in terms of local data (see Section
\ref{bd3}). The moduli space of flat bundles with the
quasi-parabolic structures at the marked points is given as the
double coset space:
  \beq{int08}
 Bun(\Si_{g,n},G)=
G(\Si_{g,n})  \setminus G(D^\times)/G(D)\,,
  \eq
where  $G(X)$ are the holomorphic maps from $X\subset\Si_g$ to $G$,
$D^\times$ is the disjoint union of punctured small disks around the
marked points (where the bundle is trivialized) and $G(D)$ are
special maps preserving the flags corresponding to the
quasi-parabolic structures at the marked points. At the level of
$FBun(\Si_g,G)$ this construction means that we consider trivial
bundles over $\Si_g$ with regular singularities at the marked
points. It is known that a regular singularity can be transform to
the Fuchsian one by some meromorphic gauge transformation on a disk.
 So, it is natural to equip a connection with the meromorphic gauge transformation,
defined up to right multiplication by holomorphic transformation on
the disk.
 Thus, the action of $G(D^\times)$ can transform the regular singularity  to the Fuchsian ones.
 We treat such meromorphic gauge transformation as a transition function for a non-trivial bundle
 with the Fuchsian singularity. In this way we come to description of $FBun(\Si_g,G)$ similar to (\ref{int08}).
%

The later description is very natural for the definition of the
Hecke operators (or modifications of bundles) \cite{AL,AL02} which
relates the linear problems (\ref{int02}) and equations
(\ref{int01}) for different characteristic classes. At the level of
the connections $L(z)$ the modification acts by gauge transformation
  \beq{int09}
L\stackrel{\Xi}\longrightarrow L^{mod}\,,~~   L^{mod}(z)\Xi(z)=
\Xi(z)L(z)-\p_z\Xi(z)\,,
  \eq
degenerated at some point $z_0$, i.e. $\det\Xi(z,z_0)\sim z-z_0$
near $z_0$.
In this way we extend the {\em Symplectic Hecke Correspondence}
introduced in \cite{LOZ1} to the case of the monodromy preserving
equations. As an example we consider the Painlev\'e VI equation. It
can be described in terms of $\SL$-bundles. Since ${\cal Z}(\SL)=
\mZ_2$, the equation
 has two representations related by the  Hecke transformation:

\vspace{1mm}

 1) elliptic form of Painlev\'e VI \cite{Painleve1906,Manin};
  \beq{i100}
\frac{d^2u}{d\tau^2}=\sum\limits_{a=0}^{3}\nu_a^2\wp'(u+\om_a);
  \eq

\vspace{1mm}

2) as  the non-autonomous Zhukovsky-Volterra gyrostat \cite{LOZ2}:
  \beq{int10}
\p_\tau S=[S,J(S)]+[S,\nu']\,,
  \eq
where $S$ is ${\rm sl}^*(2,{\mathbb C})$-valued dynamical variable
while $J$ and $\nu'$ are non dynamical but $\tau$-dependent. The
$2\times 2$ linear problems for (\ref{i100}) and (\ref{int10}) were
described in \cite{Zotov04} and \cite{LOZ2} respectively. See
Section \ref{p6} for details. In general case the Hecke
transformation (\ref{int09}) can be considered as a B\"acklund
transformation of the monodromy preserving equations. It can be
considered as a discrete time, or the Schlesinger transformation
\cite{JM2}.

The monodromy preserving equations can be considered as a deformation of integrable systems.
The integrable hierarchies corresponding to the systems considered
here were constructed in \cite{LOSZ1,LOSZ2}. The correspondence
between integrable systems and isomonodromic deformations, which was
called in \cite{LO97} the Painlev\'{e}-Calogero correspondence, was
used by P. Boutroux to investigate solutions of the Painlev\'{e}
equations \cite{Bout}. Similarly, R. Garnier constructed an
autonomous analogue of the Schlesinger equations \cite{Ga}, and came
in this way to the isospectral problem.
 H. Flaschka and A. Newell in \cite{FN} and recently I. Krichever in
\cite{Kr3} developed the Boutroux-Garnier program and found that
the WKB approximation with respect to the "Planck constant" $\ka$
converts the isomonodromy problem into a isospectral problem.

\vspace{0.3cm}

The paper is organized as follows. It consists of three parts. The
first one is devoted to the general approach to the isomonodromy
problem on $\Sigma_{g,n}$. In Sections 2 and 3 we describe the
moduli space of flat connections in two ways. First, in terms of
connections (\ref{int07}) and the gauge group. Second, in terms of
local data, as the double coset space (\ref{int08}). The later
description is used for definitions of the characteristic classes
and the Hecke transformations in the end of Section 3. In Section 4
the isomonodromy problem is described and the Hamiltonian approach
is given. In the second part we consider the case of elliptic curve
in detail. We extend the previously obtained classification of the
elliptic integrable systems to the monodromy preserving equations.
We also briefly review the related KZB equation equations and field
theory generalizations. The classification is demonstrated with the
example of Painlev\'e VI equation (Section \ref{p6}). In Appendix
 we give the necessary short summary of Lie groups and elliptic
functions. We also describe the generalized sin-algebra basis in Lie
algebras. It is convenient for classification by characteristic
classes. At last, we describe conformal versions of Lie groups in
order to relate the characteristic classes with degrees of bundles.

\vspace{0.3cm}
{\small {\bf Acknowledgments.} We are grateful to V. Poberezhny for
useful discussions. The work was supported by RFBR grants
12-02-00594 and 12-01-33071 mol$\_$a$\_$ved, by Russian President
grant NSh-4724.2014.2 for support of leading scientific schools. The
work of A.L. was partially supported by AG Laboratory GU-HSE, RF
government grant, ag. 11 11.G34.31.0023. The work of A.Z. was
partially supported by the D. Zimin's fund "Dynasty".
 }


\part{General approach to the Isomonodromy problem}

Let $G$ be a complex simple Lie group and $E_G$ be a flat $G$-bundle over a curve $\Si_{g,n}$ of genus $g$
with $n$ marked points. A complex structure on $\Si_{g,n}$ defines a polarization of connections acting on
sections $\G(E_G)$. Locally, in complex coordinates $(z,\bz)$ $\,d+\clA=((\p+A)\otimes dz,(\bp+\bA)\otimes d\bz)$.
Due to the flatness the following system is consistent
$$
\left\{
\begin{array}{c}
  (\p+A)\psi=0 \\
   (\bp+\bA)\psi=0
\end{array}
\right.~~~ \psi\in\G(E_G)\,.
$$
Assume that the monodromies of solutions $\psi$ are independent of
complex structure on $\Si_{g,n}$. The independence conditions are
differential equations that in some cases can be written down
explicitly. Our main objects are flat $G$-bundles over elliptic
curves $\Si_\tau=\mC/(\mZ+\tau\mZ)$.  But before we will consider
  general curves.


\section{Flat bundles. General case}
\setcounter{equation}{0}

Let $\clP$ be a principal  $G$-bundle over $\Si_{g,n}$, $V$ is a finite-dimensional module over $G$,
 and $E_G=\clP\times_GV$.

We will consider two cases:\\
1) smooth proper (compact)  curves ($n=0$);\\
2) smooth proper curves with punctures (marked points) ($n\neq 0$).

\subsection{Moduli space of flat bundles over smooth curves}

For smooth curves define the space $Conn(\Si_g,G)=\{d+\clA\}$ of smooth connections  on $E_G$.
The group of  automorphisms $\clG$ of $E_G$ (the gauge group) acts on connections
by the affine transformations
 \beq{ggr}
\clG\,:\,\clA\to f^{-1}df+f^{-1}\clA f\,.
 \eq
 Let $FConn(\Si_g,G)$ be the space of flat connections
  \beq{fla2}
FConn(\Si_g,G)=\{d+\clA\,|\,d\clA+\oh\clA\wedge\clA=0\}\,.
  \eq
 The group $\clG$ preserves the flatness.
  The moduli space of the flat connections
 is the quotient
 \beq{fba}
FBun(\Si_g,G)=FConn(\Si_g,G)/\clG\,.
 \eq
On the other hand $FBun(\Si_g,G)$ can be described as a result of
the Hamiltonian reduction of the symplectic space $Conn(\Si_g,G)$ of
all smooth connections by action of the gauge group $\clG$. The
symplectic form on $Conn(\Si_g,G)$ is the form
 \beq{G.2}
\om=\oh\int_{\Si_g}\lan\delta {\cal A}\wedge \delta{\cal A}\ran,
 \eq
 where $\lan\cdot,\cdot\ran$ denotes the Killing form,
$\delta{\cal A}$ is a Lie$(G)$-valued one-form on $\Si_g$. So,
$\lan\delta{\cal A}, \delta{\cal A}\ran$  is a two-form,
and the integral is well defined. The form is gauge-invariant. The symplectic reduction with respect
to this action leads to the momentum map
$$
\begin{array}{c}
  Conn(\Si_g,G)\to \clG^*\sim\Om_{C^\infty}^2(\Si_g,Lie^*(G)) \\
   \clA\mapsto F_{\cal A}= d\clA + {\cal  A}\wedge{\cal A}\,.
\end{array}
$$
 Hence,
the preimage of $0$ under the momentum map is the space of flat
connections $FConn(\Si_g,G)$ (\ref{fla2}).
The choice of the complex structure on $\Si_g$ defines the polarization of $Conn(\Si_g,G)$.
Then the connection is decomposed in $(1,0)$ and $(0,1)$ parts $\clA=(A,\bA)$.
 We can write $(1,0)$ and $(0,1)$ components of the connection in local coordinates  $(z,\bz)$
  \beq{stp}
 Conn(\Si_g,G)=\{d'=(\p+A)\otimes dz\,,~~d''=(\bp+\bA)\otimes d\bz\}\,,~~(\p=\p_z\,,~\bp=\p_{\bz})\,.
  \eq
 In this description $\om$ (\ref{G.2}) assumes the form
  \beq{G.1}
 \om=\int_{\Si_g}\lan\de A\wedge\de\bA\ran\,.
  \eq
Define the moduli space of holomorphic bundles.
 A section $s\in\G(E_G)$ is holomorphic if it is annihilated by the operator $\bp+\bA$. The
 moduli space of holomorphic bundles is the quotient
  \beq{hb}
 Bun_{\Si_g,G}=\{\bp+\bA\}/\clG\,.
  \eq
%
%
%
Let $b$ be a point in  $Bun(\Si_g,G)$ and ${\cal V}_b$ is the corresponding
$E_G$-bundle on $\Si_g$. Denote by $ad_b$ the bundle of endomorphisms
of ${\cal V}_b$ as a bundle of Lie algebras. Let $Flat_b$ the space of
flat holomorphic connections on the holomorphic bundle ${\cal V}_b$. In other words, in local
coordinates
 \beq{fla}
Flat_b=\{\p+A\,|\,F(A,\bA)=0\}\,.
 \eq
 The tangent space to $Bun_{\Si_g,G}$  at the point
$b=E_G$ is canonically isomorphic to the first cohomology group $H^1(\Si_g, ad_b)$.
The space  $Flat_b$ of holomorphic connections on ${\cal V}_b$ is an
affine space over the vector space $H^0(\Si_g, ad_b\otimes \Omega ^1)$ of holomorphic
$ad_b$-valued one-forms, since a difference between any connections is
a $ad_b$-valued differential form.
The vector spaces $H^1(\Si_g,  ad_b)$ and $H^0(\Si_g,ad_b\otimes \Omega ^1)$
are dual.

Consider the map from $FBun(\Si_g,G)$ onto the moduli space  $Bun(\Si_g,G)$.
The fiber of the projection   $FBun(\Si_g,G)\to Bun(\Si_g,G)$ over a point $b$
is naturally isomorphic to $Flat_b$
 \beq{fol}
p\,:\,FBun(\Si_g,G)\stackrel{Flat_b}{\longrightarrow} Bun(\Si_g,G)\,.
 \eq
 These fibers are Lagrangian with respect
to $\om$ (\ref{G.1}).
 From the Riemann-Roch theorem we have
 \beq{dhb}
\dim\,(Bun(\Si_g,G))=\dim\,( H^1(\Si_g, ad_b))=\dim(G)(g-1)\,,
 \eq
 $$
\dim\,(FBun(\Si_g,G))=2\dim(G)(g-1)\,.
 $$
Let $K$ be a canonical class of $\Si_g$. \emph{The Higgs bundle} is
the pair $(E_G,\Om^0(\Si_g,End\,(E_G)\otimes K))$.  In local coordinates it is represented by
the pair $(\bp+\bA,\Phi)$,
where $\Phi\in\Om^0(\Si_g,End\,(E_G)\otimes K)$ is called \emph{the Higgs field}.
The Higgs bundle is a symplectic manifold equipped with the symplectic form
 \beq{hf}
\om^{higgs}=\int_{\Si_g}\lan\de \Phi\wedge\de\bA\ran
  \eq
(compare with (\ref{G.2})). The form is invariant under the action
of the gauge group
 $\Phi\to Ad_f\Phi$, $\bA\to f^{-1}\bp f+f^{-1}\bA f$.
After the symplectic reduction we come to the cotangent bundle
$T^*Bun(\Si_g,G)$ to the moduli space $Bun(\Si_g,G)$. It is the
phase space of \emph{the Hitchin integrable systems} \cite{Hitchin}.



\subsection{Moduli space of flat bundles with quasi-parabolic structure}

Now consider  curves $\Si_{g,n}$ with marked points
$\bfx=(x_1,\ldots,x_n)$. Let $P_a$ $(a=1,\ldots,n)$ be parabolic subgroups $P_a\subset G$.
Attach to the marked points $G$-flag varieties $Flag_a\sim G/P_a$.
 We say in this case
that the bundle $E_G$ has \emph{the quasi-parabolic structure}
($qp$-structure) \cite{Si}. The group of automorphisms of this
bundle $\clG_P$ preserves flags at the marked points. It means that
in the neighborhood $\clU_a$ of $x_a$
 \beq{clgp}
\clG_P=\{f\in\clG\,|\,f|_{\clU_a}=P_a+O(z-x_a)\}\,.
 \eq
Replace the moduli space $Bun(\Si_g,G)$ (\ref{hb}) by the
moduli space $Bun(\Si_{g,n},\bfx,G)$ of $G$-bundles $E_G$ with the  $qp$-structure
 \beq{mqp}
Bun(\Si_{g,n},\bfx,G)=\{\bp+\bA\}/\clG_P\,.
 \eq
%
%
%
 We have natural ``forgetting''
projection:
 \beq{fgp}
\pi\,:\,Bun(\Si_g,{\bf x},G)\stackrel{ \Pi_aFlag_a}{\longrightarrow} Bun(\Si_g,G)\,.
 \eq
The bundle of endomorphisms with this data  is the bundle
$Lie(\clG_P)=ad_{b,Fl}(-{\bf x})$ of endomorphisms which sections
behaves near the marked points as
 $$
\varphi(z-x_a)=\varphi_a^0+(z-x_a)\varphi_a^1+\ldots\,,~~\varphi_a^0\in Lie(P_a)\,.
 $$
 The tangent space to
$Bun_{\Si_g,{\bf x},G}$
is isomorphic to $H^1(\Si_g, ad_{b,Fl}(-{\bf x}))$. The dual space to
this space is  $H^0(\Si, ad^*_{b,Fl}({\bf x})\otimes \Omega ^1)$.

 Consider the coadjoint $G$-orbits in the Lie coalgebra $\gg^*$  located at the marked points
   \beq{co1}
 \clO_a=\{\bfS_a=Ad_gS^0_a\,,~g\in G\,,~\bfS^0_a\in\gg^*\}\,, ~~(a=1,\ldots,n)\,.
   \eq
  The variables $\bfS_a$ are called \emph{the spin variables}. The
coadjoint orbit $ \clO$ is fibered over the flag variety  $Flag$ and the fibers are
the principal  homogeneous spaces  $PT^*Flag$ over the cotangent bundles  $T^*Flag$.
 It is a symplectic variety  with the Kirillov-Kostant symplectic
 forms:
   \beq{KK}
  \om^{KK}=\de\lan S^0,\de gg^{-1}\ran=\lan S, g^{-1}\de g\wedge  g^{-1}\de g\ran\,.
   \eq
Consider the space of smooth connection $Conn_{\Si_{g,n}G}$ with
singularities at the marked points. In small neighborhoods $\clU_a$
of the marked points we can gauge away $\bA$ $(\bA=0)$.
 Assume that $A$ has holomorphic first order poles at $x_a$ with the residues
taking values in the orbits $\clO_a$. Then the space of connections is defined as
 \beq{cos}
 Conn(\Si_{g,n}G)=\left\{(A,\bA)\,|\,
 \begin{array}{lc}
   \bA\,|_{\clU_a}=0\,, &  \\
   A= \bfS_a(z_a-x_a)^{-1}+O(1) &\bfS_a\in\clO_a \,,
 \end{array}
\right\}
 \eq
The flatness condition (\ref{fla2}) takes into account the poles
 \beq{fla1}
F Conn(\Si_{g,n},G)=\{\clA\in Conn(\Si_{g,n},G)\,|\,
\sum_{a=1}^nF_{\clA}=\bfS_a\de^{(2)}_{x_a}\}\,.
 \eq
Consequently, in order to get the symplectic variety we must replace
the affine space $Flat_b$ of flat holomorphic connection by the
affine space $Flat_b(\log)$ of flat connections with logarithmic
singularities. As a result, we get the moduli space (see
(\ref{fba})) of pairs (holomorphic bundle, holomorphic connection
with logarithmic singularities with residues at marked points with
orbits ${\cal O}_a$):
 \beq{fba1}
FBun_{\Si_g,{\bf x},G}=F Conn(\Si_{g,n},G)/\clG_P=Conn(\Si_{g,n},G)//\clG_P\,.
 \eq
 Now the symplectic form (\ref{G.1}) is
 \beq{G.3}
\om=\int_{\Si_g}\lan\de A\wedge\de\bA\ran+\sum_{a=1}^n\om_a^{KK}\,.
 \eq
%
%
Similarly to (\ref{fgp}) we have the projection
 \beq{fgp1}
\pi\,:\,FBun(\Si_g,{\bf x},G)\stackrel{ \Pi_a\clO_a}{\longrightarrow} FBun(\Si_g,G)\,.
 \eq
and the Lagrangian projection (compare with (\ref{fol}))
 \beq{fol1}
p\,:\,FBun(\Si_g,\bfx,G)
\longrightarrow Bun(\Si_g,\bfx,G)\,.
 \eq
The fiber of this projection is an affine space over
 $H^0(\Si, ad^*_{b,Flag}({\bf x})\otimes \Omega ^1)$.
 From (\ref{dhb}) we obtain:
$$
\dim\,(Bun(\Si_g,{\bf x},G))=\dim(G)(g-1)+\sum_{a=1}^n\dim(Flag_a)\,,
$$
 \beq{dhb1}
\dim\,(FBun(\Si_g,{\bf x},G))=2\dim(G)(g-1)+\sum_{a=1}^n\dim(\clO_a)\,.
 \eq
Therefore, to come to non-trivial moduli space of bundles over
 elliptic curves we should have at least one marked point.


\section{Flat bundles and characteristic classes}\label{bd3}
\setcounter{equation}{0}

\subsection{Moduli space of holomorphic bundles  via double coset construction}

We also need another descriptions of the moduli space
$Bun(\Si_{g,n},\bfx,G)$ and
 $FBun(\Si_{g,n},\bfx,G)$.
They can be defined in the following way.
A $G$-bundle can be trivialized over small disjoint disks $D=\cup_{a=1}^nD_a$ around the
marked points and over $\Si_{g,n}\setminus \bfx$. In this way,
 $E_G$ is defined by the transition holomorphic functions
 on  $D^\times=\cup_{a=1}^n(D_a^\times)$, where
 $D_a^\times=D_a\setminus x_a$. If $G(X)$ are the holomorphic maps from $X\subset\Si_g$ to $G$,
 then the  isomorphism classes of holomorphic bundles are defined as the double coset space
$$
Bun(\Si_g,G)= \clG_{out}  \setminus G(D^\times)/\clG_{int}\,,~~~
 \clG_{out}=G(\Si_{g}\setminus \bfx)\,,~~\clG_{int}=G(D)\,.
$$
Here $G(D^\times)$ and $G(D)$ are defined as the product
 of the loop groups.
 Let $t_a$ be a local coordinate in  the disks $D_a$. We replace $G(D)$ and $G(D^\times)$ by the formal series
 \beq{lre}
G(D)=\prod_{a=1}^nG(D_a)\to\prod_{a=1}^nG\otimes\mC[[t_a]]\,,
 \eq
 \beq{ma}
G(D^\times)\to\prod_{a=1}^nL_a(G)\,,~
~L_a(G)=G\otimes\mC[t^{-1}_a,t_a]]\,,~(\ref{logr}) \,.
  \eq
  The quotient is isomorphic to $Bun(\Si_g,G)$ again
 \beq{dcc}
 Bun(\Si_g,G)= \clG_{out}  \setminus \prod_{a=1}^nL_a(G)/\clG_{int}\,,~~~
 \clG_{out}=G(\Si_{g}\setminus \bfx)\,,~~\clG_{int}=\prod_{a=1}^nG\otimes\mC[[t_a]]\,.
   \eq
The connection $d''$ (\ref{stp}) can be reconstructed from the holomorphic transition
functions $g_a\in L_a(G)$. Then the double coset (\ref{dcc})  is equivalent to
the  definition (\ref{hb}).

Let us fix $G$-flags at fibers over the marked points. Remember that
 the $qp$-structure of the $G$-bundle means that $G(D)$ preserves these $G$-flags.
 In other words, $\clG_{int,P}=\prod_{a=1}^nL^+_a(G)$, where
 \beq{lpl}
L^+_a(G)=\{g_0 + g_1t + \ldots\,,~~  g_0 \in P\}{\rm ~~(\ref{bpl})}\,.
 \eq
The moduli space of holomorphic bundles with the $qp$-structures at the marked points
is the double coset
 \beq{dco}
Bun(\Si_{g,n},\bfx,G)=   G(\Si_{g}\setminus \bfx)  \setminus
G(D^\times)/\clG_{int,P}\,.
  \eq
We prove below that this definition is equivalent to the previous one (\ref{mqp}).



\subsection{Characteristic classes}{\label{cc}}

Consider first for simplicity in (\ref{ma}) the one-point case $\bfx=x_0$.
Let $t$ be a local coordinate in the disk $D_{x_0}$ $(t=0\sim x_0)$.
In the representation (\ref{dco}) replace $g(t)\in G(D^\times)$ by  $g(t)h(t)$,
where $h(t)\in D_{x_0}$.
Due to (\ref{dco}), $\,h(t)$ is defined up to the multiplication by
 $f(t)\in G(\mC[[t]])$ from the right.
On the other hand, since the original transition function $g(t)$ is
defined up to the multiplication
 from the right by an element from $G(\mC[[t]])$, $h(t)$ is an element from the double coset
  \beq{affl}
L^+(G)\setminus L(G)/L^+(G)\,.
 \eq
 For $g(t)=\hat w=wt^\ga$ this double coset space is the
affine Schubert cell $C_{\hat w}$ (\ref{spm}) in the affine flag variety
$Flag^{aff}=L(G^{ad})/L^+(G^{ad})$ (\ref{affl}).
The dimension of $C_{\hat w}$  is $l(\hat w)$ (\ref{dsm}).

Due to (\ref{PS}), (\ref{PS1}), (\ref{PS2}) the moduli space
(\ref{dco}) is the union of sectors defined by the affine Weyl
groups. In particular, for  $G^{ad}$-bundles
 \beq{dco1}
Bun(\Si_{g,n},\bfx,G^{ad})=\bigcup_{\hat w\in \ti W_P}Bun_{\hat w}(\Si_{g,n},\bfx,G^{ad})\,,
\eq
\beq{dcc2}
Bun_{\hat w}(\Si_{g,n},\bfx,G^{ad})=G^{ad}(\Si_{g}\setminus \bfx)\setminus G^{ad}(D^\times)\hat
w/G^{ad}(D)\,.
\eq
We can write this representation for the groups $\bG$ and $G_l$ (\ref{fgl2}).
\begin{predl}
The double-coset construction of the moduli space (\ref{dco1})
is equivalent to its Dolbeault  construction (\ref{mqp}).
\end{predl}
\emph{Proof}\\
Let $g(t)\in G\otimes[t^{-1},t]]$ be a transition function  defining
the bundle $E_G$. Consider the decomposition (\ref{PS2}) of $g(t)$
on a small disk
 $\ti D_{x_0}\supset D_{x_0}$
 \beq{got}
g(t)=g_{-}\hat wg_{+}(t)\,,~~g_{-}\in N^-(G)\,,~(\ref{bne})\,,~~
g_{+}(t)\in G(D_{x_0})=L^+(G) \,, ~(\ref{lre})\,.
 \eq
 For $\mC P^1$ it coincides with
   the  Birkhoff  decompositions (\ref{PS}), (\ref{PS1}),
  (\ref{PS2})  \cite{PS}.
  It means that any vector bundle $E_G$ over ${\mathbb{CP}}^1$
  is isomorphic to the direct sum of the line bundles $\oplus_{i=1}^l \clL_{\ga_i}$, where $\clL_{\ga_i}$ is
  defined by the transition function $t^{\ga_i}$,
  $\ga=(\ga_1,\ldots,\ga_l)$
  \footnote{In fact, the bundles with $\ga\neq 0$ are  unstable.}.

To describe the moduli space in a general case we modify the construction from \cite{LO2},
 where it was applied to the \^{C}ech description of $Bun(\Si_g)$.
Represent $g(t)$ (\ref{got}) as
 \beq{dcr00}
g(t)=h_{out}^{-1}\hat wh_{int}\,.
 \eq
 Any $g(t)\in L(G)$ (\ref{logr}) can be represented in this form. As an example,
 consider the Bruhat representation (\ref{PS2}) of $g(t)=g_{-}\hat wg_{+}(t)$.
 Since $N^-(G)$ is a unipotent group its logarithm $\ln(g_{-})$ is
well defined. Note that $g_{-}$ as well as $\ln(g_{-})$ are true
holomorphic functions on the  punctured disk $\ti D_{x_0}^\times$.
By means of the smooth function $\chi(t,\bar t)$
 \beq{shap}
\chi(t,\bar t)=\left\{
\begin{array}{cc}
  0\,, & t\notin \ti D_{x_0}\,,  \\
  1\,, &  t\in  D_{x_0}
\end{array}
 \right.
 \eq
define the (nonholomorphic) continuation of (\ref{got}) from $\ti D^\times_{x_0}$ to $\Si_g$
 \beq{dcr}
g(t)=g_{-}\hat wg_{+}(t)\to g(t)=h_{out}^{-1}\hat wh_{int}\,,
~~
 h_{int}=g_{+}\,,~~
h_{out}=\exp(\chi(t,\bar t)\ln(g_{-}))\,.
 \eq
 This representation has the following interpretation.
 Let $({\bf e}^{hol})$ be a basis of the space of sections of a trivial holomorphic $G$-bundle
over $D_{x_0}$, and $({\bf e}^{C^\infty})$ is a basis of the space of sections of a
 trivial $C^\infty$ $G$-bundle over $D_{x_0}$. The transformation $h_{int}$
can be considered as a transformation from $({\bf e}^{hol})$ to  $({\bf e}^{C^\infty})$.
Therefore, $h_{int}$ is defined in (\ref{dcr}) up to the multiplication from right by $g_+$ and
from left by $f_{int}$.
The similar role plays the transformations $h_{out}$.
 We call this representation \emph{the nonholomorphic Birkhoff
decomposition}.
The multiplications from right by holomorphic transformations
\beq{lac}
h_{int}\to h_{int}g_+(t)\,,~g_+(t)\in  G(D_{x_0})\,,~~
h_{out}\to h_{out}g_-\,,~g_-\in  G(\Si_g\setminus x_0)
\eq
corresponds to the actions of $G(D_{x_0})$ and $G(\Si_g\setminus x_0)$ in the double
coset representation (\ref{dco1}).


Define trivial connections on $D_{x_0}$ and $\Si_g\setminus x_0$
 \beq{bad} \bA_{int}=\bp h_{int}h^{-1}_{int}\,,~~\bA_{out}=\bp
h_{out}h^{-1}_{out}\,.
 \eq
 It follows from  the holomorphicity of
$g(t)$ that
 $$
\bA_{int}=\hat w\bA_{out}\hat w^{-1}\,.
 $$
It means that by means of $(h_{int},h_{out})$ we define connection
$d''$ on sections of nontrivial $G$-bundle. The multiplications from
right by holomorphic transformations (\ref{lac}) do not change
$\bA_{int}$ and $\bA_{out}$. On the other hand, the multiplications
from left by smooth transformations \beq{rac} h_{int}\to f_{int}
h_{int}\,,~f_{int}\in  \clG(D_{x_0})\,,~~ h_{out}\to
f_{out}h_{out}\,,~f_{out}\in \clG(\Si_g\setminus x_0) \eq such that
\beq{fwf} f_{out}=\hat w f_{int}\hat w^{-1} \eq do not change the
$g(t)$ but acting as the gauge transform on connections
 $$
\bA_{int}\to\bp f_{int}f^{-1}_{int}+f_{int}\bA_{int}f^{-1}_{int}\,,~~
 \bA_{out}\to\bp f_{out}f^{-1}_{out}+f_{out}\bA_{out}f^{-1}_{out}\,.
 $$
Summarizing, we introduced the space
 of pairs $(h_{out},h_{int})$ in the decomposition (\ref{dcr})
 $$
\clT(\Si_g,G)=\{(h_{out},h_{int})\in \clG(D_{x_0})\times
 \clG(\Si_g\setminus x_0)\}\,.
 $$
There are groups of holomorphic and smooth automorphisms acting on $\clT(\Si_g,G)$
 $$
\clG^{hol}\sim G(D_{x_0})\times G(\Si_g\setminus x_0)
 $$
with the action (\ref{lac}), and
 $$
\clG^{C^\infty}=\{(f_{int},f_{out})\,,~  f_{int}\in  \clG(D_{x_0})\,,~
f_{out}\in \clG(\Si_g\setminus x_0)\,,~f_{out}=\hat w f_{int}\hat w^{-1}\,.
 $$
with the action (\ref{rac}).
Therefore starting with the space of pairs $\clT(\Si_g,G)$
we obtain the moduli space in Dolbeault (\ref{hb}) or in the Double Coset (\ref{dcc2})
descriptions:
 $$
\begin{array}{ccccc}
   &  & \clG(D_{x_0})\times \clG(\Si_g\setminus x_0) &  &  \\
  \clG^{hol}& \swarrow &  & \searrow &\clG^{C^\infty} \\
  &&&&\\
{\rm  Dolbeault}  &  &  &  &{\rm Double~Coset }\\
&&&&\\
 \clG^{C^\infty}& \searrow & &\swarrow &\clG^{hol}  \\
 & & Bun_{\hat w}(\Si_{g}G) & &
\end{array}
 $$
 Let $P$ be a parabolic subgroup and $f_{int}=P+O(t,\bar t)$. Then we come to the Dolbeault
description of the moduli space of bundles with the $qp$-structures (\ref{mqp}).
This construction can be generalized to multi-point case leading to $Bun(\Si_g,\bfx,G)$.
$\Box$

 \bigskip

Now pass to a  coarse scale of the moduli space $Bun(\Si_{g,n},x_0,G)$.
Consider the monodromies of the transition functions $g(t)\in G(D^\times)$ around
$x_0$: $\,g(t\exp(2\pi\imath)=\zeta g(t)$. For the component enumerated by $\hat w=wt^\ga$
the monodromy is $\zeta=\exp(2\pi\imath\ga)$. If $\ga\in Q^\vee$ the monodromy is trivial
$\zeta=1$. For general $\ga\in P^\vee$, $\zeta=\exp(2\pi\imath\ga)$ is an element of the
center $\clZ(\bG)$ (\ref{center}).
Then passing to the quotient $P^\vee/Q^\vee\sim\clZ(\bG)$ in (\ref{dco1}) we come to
the decomposition
 \beq{dco2}
Bun(\Si_{g,n},x_0,G)=
\bigcup_{{\zeta}\in\clZ(\bG)}Bun_{\zeta}(\Si_{g,n},x_0,G)\,.
 \eq
 If $\zeta=1$ then $g(t)$ can serve as a transition function for the $\bG$-bundle $E_{\bG}$.
In this way non-trivial monodromies are obstructions to lift
$G$-bundles to $\bG$-bundles.

We can write the similar decomposition  for other quotient groups
$G$ (\ref{fgl}).
In the cases $A_{n-1}$ ($n=pl$ is non-prime), and $D_n$ the center
$\clZ(\bar G)$ has non-trivial subgroups $\clZ_l\sim\mu_l=\mZ/l\mZ$.
Then there exists the quotient groups
 \beq{fgl2}
G_l=\bG/\clZ_l\,,~~~G_p=G_l/\clZ_p\,,~~~G^{ad}=G_l/\clZ(G_l)\,,
 \eq
where $\clZ(G_l)$ is the center of $G_l$ and
$\clZ(G_l)\sim\mu_p=\clZ(\bar G)/\clZ_l$. The group $\bar
G=Spin_{4n}(\mC)$ has a non-trivial center
 $$
 \clZ(Spin_{4n})=(\mu^L_2\times\mu^R_2)\,,~~\mu_2=\mZ/2\mZ\,,
 $$
and the subgroups of $Spin_{4n}(\mC)$ are described by the diagram (\ref{spin4}).
Therefore, in general case the following monodromies are obstructions to lift the bundles
 $$
\begin{array}{l}
\zeta\in \clZ(\bG) -{\rm ~obstructions ~to~ lift~}
 E_{G^{ad}}-{\rm bundle~to~}E_{\bG}-{\rm bundle}\,,\\
  \zeta\in \clZ_l -{\rm ~obstructions ~to~ lift~}
 E_{G_l}-{\rm bundle~to~}E_{\bG}-{\rm bundle}\,, \\
    \zeta\in \clZ(G_l) -{\rm ~obstructions ~to~ lift~}
 E_{G^{ad}}-{\rm bundle~to~}E_{G^l}-{\rm bundle}\,.
  \end{array}
 $$
To define the cohomological interpretation of the obstructions
we write three exact sequences
 $$
\begin{array}{l}
  1\to\clZ(\bG))\to\bG(\clO_\Si)\to G^{ad}(\clO_\Si)\to 1\,, \\
    1\to\clZ_l\to\bG(\clO_\Si)\to G_l(\clO_\Si)\to 1\,, \\
    1\to\clZ(G_l)\to G_l(\clO_\Si)\to G^{ad}(\clO_\Si)\to 1\,.
\end{array}
 $$
They  lead to the long exact sequences of cohomologies  with
coefficients in analytic sheaves
 \beq{coh1}
\to H^1(\Si_g,\bG(\clO_\Si))\to H^1(\Si_g,G^{ad}(\clO_\Si))\to H^2(\Si_g,\clZ(\bG))\sim\clZ(\bG))\to 0 \,,
 \eq
 \beq{coh2}
\to H^1(\Si_g,\bG(\clO_\Si))\to H^1(\Si_g,G_l(\clO_\Si))\to H^2(\Si_g,\clZ_l)\sim\mu_l\to 0\,,
 \eq
 \beq{coh3}
\to H^1(\Si_g,G_l(\clO_\Si)\to H^1(\Si_g,G^{ad}(\clO_\Si))\to H^2(\Si_g,\clZ(G_l))\sim\mu_p\to 0\,.
 \eq
The first cohomology group $H^1(\Si_g,G(\clO_\Si))$ defines the tangent space to  $Bun(\Si_g,G)$.
The elements from $H^2$ are obstructions to lift bundles, namely
 $$
\begin{array}{l}
 H^2(\Si_g,\clZ(\bG)) -{\rm ~defines~obstructions ~to~ lift~}
 E_{G^{ad}}-{\rm bundle~to~}E_{\bG}-{\rm bundle}\,,\\
 H^2(\Si_g,\clZ_l) -{\rm~defines ~obstructions ~to~ lift~}
 E_{G_l}-{\rm bundle~to~}E_{\bG}-{\rm bundle}\,, \\
H^2(\Si_g,\clZ(G_l)) -{\rm ~defines~obstructions ~to~ lift~}
 E_{G^{ad}}-{\rm bundle~to~}E_{G^l}-{\rm bundle}\,.
  \end{array}
 $$
\begin{defi}
The images $\zeta(E_G)$ (as elements of $H^1(\Si_g,G(\clO_\Si))$) in
$H^2(\Si_g,\clZ)$ are called the characteristic classes of
$E_G$-bundles.
\end{defi}
Now consider the multipunctured case. Attach to the marked points
the transformations of the affine Weyl group $W_P$ (\ref{q25})
  $$
\vec{\hat w}=(w_1t^{\ga_1},\ldots,w_nt^{\ga_n})\,,~~\ga_a\in P^\vee\,.
  $$
Using the decomposition (\ref{dco1}), define the sector of the moduli space (\ref{dco})
   $$
  Bun_{\vec\ga}(\Si_{g,n},\bfx,G^{ad})=G^{ad}(\Si_g
  \setminus \bfx) G^{ad}(D^\times)\vec{\hat w}/G^{ad}(D)
 $$
and the decomposition of the moduli space
 \beq{btr}
Bun(\Si_{g,n},\bfx,G^{ad})=
\bigcup_{{\vec\ga}\in\oplus P^\vee}Bun_{\vec\ga}(\Si_{g,n},\bfx,G^{ad})\,.
 \eq
In the many points case  the sector
$Bun_{\zeta}(\Si_{g,n},\bfx,G^{ad})$ as in (\ref{dco2}) is defined by the local transition functions $g(t_a)$
with monodromies $\zeta_a$ around the points $x_a$, such that
$\zeta=\Pi_{a=1}^n\zeta_a\in\clZ(\bG)$. It means that in (\ref{btr})
 we identify components $Bun_{\vec\ga}$ and $Bun_{\vec\ga'}$,
 if $\sum_{a=1}^n(\ga_a-\ga'_a)\in Q^\vee$. Then
we come to the decomposition of the moduli space (\ref{dco2}) into
topological sectors:
     \beq{grm}
Bun(\Si_{g,n},\bfx,G^{ad})=
\bigcup_{{\zeta}\in\clZ(\bG)}Bun_{\zeta}(\Si_{g,n},\bfx,G^{ad}) \,.
  \eq


 \subsection{Flat bundles}

 Consider first  the double coset construction for the moduli space of  Higgs bundles
 $T^*Bun(\Si_{g,n},\bfx,G^{ad})$.
The Higgs bundles are defined as the set of  pairs
 $(\Phi_a,g_a)$, $(a=1,\ldots,n)$,
  where $g_a\in L_a(G)$ and $\Phi_a\in Lie^*(L_a(G))\otimes dt_a$ are the Higgs fields.
It is a symplectic manifold with the symplectic form
 \beq{s11}
 \om=\sum_a\oint_{\G_a}\lan\de (\Phi_ag_a^{-1})\wedge\de g_a\ran\,,
 \eq
where the contour $\G_a\subset D_a^\times$.
  The action of $\clG_{in}=G(D)$ and $\clG_{out}=G(\Si_g\setminus\bfx)$ on these pairs
  is lifted to   $T^*Bun(\Si_{g,n},\bfx,G^{ad})$ as
 \beq{gah}
\clG_{a,int}\,:\,\Phi_a\to f_{a,int}^{-1}\Phi_af_{a,int}\,,~~g\to gf_{a,int}\,,
 \eq
 \beq{gah1}
\clG_{out}\,:\,\Phi_a\to \Phi_a\,,\,,~~g\to f_{out}g\,.
 \eq
By the symplectic reduction we define the moduli space
 $$
\clG_{out} \setminus\setminus\bigcup_{a=1}^n(\Phi_a,g_a)//\clG_{a,int}
 $$
It can be proved that this double symplectic quotient is isomorphic to
 $T^*Bun(\Si_{g,n},\bfx,G)$. Since the moduli  space
 of the Higgs bundles  $Bun(\Si_{g,n},\bfx,G)$ is the unions of sectors (\ref{grm}),
 the cotangent bundle is also the union
 $$
T^*Bun(\Si_{g,n},\bfx,G)=
\bigcup_{{\vec\ga}\in\oplus P^\vee}T^*Bun_{\vec\ga}(\Si_{g,n},\bfx,G^{ad})
=\bigcup_{\zeta\in\clZ(G)}T^*Bun_{\zeta}(\Si_{g,n},\bfx,G)\,.
 $$
Our main interest is the moduli space of flat bundles. We will prove
that is the unions
    \beq{grm1}
FBun(\Si_{g,n},\bfx,G)
=\bigcup_{{\vec\ga}\in\oplus P^\vee}FBun_{\vec\ga}(\Si_{g,n},\bfx,G^{ad})
=\bigcup_{\zeta\in\clZ(G)}FBun_{\zeta}(\Si_{g,n},\bfx,G)  \,.
  \eq
As above, we  replace the space of connections $Conn(\Si_g,G)$
 (\ref {stp}), by
 the set of pairs $\clR$ on $G(D^\times)$
  \beq{msfb}
 \clR= \{((\p_{t_a}+X_a)\otimes dt_a,g_a))\,,~a=1,\ldots,n\}\,.
 \eq
where $g_a\in G(D_a^\times)$ (\ref{lre}) and can have a non-trivial
monodromy, $X_a\in \gg(D_a^\times)=\gg\otimes C[t_a^{-1},t_a]]$
(\ref{la}). The component
 $(\p_{t_a}+X_a)\otimes dt_a$ belongs to
the principal homogeneous space $PH/T^*G(D_a^\times)$ over the cotangent bundle
$T^*G(D_a^\times)$.

The form $\om$ (\ref{G.1}) in this parametrization can be rewritten as (compare with (\ref{s11}))
   \beq{sfdb}
  \om=\sum_a\oint_{\G_a}\lan\de (X_ag_a^{-1})\wedge\de g_a\ran+
  \oh\oint_{\G_a}\lan g_a^{-1}\de g_a\wedge\p_t (g^{-1}\de g_a)\ran\,.
   \eq
  \begin{rem}
In fact, the pairs $((\p_{t_a}+X_a)\otimes dt_a,g_a))$ should be
replaced by the pairs $((\ka_a\p_{t_a}+X_a)\otimes
dt_a,(g_a,\la_a)))$, where $\la_a$ is a central extension of the
loop group $G(D_a^\times)$ and $\ka_a\in\mC$ are the dual co-central
extensions \cite{Ka}. In this case $\om$ acquires additional terms
$\sum_a\de\ka_a\wedge\de\ln\la_a$.  It can be proved that these
scalar fields are non-dynamical and do not contribute to the
monodromy preserving equations. For this reason will not consider
the central extensions $\la_a$. Nevertheless, we leave the parameter
$\ka$ because in the limit $\ka\to 0$ in Section \ref{idi} we come
to integrable systems.
\end{rem}
 The symplectic form (\ref{sfdb}) is invariant under the gauge transformations
from $\clG_{a,int}= L_a^+(G)$ (\ref{lpl}) and
 $\clG_{out}=G(\Si_{g}\setminus \bfx)$
 \beq{ag}
\clG_{a,int}\,:\,X_a\to Ad^*_{f_{a,int}}(X_a)=
 f_{a,int}^{-1}\p_{t_a}f_{a,int}  +f_{a,int}^{-1}X_af_{a,int}\,,~~g_a\to g_af_{a,int}\,,
 \eq
 \beq{gg}
\clG_{out}\,:\,,X_a\to X_a\,,\,,~~g_a\to f_{out}g\,.
 \eq
They are generated by the Hamiltonians
 \beq{hint}
F_{a,int}=\oint_{\G_a}\lan\ep_{a,int},X_a\ran\,,~~
F_{out}=\sum_a\oint_{\G_a}\lan\ep_{out},(gX_ag^{-1}-\p_{t_a} gg^{-1})\ran\,.
 \eq
Here $\ep_{out}\in Lie(G(\Si_{g}\setminus \bfx)$ and
 \beq{lin}
\ep_{a,int}\in Lie(L_a^+(G))\,,~~\ep_{a,int}=x_{a,0}+t_ax_{a,1}+\ldots\,,~x_{a,0}\in\gp_a
=Lie(P_a)\,.
 \eq


\subsubsection{$\clG_{int}$ action}

Let us repeat the finite-dimensional construction of the coadjoint
orbits (\ref{sfcb})-(\ref{kkf}). The moment map corresponding to the
$\clG_{a,int}$-action is equal to $\mu_{int}=\sum_a\mu_a$,
$\mu_a=Pr|_{Lie^*(L_a^+G)}(X_a)$, where  the dual space is defined
as
 $$
 Lie^*(L_a^+G)=\{y_{a,0}t_a^{-1}+y_{a,1}t^{-2}_a+\ldots\,,~y_{a,0}\in\gp^*_a\}\,.
 $$
From (\ref{psa}) and (\ref{pns}) we find that $\gp^*_a=\gg'_a\oplus\ti\gh^*_a\oplus\gn_a^-$.
Take the moment value as
 $$
\mu_a =Pr|_{Lie^*(L_a^+G)}(X_a)=\nu_at_a^{-1}\,,~~\nu_a\in\gs_a^*-{\rm Levi~subalgebra~}
(\ref{psa})\,.
 $$
In other words, the field $A$ has simple poles and
 \beq{spo}
X_a(t_a)= (\nu_a+\xi_{a,-1})t_a^{-1}+\xi_{a,0}+O(t_a)=\nu_a t_a^{-1}+\xi(t_a)\,,
 ~~\xi_{a,-1}\in\gn^+\,.
 \eq
It means that $\xi(t_a)$ is an arbitrary element of the space $\gb(t_a)=\gn^+t_a^{-1}+
\gg\otimes\mC[t_a]]$.
Note, that $\clG_{a,int}$ preserves the term $\nu_at_a^{-1}$ and for $\nu_a\neq 0$ acts
freely on $\gb(t_a)$:
  $$
Ad^*_{L_a^+(G)}(X_a(t_a))=\nu_a t_a^{-1}+\xi'(t_a)\,,~~
\xi'(t_a)=\xi'_{a,-1}t_a^{-1}+\xi'_{a,0}+O(t_a)\,.
  $$
%
%
 From (\ref{ag}) and (\ref{spo}) we find that the symplectic quotient
is defined as the set of pairs
 $$
(g_a(t_a),\nu_at_a^{-1}+\xi(t_a))\,,~~(a=1,\ldots,n)
 $$
with the equivalence relation
 \beq{era}
(g_a(t_a),(\nu_at_a^{-1}+\xi(t_a)))\sim(g_a(t_a)f(t_a),Ad^*_{f(t_a)}(\nu_at_a^{-1}+\xi(t_a)))
\,,~~f(t_a)\in G(D_a)\,.
 \eq
%
Let us fix the gauge of the $G(D_a)$-action in (\ref{era}) by
putting $\xi(t_a)=0$. The Levi subgroup $L_a\subset \clG_{a,int}$
(\ref{les})  preserves the gauge
 $$
L_a=\{f(t_a)\in G(D_a^\times)\,|\,(Ad^*)^{-1}_{f(t_a)}\nu_a=\nu_a\}\,.
 $$
It follows from (\ref{era}) that $g_a(t_a)$ is defined up to multiplication from right by
$L_a$. Therefore, the reduced space in this case is  the coadjoint orbit
 $$
\clO_{a}^{aff}=G(D_a^\times)/L_a=\{g(t_a)\in G(D_a^\times)\,|\,X_a(t_a)=
(Ad^*)^{-1}_{g(t_a)}\nu_at^{-1}_a
 $$
 $$
=-\p_{t_a}g(t_a)g(t_a)^{-1}+g(t_a)\nu_a t^{-1}_a g^{-1}(t_a)\}\,.
 $$
This space is the principal homogeneous space over the cotangent
bundles to the affine flags varieties $PH/T^*Flag^{aff}_a$,
($Flag^{aff}_a\sim G(D_a^\times)/G(D_a)$).
%
%
The form (\ref{sfdb}) on $\clO_{p_a}$ becomes
 \beq{fao}
\om_{\clO^{aff}_{a}}=\oint_{\G_a}\lan\de (\nu_at_a^{-1}g_a^{-1})\wedge\de g_a\ran+
  \oh\oint_{\G_a}\lan( g_a^{-1}\de g_a)\wedge \p_{t_a}(g^{-1}\de g_a)\ran\,.
 \eq
In this way upon the symplectic
reduction by $\clG_{int}=\prod_a\clG_{a,int}$ we come from $\clR$ (\ref{msfb})
to the space
 \beq{afor}
\bigcup_{a=1}^n\clO^{aff}_{a}=\clR//\prod_a\clG_{a,int}\,.
 \eq

\subsubsection{$\clG_{out}$ action}

The $\clG_{out}$ action on the symplectic space (\ref{fao}), (\ref{afor})
defines the moment
 \beq{muo}
\mu_{out}=Pr_{Lie^*(G(\Si_{g}\setminus x_0)})X(t)=0\,,~~
X(t)=g\frac{\nu}tg^{-1}-\p_{t} gg^{-1}\,.
 \eq
Let $G(t,z)$ be the Cauchy kernel on the punctured curve
$\Si_{g}\setminus x_0$. It is a function in the first variable and
one-form in the second variable. It is regular outside the diagonal,
at which it has a first order pole. It exists because the punctured
curve is affine.
\begin{predl}{\label{ap}}
Let $\G_0$ be a small contour around $x_0$.
The form
 \beq{ta}
\tilde{A}(z)={\rm Res}_{t=0}(X(t)G(t,z))\,,
 \eq
where $X(t)$ satisfies (\ref{muo}), is holomorphic on the $\Si_{g}\setminus \G_0$.
Its asymptotic
 at $z\to \G_0$ coincides
  with $X(t)$.
\end{predl}
\emph{Proof}\\
   Let us calculate the $k$-th coefficient of the  Laurent expansion $\tilde{A}$.
It is equal to
  $$
\frac1{2\pi i}\oint_{\G_0} s^{-k-1} \tilde{A}(s)=
\frac1{2\pi i} \oint_{\G_0} s^{-k-1}({\rm Res}_{t=0}(X(t)G(t,s)))=
{\rm Res}_{t=0}\left(X(t)\frac1{2\pi i}\oint s^{-k-1}
G(t,s)\right)\,.
  $$
Here the interchanging of the integrations and
 taking residue is correct.  We  take the residue at zero with respect to
 $t$, hence we calculate the integral under assumption $|t|<|s|$.
 The integrand
has singularities at $s=t$ and $s=0$. The contribution from the
first singularity equals
 $t^{-k-1}$ and the contribution from the second
is equal to $\frac1{k!} (\partial_s)^kG(t,s)_{s=0}$. The residue of
product of $\mu_{out}$ with the first summand give the $k$-th
coefficient of $\mu_{out}$, the residue of the product with second
vanishes, since   $\frac1{k!} {\partial^k G(p,s)}/{\partial
s^k}|_{s=0}$ is regular outside puncture. $\Box$

Define on $\Si_g\setminus x_0$ the holomorphic connection
  $$
A_{out}=h_{out}(\p+\ti A)h^{-1}_{out}=   h_{out}\ti Ah^{-1}_{out}-\p_zh_{out}h^{-1}_{out}\,,
  $$
where $h_{out}$ is defined from the decomposition (\ref{dcr}). On
$D^\times_{x_0}$ \beq{aou}
A_{out}=h_{out}X(t)h^{-1}_{out}-\p_th_{out}h^{-1}_{out}\,, \eq where
$X(t)$ is from (\ref{muo}).
%
It follows from (\ref{bad}) and Proposition \ref{ap} that on $\Si_g\setminus x_0$ the curvature vanish
 \beq{kri}
F(A_{out},\bA_{out})=[\p+A_{out},\bp+\bA_{out}]=h_{out}[\p+\ti A,\bp]h^{-1}_{out}=0\,.
 \eq
 Define $A_{int}$ as (see (\ref{bad}) and (\ref{muo}))
  $$
 A_{int}=h_{int}\nu t^{-1}h^{-1}_{int}-\p_t h_{int}h^{-1}_{int}\,.
  $$
 It follows from (\ref{aou}) that
  $$
 A_{out}=\hat w A_{int}\hat w^{-1}\,.
  $$
 The curvature on $D_{x_0}$  takes the form
  $$
 F(A_{int},\bA_{int})=(h_{int}\nu h^{-1}_{int})\bp t^{-1}\,.
  $$
 Since $h_{int}(t,\bar t)=h_0+O(t,\bar t)$ we obtain
  $$
 F(A_{int},\bA_{int})=\bfS\de(t)\,,~~\bfS=h_0\nu h_0^{-1}\,.
  $$
 Thus, starting with $Conn(\Si_g,G)=\clR$ (\ref{msfb}) we come to the flat bundles
 with the Fuchsian singularity.

%


\subsection{Hecke transformations}

The Hecke transformation is a  singular gauge transformation that intertwines sections of
bundles corresponding to different components defined by the Weyl elements in  (\ref{btr}),
or elements of the center in (\ref{grm}).

 To avoid complications we consider  a curve with one marked point $\Si_g\setminus x_0$.
Consider a holomorphic bundle $E_G$ defined by transition function
$g(t)$ in the double coset construction (\ref{btr}) or (\ref{grm}).
Replace the transition function $g(t)$ by  $g(t)h(t)$, where
$h(t)\in L(G)$ can have a non-trivial monodromy around $t=0$. Due to
(\ref{dco}), $\,h(t)$ is defined up to the multiplication from the
right by
 $f(t)\in L^+(G)$. As it was explained in Section \ref{cc} this transformation can change the
 characteristic class of $E_G$.
It follows from (\ref{PS2}) that we can take
 as a representative
of this double coset an element $h(t)=\hat w\in W_{t(G)}$.
 \beq{mtm}
  g(t)\to g(t)\hat w\,,~~\hat w=w t^{\ga}\,,~~
t^{\ga}=\bfe(\ln\,(t\ga)))\,,~~\ga\in t(G)\,,
  \eq
 where $t(G)$ is the
coweight lattice (\ref{coch1}). The monodromy of $t^{\ga}$ is
$\exp\,-(2\pi\imath\ga)$. Since $\lan\al,\ga\ran\in\mZ$
  for any $\bfx\in\gg\,$ we have Ad$_{\exp\,-(2\pi \imath\ga)}\bfx=\bfx$.
  Then $\exp\,-(2\pi\imath\ga)$ is
 an element of $\clZ(\bG)$ (\ref{center}). If the transition matrix $g(t)$
  defining  $E_G$ has a trivial monodromy, the new transition matrix (\ref{mtm}) acquires
   a non-trivial monodromy.  In this way we come to a new bundle $\ti E_G$
    with a non-trivial characteristic class $\zeta(\ti E_G)$.
The bundle $\ti E_G$ is called \emph{the modified bundle}.
 Another name of the modification is \emph{the Hecke transformation}.
It is defined by the new transition matrix (\ref{mtm}). If $\ga\in
Q^\vee$, then $\zeta=1$ and the modified bundle $\ti E_G$  has the
same characteristic class as $E_G$. It follows from (\ref{PS2}) and
(\ref{affl}) that the space of Hecke modifications of type $\ga$
coincides with the Schubert cell $C_{\hat w}$ (\ref{spm}). Its
dimension is equal to $l(\hat w)$ (\ref{dsm}).

   For the transformation of sections we use
 the notation
 \beq{ksi}
 \Xi(\zeta)\,:\,\G(E_G)=\G(E(\zeta=1))\to\G(\ti E_G)= \G( E(\zeta))\,.
 \eq
If $d+\clA$ is a connection on $E_G$ and $d+\ti\clA$ on $\ti E_G$ then
\beq{moc}
(d+\ti\clA)\,\Xi(\zeta)=\Xi(\zeta)\,(d+\clA)\,.
\eq

\bigskip
We formulate some restriction on the choice $\ga\in P^\vee$ for the bundles with the $qp$-structure
at $x_0$.
 Assume that the modification  preserves the $qp$-structure at the marked point $x_0$. In this case
 $\hat w$ should normalizes $G(D)=L^+(G)$ in the decomposition (\ref{dco1}):
  \beq{norm}
 \hat w^{-1}L^+(G)\hat w=L^+(G)\,.
  \eq
  Remind that $\ga$ defines the parabolic subalgebra $\gp_\ga\subset\gg$ (\ref{pist}) and,
  in this way, the $qp$-structure.
\begin{predl}
Let $\ga\in\hat\Upsilon^\vee$ is an admissible fundamental coweight (\ref{afc}),
and $\gp_\ga=\gs+\gn^+$ is the admissible parabolic
subalgebra with respect to $\ga$ (\ref{adm}) in the decomposition $Lie(L^+(G))=\gp+t\gg+o(t)$. If
\beq{pqs}
Ad_w\gn^+=\gn^-\,,
\eq
 then $\hat w$ normalizes $L^+(G)$.
\end{predl}
\emph{Proof}\\
Let $x(t)\in Lie(L^+(G))$, $x(t)=a+b+t(a_1+b_1 +c_1)+o(t)$, where $a,a_1\in\gs$, $b,b_1\in\gn^+$
and $c_1\in\gn^-$. Then due to (\ref{adm})
 $$
Ad_{\hat w}=Ad_w\left((a+c_1)+ta_1+t^2b_1+\ldots\right)\,.
 $$
If $Ad^*_w\gn^+=\gn^-$ then $Ad_w(a+c_1)\in \gp=\gs+\gn^+$  and $=Ad_wb_1\in\gn^-$, then
$qp$-structure is restored.$\Box$

 For generic maximal parabolic subgroups
 such transformations do not exist. For the classical groups it exists for SL$(2n,\mC)$
 and the Levi subgroup S(GL$(n,\mC)\times$GL$(n,\mC))$, in the B, C cases and in the D case for
 $\ga=\varpi^\vee_1$ (see Table 4 in Appendix A).
However, if $P$ is not maximal $(P\subset P_\ga)$ then  the Hecke transformation can
preserve the $qp$-structure in more cases. For example if $P=B$ is a Borel subgroup
taking in (\ref{mtm}) $w\,:\,\al\to -\al$
we come to (\ref{norm}).
\bigskip

The Hecke transformations of the holomorphic bundles can be lifted
to the Hecke transformations of the flat bundles. In this case
instead of flag varieties attached to the marked points we have the
coadjoint orbits defined by the Levi subalgebras $\gs(\Pi')$
(\ref{psa}). For this reason, instead of (\ref{pqs}) we need
 $$
Ad_{ \hat w}\gs=\ti\gs\,,
 $$
  Here $\hat w=wt^\ga$ and $\ga\in\hat\Upsilon^\vee$. Then $\ti\gs=Ad_w\gs$. Though
  this transformation changes the Levi subgroup $L\to\ti L$, it preserves the orbit $\clO=G/L$.
  Therefore, for  flat bundles we only need $\ga\in\hat\Upsilon^\vee$.




\section{Hamiltonian approach to the isomonodromy problems}
\setcounter{equation}{0}

\subsection{Deformation of complex structures of curves}

 The complex structure on $\Si_{g}$ is defined
 by the $\bp$ operator. Consider its  deformations defined by the change of variables
  \beq{dc}
w=z-\ep(z,\bz)\,,~~\bar  w=\bz-\overline{\ep(z,\bz)}\,,
  \eq
where $\ep(z,\bz)$ is small. Up to a common multiplier the partial derivatives assumes the form
  $$
\left\{
 \begin{array}{l}
\p_w=\p_z+\bar{\mu}\p_{\bz}\,,\\
\p_{\bw}=\p_{\bz}+\mu\p_z\,,
 \end{array}
\right.
  $$
where
  \beq{db}
\mu=\frac{\bp\ep}{1-\p\ep}\sim\bp\ep
  \eq
 is the Beltrami differential $\mu\in\Om^{(-1,1)}(\Si_{g})$.
We pass from the local coordinates  $(w,\bw)$ (\ref{dc}) to the chiral coordinates $(w,\ti w)$
 \beq{deco}
w=z-\ep(z,\bz)\,,~~\ti  w=\bz\,,
 \eq
 because the $\bar\mu$ dependence is nonessential in our construction.
  The pair $(w, \ti w)$ is just a pair of local coordinates on $\Si_g$. In these coordinates
 the partial derivatives assume the form
  \beq{nv}
\left\{
 \begin{array}{l}
\p_w=\p_z\,,\\
\p_{\ti w}=\p_{\bz}+\mu\p_z\,.
 \end{array}
\right.
  \eq
 What is important is that $\p_{\ti w}$ annihilates  holomorphic functions  $\p_{\ti w}f(w)=0$.
A smooth function $w(z,\bz)\in C^{\infty}(\Si_g)$ defines a global diffeomorphism
of $\Si_g$.
We say that the Beltrami differential $\mu(z,\bz)$ is equivalent to
$\mu'(z,\bz)$, if
$\mu'(z,\bz)=\mu(w(z,\bz),\bw(z,\bz))$. The equivalence relations in
 $\Om^{(-1,1)}(\Si_{g})$ under the action
 of $Diff_{C^\infty}(\Si_g)$, is the
moduli space $\gM(\Si_{g,n})$ of complex structures on $\Si_{g}$.
 The tangent space to the
moduli space is the Teichm\"{u}ller space $\gT_{g}\sim
H^1(\Si_{g},\G)$, where $\G\in T\Si_g$. From the Riemann-Roch
theorem one has
  \beq{2.6}
\dim(\gT_{g})=3(g-1)\,.
  \eq
Let $(\mu_1^0,\ldots,\mu_l^0)$ be a basis in the vector space
$H^1(\Si_{g},\G)$. Then
   \beq{3}
 \mu=\sum_{l=1}^{3g-3}\tau_l\mu^0_l\,,
   \eq
  where the local coordinates $\tau_l$ will play the role of times in the isomonodromic
  deformation problem.


  \subsubsection*{Moving marked points}

  Consider the moduli space $\gM(\Si_{g,n})$ of complex structures of
  curves with marked points $\Si_{g,n}$. This space is
 foliated over the moduli space of complex structures of compact curves $\gM(\Si_{g})$
  with fibers $\clU\subset\mC^n$ corresponding to the moving  marked points.
 The moduli space $\gM(\Si_{g,n})$ is the classes of equivalence relation in
 the space of differentials $\Om^{(-1,1)}(\Si_{g,n})$
 under the action of the group of diffeomorphisms   $Diff_{C^\infty}(\Si_{g,n})$
  vanishing at the marked points.

 Consider local coordinates on a fiber.
  Let $(z,\bz)$ be the local coordinates in
the neighborhood of the marked points $x^0_a$ $(a=1,\ldots,n)$ and
${\cal U}_a$ are neighborhoods of $x^0_a$
 such that $\clU_b\cap{\cal U}_a=\emptyset$.
Define the $C^\infty$ function
  \beq{chi}
\chi_a(z,\bar z))=\left\{
 \begin{array}{cl}
1,& z\in {\cal U}_a'\subset {\cal U}_a\\
0,& z\not\in {\cal U}_a.
 \end{array}
\right.
  \eq
The moving points $(x_a^0\to x_a)$ correspond to  the following local deformation of
local coordinates (\ref{dc})
  \beq{mpt}
  w=z-\sum_{a=1}^n\ep_a(z,\bz)\,, ~~
\ep_a(z,\bz)=-t_a\chi_a(z,\bar z)+\sum_{j>0}t_a^{(j)}(z-x_a^0)^j\chi_a(z,\bar z)\,,~~t_a=x_a-x^0_a\,.
  \eq
The action of  $Diff_{C^\infty}(\Si_{g,n})$ allows one to put $t_a^{(j)}=0$ for $j>0$.
Thus, in general we have only $n$ times $t_a$.
The part of the Beltrami differential related to the marked points assumes the form
  \beq{mp}
\mu=\sum_{a=1}^nt_a\mu_a^{(0)}\,,~~~
\mu_a^{(0)}=\bp\chi_a(z,\bz)\,.
  \eq
 Thus the local coordinates on $\gT(\Si_{g,n})$ are
  \beq{lct}
 \gT(\Si_{g,n})=\{(\tau_1,\ldots,\tau_{3(g-1)},t_1,\ldots,t_n)\}\,,
  \eq
and
  \beq{dmp}
\dim(\gT(\Si_{g,n}))=3(g-1)+n\,.
  \eq


\subsection{Equations of motion and the isomonodromy problem}

Fix a complex structure on $\Si_{g,n}$ and replace the connections (\ref{stp}) by the pair
 $$
((\ka\p_z+A)\otimes dz,(\bp+\bA')\otimes d\bz)\,.
 $$
Here we following P. Deligne introduce the notion of
$\ka$-connections, where $\ka$ is a small parameter  (in a certain
sense it resembles the Planck constant). The moduli spaces of
$\ka$-connections was investigated in \cite{Arinkin}. The
$\ka$-connections allow one to pass in the quasi-classical limit
 to the Higgs field $\Phi=\lim_{\ka\to 0}(\ka\p_z+A)\otimes dz\in\Om^0(\Si_g,Lie G\otimes K)$,
where $K$ is a canonical class on $\Si_g$. In this limit the
monodromy preserving equations becomes equations of motion for
integrable systems of Hitchin type. The corresponding linear problem
for the latter systems is a Isospectral Problem instead of the
Isomonodromy Problem in the former case. We consider this procedure
below in detail.

Consider the polarization (\ref{deco}) in the deformed coordinates.
  $$
(\ka\p_w+A)\otimes dw\,,(\p_{\ti w}+\bA)\otimes d\ti w\,.
  $$
The component of connection $\bA'$ in the deformed coordinates is defined as
  \beq{nba}
 \bA'=\bA-\f1{\ka}\mu A\,,~~ (\bp+\mu\p+\bA)\otimes d\ti w =(\p_{\bar w}+\bA)\otimes d\ti w\,.
   \eq
 The form $\om$ (\ref{G.1}) is rewritten as
  $$
 \om=\int_{\Si_g}(\de A\wedge \de\bA)-\f1{\ka}\int_{\Si_g}(A,\de A)\de\mu\,.
  $$
 The form $\om$ can be considered as the  differential of the Poincar\'e-Cartan one-form
  $(\om=$$\om^{PC}=\de\vartheta^{PC})$ \cite{Ar} on the extended space $(A,\bA,\mu$.
 In the canonical coordinates on the phase space $(p_j,q_j)$ and for
 Hamiltonians $H_l(\vec p,\vec q;t_1,\ldots,t_n)$ it takes the form
  $$
 \vartheta^{PC}=\sum_jp_j\de q_j-\sum_lH_l(\vec p,\vec q;t_1,\ldots,t_n)\de t_l\,.
  $$
The connections $(A,\bA)$ play the role of the canonical coordinates
on $Conn_{(\Si_g,G)}$, while the second term can be considered as
differentials of quadratic Hamiltonians pairing with the
corresponding times. More concretely, taking into account (\ref{3}),
we rewrite $\om$  as
   \beq{2}
 \om^{PC}=\om_0-\f1{\ka}\sum_{l=1}^{3g-3}\de H_l\de \tau_l\,,~~H_l=\oh\int_{\Si_g}(A,A)\mu^0_l\,,~~
 \om_0=\int_{\Si_g}(\de A\wedge \de\bA)\,.
   \eq
 The  Poincar\'e-Cartan form gives rise to the action functional
   $$
 S=\sum_{l=1}^{3g-3}\int_0^\infty\Bigl(
 \int_{\Si_g}(A,\p_l\bA)-\f1{2\ka}(A,A)\mu^0_l\Bigr)d\tau_l\,,~~
 (\p_l=\p_{\tau_l})\,.
   $$
 The equations of motion following from the action (or from the Hamiltonians) are
   \beq{em}
 \p_l\bA=\f1{\ka}A\mu^0_l\,,~~ \p_lA=0\,.
  \eq
These equation are the compatibility conditions for the following
linear system:
  \beq{2.15}
\left\{
 \begin{array}{ll}
  1. & (\ka\p_w+A)\psi=0\,, \\
  2. & \bigl(\p_{\ti w}+\bA\bigr)\psi=0\,, \\
  3. & \ka\p_{\tau_l}\psi=0\,,
 \end{array}
\right.
 \eq
where $\p_{\bar w}=\bp+\mu\p$ and
$\psi\in \Om^{(0)}(\Si_{g},{\rm Aut}\,E_G)$. The equations of motion (\ref{em}) for $A$ and
$\bA$ are the consistency
conditions of (1.$\&$3.) and (2.$\&$3.) in (\ref{2.15}). The monodromy of $\psi$ is  the transformation
  $$
 \psi\rar\psi {\cal Y}, ~~{\cal Y}\in {\rm Rep}(\pi_1(\Si_{g})\to  G).
  $$
The equation 3.(\ref{2.15}) means that the monodromy is independent on the times. The consistency condition of 1. and
2. is the flatness constraint (\ref{fla2}).
The form $\om^{PC}$ is defined on the bundle ${\cal P}(G)$ over the Teichm\"{u}ller space
${\cal T}_{g}$. Its fibers is the space of smooth connections $Conn_{\Si_{g},G}$
  \beq{pcu}
 \begin{array}{cc}
{\cal P}(G)& \\
\downarrow& Conn(\Si_{g},G)\\
\gT_{g} &
 \end{array}
  \eq

%
The linear equations (\ref{em}) describe a free motion on
$Conn(\Si_g,G)$. They become non-trivial on $FBun(\Si_g,G)$.


\subsection{Contribution of  marked points}

Assume that  the $(1,0)$ component (as in (\ref{cos})) of the
connection has simple poles at the marked points
 $$
Res A_{z=x_a}=\bfS^a\,,~~\vec\bfS=(\bfS_1,\ldots,\bfS_n)
 $$
The bundle ${\cal P}(G)$ is defined now  over the Teichm\"{u}ller
space $\gT_{g,n}$. The set of its fibers is the space of smooth
connections $Conn_{\Si_{g,n},G}$ with singularities at the marked
points. Therefore, the local coordinates on ${\cal P}(G)$
  are
  $$
(A,\bA,\vec\bfS,\bft)\,,~~
\bft=(\tau_1,\ldots,\tau_{3g-3};t_1,\dots,t_n)\,.
  $$
The bundle ${\cal P}(G)$ plays the role of the extended phase space
while the space $Conn(\Si_{g,n},G)$ is the standard phase space
 and $\bft$ is the set of times. The  form (\ref{2}) acquires additional
 terms:
  \beq{4}
\om^{PC}=\om^{Conn}- \f1{\ka}\Bigl(\sum_{l=1}^{3g-3}\de H_l\de
\tau_l+\sum_{a=1}^{n}\de H_a\de t_a\Bigr)\,,~~
\om^{Conn}=\om_0+\sum_{a=1}^n\om^{KK}_a\,,
  \eq
  $$
H_a=\oh\int_{\clU_a}(A,A)\bp\chi_a(z,\bz)\,.
 $$
The form $\om^{PC}$ degenerates on $(3g-3+n)\,$ vector fields
$D_s$:~ $\om(D_s,\cdot)=0$, where
  $$
D_l=
\p_{\tau_l}+\f1{\ka}\{H_l,\cdot\}_{\om_0}\,,~~(l=1,\dots,3g-3)\,,
~~~ D_a= \p_{t_a}+\f1{\ka}\{H_a,\cdot\}_{\om_a}\,,~(~a=1,\dots,n)\,.
  $$
The Poisson brackets corresponding to $\om_0$ are the Darboux brackets,  and those
corresponding to $\om^{KK}_a$ are the Lie
brackets. They are non-degenerate on the fibers.
 The vector fields $D_s$ define the
equations of motion for any function $f$ on ${\cal P}(G)$
  $$
\frac{df}{dr_s}=\p_{r_s}f +\f1{\ka}\{H_s,f\}\,,~~r_s=\tau_s\,,~{\rm
or}~r_s=t_s\,.
  $$
 Equations (\ref{em}) are particular examples.
The compatibility conditions are the so-called the Whitham equations
\cite{Krich2}:
  \beq{WE}
\ka\p_sH_r-\ka\p_rH_s+\{H_r,H_s\}=0.
  \eq
In the rational case
the $\tau$-functions
were investigated in \cite{JM1}. For the Painlev\'{e} I-VI
they were considered in \cite{O2}.

The Hamiltonians are the Poisson commuting quadratic Hitchin Hamiltonians.
It means that  there exists the generating
function (the tau-function)
 \beq{tf}
H_s=\frac{\p}{\p t_s}\log \tau, ~~
\tau=\exp\oh\sum_{s=1}^l\int_{\Si_{g,n}}<A^2>\mu_s.
 \eq



\subsection{Symplectic reduction}

Up to now the equations of motion, the linear problem, and the tau-function are
trivial.
The meaningful equations arise after imposing the corresponding constraints (\ref{fla1}) and
 the gauge fixing.
The form $\om^{PC}$ (\ref{4}) is invariant under the action of  the gauge group $\clG_P$ (\ref{clgp})
  \beq{gtra}
A\rar f^{-1}\ka\p f+f^{-1}Af,~~\bA\rar f^{-1}\bp f+f^{-1}\bA f\,.
  \eq
%
Let us fix $\bA$ in such a way that $\bL$ parameterizes generic
orbits of the $\clG_B$-action
  \beq{2.30}
\bA=f(\bp+\mu\p)f^{-1} +f\bL f^{-1}\,.
  \eq
Then the dual to $\bL$ field $L$ is obtained from $A$ by the same element $f$, defined up to the left multiplication
on elements preserving $\bL$
  \beq{2.31}
L=f^{-1}\ka\p f+f^{-1}Af\,.
  \eq
Thus, in local coordinates  the moment equation takes the form (see (\ref{fla1}))
  \beq{2.18}
(\bp+\p\mu)L-\ka\p\bL+[\bL,L]=2\pi i\sum_{a=1}^n\bfS^a\de(x_a)\,.
  \eq
%
The gauge fixing (\ref{2.30}) and the moment constraint define the
reduced space $FBun(\Si_{g,n},G)=\{L,\bL,{\bf S}\}$. It becomes
finite-dimensional (\ref{dhb1}) as well as  the bundle ${\cal P}(G)$
  $$
\dim{\cal P}(G)=\dim\,(FBun(\Si_{g,n},\bfx,G))+\dim(\gT_{g,n})=
2\dim(G)(g-1)+\sum_{a=1}^n\dim(\clO_a)+3g-3+n\,.
  $$
Due to the invariance of $\om$ (\ref{4}) it preserves its form  on $FBun(\Si_{g,n},\bfx,G)$
  \beq{qha}
 \om=\om^{FBun}-
\f1{\ka}\sum_{s=1}^{3g-3+n}\de H_s\de \tau_s\,,
~~H_s(L)=\oh\int_{\Si_{g,n}}\lan L^2\ran\mu_s^0\,
  \eq
   \beq{red}
\om^{FBun}=\om_0+\sum_{a=1}^n\om^{KK}_a\,,~~\om_0=\int_{\Si_{g,n}}\lan dL,d\bL\ran\,.
  \eq
Due to (\ref{2.18}), the system is no longer free because $L$
depends on the moduli space of flat bundles and ${\bf S}$. Moreover,
since $H_s$ depends explicitly on the times, the  system  is
non-autonomous.

Let $M_s=\p_sff^{-1}$.  It follows from (\ref{2.30}) and
(\ref{2.31}) that the equations of motion (\ref{em}) on the space
$FBun(\Si_{g,n},\bfx,G)$  take the form:
  \beq{2.19}
\left\{
 \begin{array}{ll}
  1. & \ka\p_sL-\ka\p M_s+[M_s,L]=0,~~s=1,\ldots,l\,, \\
  2. & (\bp+\p\mu)M_s=-L\mu_s^0\,. \\
 \end{array}
\right.
  \eq
The equations  1. from (\ref{2.19}) are the Lax equations. The
essential difference with the integrable systems is the additional
term $\ka\p M_s$.
On  $FBun(\Si_{g,n},G)$ the linear system (\ref{2.15}) assumes the
form:
  \beq{4.36}
\left\{
 \begin{array}{ll}
1.&(\ka\p+L)\psi=0\,,\\
2.&(\p_{\ti w}+\bL)\psi=0\,,\\
3.&(\ka\p_s+M_s)\psi=0,~~(s=1,\ldots,l_2)\,.
 \end{array}
\right.
  \eq
 The equations 1. and  2. from (\ref{2.19})  are consistency conditions for the linear problems (1.$\&$3.) and (2.$\&$3.),
  while (1.$\&$2.) is the flatness condition (\ref{2.18}).
The equations 3.(\ref{4.36}) provides the isomonodromy property of
the system 1.$\&$2. from (\ref{4.36}) with respect to variations of
the times $t_s$. We refer to the nonlinear equations (\ref{2.19}) as
the {\em Hierarchy of the Isomonodromic Deformations}.

The Hecke transformations act on the sections $\psi$ (\ref{ksi})
$\ti\psi=\Xi(\zeta)\psi$. Then we come from the system (\ref{4.36})
to the modified system
 $$
\left\{
 \begin{array}{ll}
1.&(\ka\p+\ti L)\ti\psi=0\,,\\
2.&(\p_{\ti w}+\ti\bL)\ti\psi=0\,,\\
3.&(\ka\p_s+\ti M_s)\ti\psi=0,~~(s=1,\ldots,l_2)\,.
 \end{array}
\right.
 $$
The transformation \beq{4.3612} \Xi(\zeta)\,:\,(L,M_s)\to (\ti L,\ti
M_s) \eq is a generator of discrete-time transformations. In this
sense it is the B\"acklund transformations for the monodromy
preserving equations.


\subsection{Isomonodromic deformations and integrable systems}\label{idi}

As it was remarked above, the isomonodromy preserving equations can
be considered as a deformation ({\it Whitham quantization}) of
integrable equations. The deformation parameter is  $\ka$. For the
Schlesinger systems this approach was developed in \cite{Tak1}.

Introduce the independent times $t^0_s=(\tau_l^0\,,x_a^0)$ as
$\tau_l=\tau_l^0+\ka t_l$, $t_a=\ka\ti t_a$ for $\ka\to 0$. It means
that $t_s=(t_l,\ti t_a)$ play the role of a local coordinates in the
neighborhood of the point $(t_l^0,x_a^0)\in \gT_{g,n}$. In this
limit  the equations of motion
 1.(\ref{2.19}) are the standard Lax equation
  \beq{lae}
\p_sL^{(0)}+[M^{(0)}_s,L^{(0)}]=0,~~s=1,\ldots,l\,,
  \eq
where $L^{(0)}=L(t^0_s)$, $(M^{(0)}_s=M_s(t^0_s))$.
The linear problem for this system is obtained from the linear
problem for the isomonodromy problem (\ref{4.36}) by the analogue of
the quasiclassical limit in quantum mechanics. Represent the
Baker-Akhiezer function in the WKB form
  \beq{WKB}
\psi=\Phi\exp\bigl(\frac{{\cal S}^{(0)}}{\ka}+{\cal S}^{(1)}\bigr)
 \end{equation}
and substitute (\ref{WKB})
 into the linear system (\ref{4.36}). If
$\p_{\bar z}{\cal S}^{(0)}=0$ and $\p _{\tau^0_l}{\cal S}^{(0)}=\p _{x_a^0}{\cal S}^{(0)}=0$,
 then the terms of order $\ka^{-1}$ vanish. In the quasiclassical
limit we put $\p{\cal S}^{(0)}=\la$. In the zero order approximation
we come to the linear system:
  \beq{lis}
\left\{
 \begin{array}{ll}
 i. & (\la+L^{(0)}(z,\tau_0))Y=0\,,\\
  ii. & \p_{\bz}Y=0\,,\\
iii. & (\p_{t_s}+M_s^{(0)}(z,\tau_0))Y=0\,.
 \end{array}
\right.
  \eq
The Baker-Akhiezer function $Y$ takes the form:
  $$
Y=\Phi e^{\sum_s t_s\frac{\p}{\p t^0_s}{\cal S}^{(0)}}\,.
  $$
The consistency condition of $i$ and $ii$  is the Lax equation (\ref{lae}).


\part{Isomonodromy problems on elliptic curves}


\section{Moduli space of flat bundles over elliptic curves}

\setcounter{equation}{0}

The moduli space of flat $G$-bundles plays the role of the phase
space in the isomonodromy problems. For trivial bundles this space
was described in \cite{Lo,BS,BS02} and for non-trivial bundles in
\cite{Sch,MF}.

\subsection{Holomorphic bundles over elliptic curves}

Hereinafter we consider the moduli spaces of
holomorphic $Bun(\Si_\tau,\bfx,G)$ and flat  $FBun(\Si_\tau,\bfx,G)$ bundles
over an the elliptic curve, described as the quotient
 $\Si_\tau\sim\mC/(\tau\mC\oplus\mC)$, $\,(Im\,\tau>0)$.
There are two generators $\rho_1$ and $\rho_\tau$ of the fundamental group
$\pi_1(\Si_\tau)$ corresponding to
the shifts $z\to z+1$ and $z\to z+\tau$ satisfying the relation
 \beq{fg}
\rho_1\rho_\tau\rho^{-1}_1\rho^{-1}_\tau=1\,.
 \eq
The sections of a $G$-bundle $E_G(V)$ over $\Si_\tau$ satisfy the
quasi-periodicity conditions
 \beq{s1}
\psi(z+1)=\clQ\, \psi(z)\,,~~~\psi(z+\tau)=\La\, \psi(z)\,,
 \eq
where the transition operators $(\clQ,\La)$ should respect (\ref{fg})
 \beq{gce2}
\clQ(z)\La^{-1}(z)\clQ^{-1}(z+\tau)\La(z+1)=1\,.
 \eq
 By definition, these transition operators define a topologically trivial bundle.
Let  $\zeta$ be an element of $\clZ( G)$. To come to a non-trivial
bundle replace (\ref{gce2}) by equation
 \beq{gce}
\clQ(z)\La^{-1}(z)\clQ^{-1}(z+\tau)\La(z+1)=\zeta\,.
 \eq
It follows from (\ref{cG}) that the r.h.s. can be represented as
$\zeta=\bfe(\ga)$, $(\bfe\,(x)=\exp\,(2\pi ix))$, where $\ga\in
P^\vee$ (\ref{cwl}). Then (\ref{gce}) takes the form:
 \beq{gce1}
\clQ(z)\La^{-1}(z)\clQ^{-1}(z+\tau)\La(z+1)=\bfe(\ga)\,.
 \eq
A bundle $\ti E$ is equivalent to $E$ if its sections $\ti s\psi$
are related to $\psi$ as $\ti \psi(z)=f(z)\psi(z)$, where $f(z)$ is
invertible operator acting in $V$. It follows from (\ref{gce1}) that
transformed transition operators have the form:
 \beq{gtt}
\clQ^f=f(z+1)\clQ f^{-1}(z)\,,~~~~\La^f=f(z+\tau)\La f^{-1}(z)\,.
 \eq
These transformations form the group of automorphisms $\clG$ (the gauge group) of the bundle $E_G$
over $\Si_\tau$.
It follows from \cite{NS} that the transition operators can be chosen as constants. Therefore, we
 come to
  \beq{gce3}
[\clQ,\La^{-1}]=\bfe(\ga)\,,~~([\clQ,\La^{-1}]=\clQ\La^{-1}\clQ^{-1}\La)\,.
 \eq
Thereby, $(\clQ,\La)$ form the projective representation of $\pi_1(\Si_\tau)$.
 The moduli space of stable holomorphic bundles over  $\Si_\tau$ can be defined as
     \beq{mhol}
    \fbox{$
 Bun_\zeta(\Si_\tau,G) =\{[\clQ,\La^{-1}]=\zeta\}/\clG\,,~~\zeta\in\clZ(G)\,.$}
     \eq
It was proved in \cite{Lo,BS,BS02} that $Bun_\zeta(\Si_\tau,G)$ for
$\zeta=1$ is isomorphic to a weighted projective space.
Let $G=\bG$ be a simply-connected group.
Let us fix a Cartan subgroup $\clH_{\bar G}\subset \bar G$. Assume that $\clQ$
is  semisimple, and therefore  it is conjugated to a generic element from $\clH_{\bar G}$.
By neglecting non-semisimple  transition operators we define a big cell $Bun^0(\Si_\tau,G)$
in the moduli space. Thus, we replace (\ref{mhol}) by
  \beq{mhol0}
    \fbox{$
 Bun_\zeta^0(\Si_\tau,G) =  \left\{[\clQ,\La^{-1}]=\zeta\right\}/\clG\,,~~\clQ\in\clH_{\bar G}\,.$}
     \eq
\begin{predl}\label{algeq}
Solutions of (\ref{mhol0}) have the following description.\footnote{See \cite{LOSZ1} for the proof.}\\
$\bullet$ The  element $\La$ has the form $\La=\La^0\bfe(\bfu)$,
where $\La^0$ is defined uniquely by the  coweight
 $\ga$ $\,(\La^0=\La^0_\ga)$. It is a special element from the Weyl group $W$
 preserving the extended coroot system  $\Pi^{\vee ext}=\Pi^\vee\cup\al^\vee_0$, and in
 this way is a symmetry of the extended Dynkin diagram. \\
 $\bullet\bullet$ Let $\ti\gh_0\subseteq\gh$ be a Cartan subalgebra, preserving by $\La^0$
 $\,(\la(\ti\gh_0)=\ti\gh_0$,   $\la=Ad_{\La^0})$. Then $\bfu\in\ti\gh_0$  \\
 $\bullet\bullet\bullet$ The element $\clQ$ has the form $\clQ=\clQ^0$, where
  \beq{ka}
\clQ^0=\exp\,2\pi i\varrho U\,,~~~ \varrho=\frac{\rho^\vee}{h}\in\gH\,,
  \eq
  where $h$  is the Coxeter number,
 $\rho^\vee=\oh\sum_{\al^\vee\in (R^\vee)^+}\al^\vee$ and $U$ commutes with $\La^0$.
 \footnote{The first statement can be found in \cite{Bou} (Proposition 5 in VI.3.2).}
 \end{predl}

 \begin{rem}
For $Spin(4n)\,$ the center $\clZ(Spin(4n))\sim\mu_2\times\mu_2$ has
two generators $\zeta_1$ and $\zeta_2$
  corresponding to the fundamental weights
$\varpi_a$, $\varpi_b$ of the left and the right spinor representations. Arguing as above we will find
two solutions $\La_a$ and $\La_b$ of (\ref{mhol0}), while $\clQ$ is the same in the both cases.
\end{rem}
\begin{rem}
If $\xi\in Q^\vee$ then $\zeta=Id$.
Then the transformation $\La_0$ is trivial $\La_0=Id$  and $\ti\gh_0=\gh$. In this case the bundle has
a trivial characteristic class, but has holomorphic moduli  defined by the vector $\bfu\in\gh$.
\end{rem}


\subsection{Decomposition of Lie algebras}\label{decS}

Let us take an element $\zeta\in\clZ(\bar G)$ of order $l$ and the corresponding $\La^0\in W$ from
Proposition\,\ref{algeq}.  Then $\La^0$ generates a cyclic group
$\mu_l=(\La^0,(\La^0)^2,\ldots,(\La^0)^l=1)$ isomorphic
to a subgroup of $\clZ_l\subseteq\clZ(\bar G)$.
Note that $l$ is  a divisor of ord$(\clZ(\bar G))$.
Consider the action of  $\La^0$ on $\gg$. Since $(\La^0)^l=Id$ we have a $l$-periodic
gradation
 \beq{gra}
\gg=\oplus_{j=0}^{l-1}\gg_j\,, ~~\la(\gg_j)=\om^j\gg_j\,,~~\om=\exp\,\frac{2\pi i}l\,,~~~\la=Ad_{\La^0}\,,
 \eq
 \beq{gra1}
[\gg_j,\gg_k]=\gg_{j+k}~(mod \,l)\,,
 \eq
where $\gg_0$ is a subalgebra $\gg_0\subset\gg$ and the subspaces
$\gg_a$ are its representations. Moreover, in this terms the Killing form on $\gg$
assumes the form
 \beq{kfg}
(\gg,\gg)=\sum_{j=0}^{l-1}(\gg_j,\gg_{l-j})\,.
 \eq
 The $\la$-invariant subalgebra $\gg_0$  contains simple subalgebra $\ti{\gg}_0$
  \beq{gd3}
\gg_0=\ti{\gg}_0\oplus V\,.
 \eq
The components of this decomposition are orthogonal with respect to the Killing form (\ref{clA})
 \beq{kfg1}
(\gg_0,\gg_0)=(\ti\gg_0,\ti\gg_0)+(V,V)
 \eq
 and $V$ is a representation of $\ti{\gg}_0$.
We find below the explicit forms of $\gg_0$ for all simple Lie
algebras from our list.
  The coroot system $\ti\Pi$ of $\ti{\gg}_0$ is constructed by averaging along
 the orbits of the $\la$-action on $\Pi^{\vee ext}$ \cite{LOSZ1}.
%
%
We summarize the information about invariant subalgebras in Table 4.

\subsection{The moduli space of holomorphic bundles}

We described a $G$-bundle $E_G(V)$  by the transition operators
$\La=\La^0\bfe\,(\bfp)$, $\clQ=\bfe\,(\frac{\rho^\vee}h+\bfq)$,
where $\La^0=\La^0(\ga)$ corresponds to the
 coweight $\ga\in P^\vee$.
The topological type of $E$ is defined by an
element of the quotient $\zeta=\bfe(\varpi^\vee)\in P^\vee/t(G)$.

Let us transform $(\La,\clQ)$ taking in (\ref{gtt})
$f=\bfe\,(-\bfq z)$. Since $f$ commutes with $\La^0$ we come to new transition operators
$\clQ=\bfe\,(\varrho+\bfq)\to\clQ=\bfe\,(\varrho)$,
$\La\to \La^0\bfe\,(\bfp-\bfq\tau)$.
Denote $\bfp-\bfq\tau=\bfu$.
Then sections of $E_G(V)$  assume the quasi-periodicities
 \beq{mon1}
\psi(z+1)=\pi(\bfe\,(\varrho))\,\psi(z)\,,~~\psi(z+\tau)=\pi(\bfe\,(\ti\bfu)\La^0)\, \psi(z)\,.
 \eq
Thus, we come to the transition operators
 \beq{qala}
\fbox{$
\clQ=\bfe\,(\varrho)\,,~~~~\La=\bfe\,(\bfu)\La^0\,.$}
 \eq
Here  $\bfu\in\ti\gh_0$
plays the role of a parameter in the moduli space $\ti\gh_0$ (Proposition \ref{algeq}).
In this subsection we describe it in details.
In fact, $\ti\gh_0$ cover the genuine moduli space, which we are going to describe.

Fixing $\clQ$ and $\La$ we still have a residual gauge symmetry
preserving $\ti\gh_0$. Let $\ti W_0$ be the Weyl group of the Lie algebra $\ti\gg_0$
and $ t(\ti G_0)$ is the group of cocharacters (\ref{tb}), (\ref{coch1})
 of the invariant subgroup $\ti G_0$ (\ref{cocharc1}):
$t(\ti G_0)=\{\mC^*\to \ti\clH_0\}=\{\chi_\ga(z)=\bfe(\ga z)\}$.
The cocharacters have the quasi-periodicities
 \beq{qpc}
\chi_\ga(z+1)=\chi_\ga(z)\,,~~\chi_\ga(z+\tau)=\chi_\ga(\tau)\chi_\ga(z)\,,~~\ga \in t(\ti G_0)\,.
 \eq
Let $\clG_0=\{f(z)\}$ the group of maps $\Si_\tau\to \ti\clH_0$
 \beq{clg0}
\ti\clG_0=\{f(z)=\sum_{\ga\in\clR^\vee} c_\ga\chi_\ga(z)\}\,,
 \eq
where $c_\ga=0$ almost for all $\ga$ and
 $$
 \ti \clR^\vee=\left\{
\begin{array}{ll}
\ti  Q^\vee & \ti G_0=\bar{\ti G}_0 \,,\\
\ti P^\vee  & \ti G_0=\ti G_0^{ad}\,, \\
\ti t(G_l) &\ti G_0=\ti G_{0,l}\,,~ (\ref{coch1})\,.
\end{array}
\right.
 $$
 It follows from (\ref{qpc}) that $f(z+1)=f(z)$.
The group of the residual gauge transformations is the semidirect product
 \beq{rgs}
\clG_{\ti\clH_0}=\ti W_0\ltimes \ti\clG_0\,.
 \eq
The transformations of the sections by $f(z)=\chi_\ga(z)$ assume the form
  \beq{gtrg}
\psi(z)\to  \bfe(\ga z)\psi(z)\,,~~\ga\in \ti R^\vee\,.
 \eq
For  the transition operators we find (see (\ref{gtt}))
 \beq{trql}
\clQ^f=\bfe(\ga z)\clQ\bfe(-\ga z)=\clQ\,,~~\La_0^f=\bfe(\ga\tau)\bfe(\ga z)\La_0\bfe(-\ga z)=\bfe(\ga\tau)\La_0\,.
 \eq
Since $\ga\tau\in\ti\gh_0$ we find that the
 action of $\clG_{\ti\clH_0}$ on $\bfu$ assumes the form
 \beq{bsm}
\bfu\to \bfu^f=
\left\{
\begin{array}{cc}
  s\bfu & s\in \ti W_0 \\
  \bfu+\tau\ga_1+\ga_2 & \ga\in \ti \clR^\vee\,.
\end{array}
\right.
 \eq
Thus, transition operators, defined by parameters $\bfu$ and $\bfu^f$ describe  equivalent bundles.
The semidirect product of the Weyl group $\ti W_0$ and the lattice
 $\tau \ti \clR^\vee\oplus\ti \clR^\vee$
is called \emph{the Bernstein-Schwarzman group} \cite{Lo,BS,BS02}.
It is elliptic analog of the affine Weyl group
 $$
W_{BS}(\clR^\vee)=\ti W_0\ltimes(\tau \ti \clR^\vee\oplus\ti \clR^\vee).
 $$
Thereby,  $\bfu$ can be taken
from the fundamental domain $C(\clR^\vee)$ of $W_{BS}(\clR^\vee)$
 \beq{csc}
\fbox{$
Bun_\zeta^0(\Si_\tau,G)=\ti\gh_0/W_{BS}=C(\clR^\vee)~{\rm is~the~
moduli ~space ~of~holomorphic} ~\bar G-{\rm bundles}\,.$}
 \eq
Let $\clZ(\bG)$ is a cyclic group of order $N$, and $\clZ_l\subseteq\clZ(\bG)$
is its subgroup of order $l$. It is the case of the groups $\bG=\SLN$, or $Spin_{2n+2}$.
Thus, for the $\bar G,G_l,G_p,G^{ad}$  bundles we have the following
interrelations between their moduli space
 \beq{msd}
\begin{array}{ccccc}
   &  & Bun_1^0(\Si_\tau,\bG) &  &  \\
   & \swarrow & \mid & \searrow &  \\
  Bun_{\zeta_l}^0(\Si_\tau,G_l) &  & \mid &  & Bun_{\zeta_p}^0(\Si_\tau,G_p) \\
   & \searrow & \downarrow & \swarrow &  \\
   &  &Bun_\zeta(\Si_\tau,G^{(ad)})  &  &
\end{array}
 \eq
Here arrows mean coverings. Note that  $Bun_1^0(\Si_\tau,\bG)$,
$Bun_\zeta(\Si_\tau,G^{(ad)})$
 and as well
 $Bun_{\zeta_l}^0(\Si_\tau,G_l)$, $Bun_{\zeta_p}^0(\Si_\tau,G_p)$ are dual to each other
 in the sense that the defining them lattices are dual.
 Similar picture we have for $Spin_{4n}$, where $\clZ(Spin_{4n})\sim(\mu_2\oplus\mu_2)$.


 \subsubsection{Holomorphic bundles with quasi-parabolic structures}

As in the general case attach to the marked points $\bfx=\{x_a\}$ $G$-flags $Flag_a=G/P_a$
 ($P_a$ is a parabolic subgroup of $G$).
 In this way we extend the space $\ti\gh_0$ as
  $$
 \ti\gh_0\times\cup_{a=1}^nFlag_a\,.
  $$
 Taking into account the action of $\clG_{\ti\clH_0}$ we find that
 a big cell $Bun_\zeta^0(\Si_\tau,\bfx,G)$ in the moduli space $Bun_\zeta(\Si_\tau,\bfx,G)$ of
 holomorphic bundles with quasi-parabolic structures is the quotient
  \beq{mhb0}
 Bun_\zeta^0(\Si_\tau,\bfx,G)=(\ti\gh_0\times\cup_{a=1}^nFlag_a)/\clG_{\ti\clH_0}\,,
  \eq
where the action of $\clG_{\ti\clH_0}$ on $\bfY_a\in Flag_a$ takes
the form
 \beq{wfl}
 \bfY_a\to \left\{
 \begin{array}{ll}
   s(\bfY_a)&s\in \ti W_0  \\
   Ad_{\chi_\ga( x_a)}(\bfY_a) &\ga \in \ti\clR^\vee
 \end{array}
\right.
 \eq
Thus, we parameterize  $Bun_\zeta^0(\Si_\tau,\bfx,G)$ as
 \beq{cbm}
Bun_\zeta^0(\Si_\tau,\bfx,G)=\{(\bfu\in C(\clR^\vee);\bfY_a\in Flag_a/\clG_{\ti\clH_0}\,,~a=1,\ldots,n)\}\,.
 \eq


\subsubsection{Flat bundles with quasi-parabolic structures}

As in the general case (\ref{fol}) we describe the moduli space of
flat bundles $FBun(\Si_\tau,\bfx,G)$ over elliptic curves $\Si_\tau$
as the
 principal homogeneous spaces $P T^*Bun_\zeta^0(\Si_\tau,\bfx,G)$
  over the cotangent bundles $T^*Bun_\zeta^0(\Si_\tau,\bfx,G)$.
Let $\bfv$ be a vector   in the leaves of the projection
 \beq{fol2}
\pi\,:\,FBun^0_\zeta(\Si_\tau,G)\stackrel{Flat}{\longrightarrow}
 Bun_\zeta^0(\Si_\tau,G)\,.
 \eq
The vector $\bfv$  dual to $\bfu$ is defined by the transition
operator $\La(\bfu)$ (\ref{qala}). In these terms the form $\om_0$
(\ref{red}) becomes $(\de\bfv\wedge\de\bfu)$.

Remind that the coadjoint orbit $\clO$ is a principal homogeneous
space $\clO=PT^*Flag$. In this way we come to the space
 \beq{clr}
\widetilde{FBun}^0_\zeta(\Si_{\tau},\bfx,G)=\{(\bfv,\bfu),\bfS^a\in\clO_a\,,~a=1,\ldots,n\}\,.
 \eq
$\widetilde{FBun}^0_\zeta(\Si_{\tau},\bfx,G)$ is a symplectic manifold with the form  (\ref{red})
 \beq{sr}
\om=(\de\bfv\wedge\de\bfu)+\sum_{a=1}^n\om_a^{KK}\,.
 \eq
It turns out that this form is annihilated by the vector fields
generated by $\ti\clG_0$ (\ref{clg0}). In this way the structure of
the moduli space of flat bundles is described.
The moduli space of flat bundles over elliptic curves with
quasi-parabolic structure is the symplectic quotient
 \beq{cbm1}
FBun_\zeta^0(\Si_\tau,\bfx,G)=\widetilde{FBun}^0_\zeta(\Si_{\tau},\bfx,G)//\ti\clG_{0}=
\{((\bfv, \bfu)\,;~\bfS^a\in\clO_a//\ti\clG_{0}\}\,.
 \eq
In these terms  $\om$ (\ref{sr}) assumes the form:
 \beq{sef}
\om^{FBun}=(\de\bfv\wedge\de\bfu)+\sum_{a=1}^n\ti\om_a^{KK}\,,
 \eq
where $\ti\om_a^{KK}$ is the form on $\clO_a//\ti\clG_{0}$.


\section{Monodromy preserving equations on elliptic curves}
\setcounter{equation}{0}

In this section we generalize the results of \cite{LOSZ1,LOSZ2} in
two directions. First, we consider the multipunctured elliptic
curve. The corresponding integrable systems are generalizations of
the elliptic Gaudin models which were introduced by N. Nekrasov in
\cite{Nekr}, where the simplest case - $\SLN$-bundle with the
trivial characteristic class was considered. Second, we describe the
isomonodromy problems instead of integrable systems. The models are
non-autonomous Hamiltonian systems generalizing the autonomous
Gaudin models. Here we use the approach of \cite{LO1,LO102,LO11} to
the isomonodromy equations, where these equations are treated as the
non-autonomous generalizations of the Hitchin systems.

 \subsection{Deformations of elliptic curves}

Let $\ti{T}^2=\{(x,y)\in\mR\,|\,x,y\in\mR/\mZ\}$ be a torus.
 Complex structure on $\ti T^2$ is defined by the complex coordinate
$z=x+\tau_0y\,$, $\Im m\,\tau_0>0$. In this way we define the elliptic curve
$\Si_{\tau_0}\sim\mC/(\mZ+\tau_0\mZ)$.
It follows from (\ref{dmp}) that $\ti{T}^2$ should have at least one moving point.
 Since there is $\mC$ action on $\Si_{\tau_0}$ $x\to x+c$,
it is possible to put the one marked point at $x_0=0$ and we
leave with one module related to the deformation
$\Si_{\tau_0}\to\Si_{\tau}\sim\mC/(\mZ+\tau\mZ)$. We assume that
$\tau-\tau_0$ is small and the marked points do not coincide.
A big cell $\gT^0_{1,n}$ in the  space of times - the Teichm\" uller space $\gT_{1,n}$ is defined as
 \beq{tec}
\gT^0_{1,n}=\left(\clH^+=\{\Im m\tau>0\}\times
\{(x_0=0,\ldots,x_n)\,,~~ x_k\neq x_j, \, {\rm mod}\,\lan\tau,1\ran\}\right)\,.
\eq
According to the general prescription introduce the new coordinate
$ w=z-\frac{\tau-\tau_0}{\rho_0}(\bz-z)$, where
$\rho_0=\tau_0-\bar\tau_0$.
 If we shift $z$ as $z+1$ and $z+\tau_0$, then $w$ is transformed as $w+1$ and $w+\tau$.
Thereby $w$ is a well defined holomorphic coordinate on $\Si_\tau$.
 But this deformation move the points $x_a$.
Instead we should use the following transformation
  $$
 w=z-\frac{\tau-\tau_0}{\rho_0}(\bz-z)(1-\sum_{a=1}^n\chi_a(z,\bz))\,,
   $$
where $\chi(z,\bz)$ is the characteristic function (\ref{chi}) and
$\chi_a(z,\bz)=\chi(z-x^0_a,\bz-\bar x^0_a)$. As in the
general case, we  use the coordinates
  \beq{dec}
\left\{
 \begin{array}{l}
  w=z-\frac{\tau-\tau_0}{\rho_0}(\bz-z)(1-\sum_{a=1}^n\chi_a(z,\bz))\,,\\
\ti w=\bz\,.
 \end{array}
\right.
  \eq
Notice that
  $$
 z+\tau_0\rightarrow (w+\tau\,,~\ti w+\bar\tau_0)\,.
  $$
Taking into account (\ref{mp}) define
  $$
\ep(z,\bz)=\frac{t_\tau}{\rho_0}(\bz-z)(1-\sum_{a=1}^n\chi_a(z,\bz))+
\sum_{a=1}^nt_a\chi_a(z,\bz)\,.
  $$
The deformed operator assumes the form $\p_{\ti{w}}=\p_{\bz}+\mu\p_z$, where (see (\ref{mp}))
$\mu$ is represented as the sum
  \beq{dmu}
\mu=t_\tau\mu^{(0)}_\tau+
\sum_{a=1}^nt_a\mu_a^{(0)}\,.
  \eq
From (\ref{db}) and (\ref{dec}) we find the form of $\mu^{(0)}_\tau$
   \beq{bel}
 \mu^{(0)}_\tau=\f1{\rho_0}\bp(\bz-z)(1-\sum_{a=1}^n\chi_a(z,\bz))\,,
 ~t_\tau=\tau-\tau_0\,,
   \eq
 and from (\ref{t1}) $\mu_a^{(0)}$
   \beq{bel1}
 \mu_a^{(0)}=\bp\chi_a(z,\bz)\,,~~t_a=x_a-x_a^0\,.
  \eq
The dual to the Beltrami-differentials basis $\mu^{(0)}_\tau$,
 $\{\mu_a^{(0)}\,,a=1,\ldots n\}$ with respect
to the integration over $\Si_\tau$ is the first Eisenstein functions (\ref{A.1})
 $E_1(z-x_a)$ and $1$. There is only one time
$t_\tau$  for the one
 marked point case (see (\ref{dmp})).

The moduli space $\gM_1$ of elliptic curves is the result of the
factorization by the action of $\SLZ$ on the upper half-plane
$\clH^+$ by the M\"obius transform
 $$
\gM_1=\clH^+/\SLZ\,,~~\left(\tau\to \frac{a\tau+b}{c\tau+d}\right)\,.
 $$
  A big cell
$\gM^0_{1,n}$ in the moduli space is defined as the quotient
  \beq{msec}
 \gM^0_{1,n}=\gT^0_{1,n}/\SLZ\,,~~(x_a\to x_a(c\tau+d)^{-1})\,,
  \eq
  where $\gT^0_{1,n}$ is (\ref{tec}).


\subsection{Symplectic reduction}
For generic configurations of connections on the bundle over elliptic curves the $(0,1)$ part
of connections can be gauged away (see (\ref{2.30}))
 \beq{gf}
\bL:=f^{-1}\p_{\ti w}f+f^{-1}\bA f=0\,.
 \eq
Then $f$ acts on $A$ as $f^{-1}\p_{ w}f+f^{-1}A f=L$.
Therefore, the moment constraint equation takes the form
 \beq{cl}
\p_{\ti w}L_\zeta=\sum_{a=1}^n\bfS^a\de(w-x_a)\,.
 \eq
It means that $L$ is meromorphic and  has the first order poles at
the marked points with the residues
 \beq{res1}
Res\,L|_{w=x_a}=\bfS^a\in\clO_a\subset\gg^*\,,~~\vec\bfS=(\bfS_1,\ldots,\bfS_n)\,.
 \eq
To define the  space $\widetilde{FBun}^0_\zeta(\Si_{\tau},\bfx,G)$ (\ref{clr})
one should take into account the quasi-periodicities
 \beq{gru}
L_\zeta (w+1)=Ad_{\clQ}L(w)\,,~~L_\zeta (w+\tau)=Ad_{\La(\bfu)}L(w)\,,~~
\left([\clQ,\La^{-1}(\bfu)]=\zeta\right)\,.
 \eq
Thus,
\beq{phe}
\begin{array}{l}
\widetilde{FBun}^0_\zeta(\Si_{\tau},\bfx,G)=\{{\rm~ solutions~of~}(\ref{cl}),~(\ref{gru})\}\,,  \\
FBun^0_\zeta(\Si_{\tau},\bfx,G)=\{{\rm~ solutions~of~}(\ref{cl}),~(\ref{gru})\} /\clG_{\ti\clH_0}\,.
\end{array}
\eq
 The linear problem (\ref{4.36}) on $FBun^0_\zeta$   assumes the
 form:
 \beq{4.361}
\left\{
 \begin{array}{ll}
1.&(\ka\p+L)\psi=0\,,\\
2.&\p_{\ti w}\psi=0\,,\\
3.&(\ka\p_s+M_s)\psi=0,~~(s=1,\ldots,n+1)\,,
 \end{array}
\right.
  \eq
 where
  $$
 \p_{\ti w}=\p_{\bz}+(t_\tau\mu^{(0)}_\tau+
\sum_{a=1}^nt_a\mu_a^{(0)})\p_z\,.
  $$
The Lax equation takes the form:
 \beq{lmp}
\ka\p_sL-\ka\p M_s+[M_s,L]=0\,.
 \eq

\subsection{Lax matrix}
To calculate the L-operator we use the GS-basis (see Appendix B).
Let $(\gt^{\,k}_{\,\beta}\,, \gh^{\,k}_{\,\alpha}\}$ be the basis
(\ref{ft}), (\ref{fth}) in the $k$-component (\ref{gra}), and
$\gh^{\,0}_{\,\alpha}$ the Cartan generators of the invariant
subalgebra $\gh'_0$. Consider the residues of $L$ (\ref{res1})
 $$
\bfS^a=(S^a)^{\gl,-k}_{-\beta} \gt^{\,k}_{\,\beta}+ (S^a)^{\gh,-k}_{\alpha} \gh^{\,k}_{\,\alpha}
+(S^a)^{\gh,0}_{\alpha}\gh^{\,0}_{\,\alpha}\,.
 $$
Represent the L-operator  as the sum of the Cartan and the root parts
  \beqn{LR123}
 \begin{array}{l}
  L(w)=+L_{\gh}(w)+L_{\gh}^{0}(w)+L_{R}(w)\,,
  \end{array}
  \eqn
Then $L$ satisfying (\ref{cl}), (\ref{gru}) takes the form
  \beqn{r123}
  \begin{array}{l}
   L_{R}(w)=\frac{1}{2}\sum\limits_{a=1}^n\,\sum\limits_{k=\,0}^{l-1} \,
   \sum\limits_{\beta\,\in\, R}\,| \beta |^2\, \varphi^{\,k}_{\,\beta} (\bfu,w-x_a) \,
   (S^a)^{\gl,-k}_{-\beta} \, \gt^{\,k}_{\,\beta}\,,\\
   \\
   L_{\gh}(w)=\sum\limits_{a=1}^n\,\sum\limits_{k=\,1}^{l-1}
   \sum\limits_{\alpha\,\in\,\Pi} \, \phi(\frac{k}l,w-x_a) \,
   (S^a)^{\gh,-k}_{\alpha}\, \gh^{\,k}_{\,\alpha}\,,\\
   \\ L_{\gh}^{0}(w)=
   \sum\limits_{\alpha\,\in\,\ti\Pi} \, \Big(v_{\,\alpha}^{\gh}+
   \sum\limits_{a=1}^n\,E_{1}(w-x_a)\,(S^a)^{\gh,0}_{\alpha}\Big) \, \,
   \gh^{\,0}_{\,\alpha}\,.
  \end{array}
   \eqn
Here functions $\varphi^{k}_{\alpha}(\bfu,z)$ are from (\ref{t303})
and $E_1(z)$ is the Eisenstein function (\ref{A.1}). The needed
properties of $L$ are provided by (\ref{A.12})-(\ref{A.14}).


\subsection*{The  Poisson structure}

Here we describe the Poisson structure on
$\widetilde{FBun}^0_\zeta(\Si_{\tau},\bfx,G)$ and
$FBun_\zeta^0(\Si_\tau,\bfx,G)$ (\ref{cbm1}). The former is the
Poisson manifold $\bfP$ with the canonical brackets for $\bfv$,
$\bfu$ and the Poisson-Lie brackets for $\bfS$:
 \beq{pmu}
\bfP=T^*C\times\cup_a\clO_a=\{\bfv\,,\bfu\,,\bfS^a\in\clO_a\}\,.
 \eq
It has dimension $\sum_a\dim\,(\clO_a)+2\dim\,(\ti\gh_0)$. We describe the Poisson structure on $FBun_\zeta^0$
as result of the Poisson reduction with respect to the $\ti\clG_{0}$ action (\ref{clg0}).

Consider  the Poisson algebra $\clA=C^\infty(\bfP)$ of smooth function
on $\bfP$.
Let $\ep\in\ti\gh_0$ and $\ga$ is a small contour around $z=0$. Consider the following function
$\mu_\ep=\oint_\ga(\ep,L(\bfv,\bfu,\bfS))=(\ep,\bfS_0^\gh)$,
$~\bfS_0^\gh=\sum_{j=1}^{p} S^{\gh}_{j}e_j$.
It generates the vector field on $\bfP~$
$V_\ep\,:\,L(\bfv,\bfu,\bfS)\to\{\mu_\ep,L(\bfv,\bfu,\bfS)\}=
[\ep,L(\bfv,\bfu,\bfS)]$.

Let $\clA^{inv}$ be an invariant subalgebra of $\clA$ under $V_\ep$ action.
Then $I=\{\mu_\ep F(\bfv,\bfu,\bfS)\,|$
$\,F(\bfv,\bfu,\bfS)\in\clA \}$ is the Poisson ideal in $\clA^{inv}$.
The reduced Poisson algebra is the factor-algebra
 $$
\clA^{red}=\clA^{inv}/I=\clA//\ti\clG_0\,,~~(\ti\clH_0=\exp\,\ti\gh_0)\,.
 $$
The reduced Poisson manifold $\bfP^{red}$ is defined by
 the moment constraint
  \beq{mcd}
 \sum_{a=1}^n(S^a)^{\gh,0}_{\alpha}=0
  \eq
  and $\dim\,\ti\gh$ gauge fixing constraints on the spin variables that we do not specify
 \beq{predu}
\bfP^{red}=\bfP//\ti\clH_0=\bfP(\ti S^{\gh}_{s}=0)/\ti\clH_0\,,~~~
\dim\,(\bfP^{red})=\dim\,(\bfP)-2\dim\,(\ti\clH_0)=\dim\,(\clO)\,.
 \eq
Due to the moment constraints  we come from (\ref{r123}) to the Lax operator
 that has the correct periodicity. It depends on variables
$\{\bfv\,,\bfu\,,\bfS^a\}\in\bfP^{red}$
 Here $\bfS^a$ are not free due to the gauge fixing.
 Thus, after the reduction we come to the
Poisson manifold that has dimension of the coadjoint orbit $\clO$, but the Poisson structure on $\bfP^{red}$ is not the
Lie-Poisson structure. The Poisson brackets on  $\bfP^{red}$ are the Dirac brackets \cite{Di,BDOZ}.

On  $\bfP^{red}$  the equations o motion acquire the Lax form
(\ref{lmp}). In the limit $\ka\to 0$  the isospectral flows becomes
completely integrable, because the number of commuting involutive
integrals  is equal to $\oh\dim(\bfP^{red})$ \cite{LOSZ1}.


\subsection{Classical r-matrix}

Let us describe
 the Poisson structure on the unreduced space $\bfP$ in terms of the classical dynamical $r$-matrix.
The Poisson brackets on the unreduced phase space $\bfP$ (see
(\ref{pmu})) are presented in the form of a direct sum of the
Poisson-Lie brackets at each marked point and the canonical one for
$\bfv$ and $\bfu$:
  \beqn{PB}
\begin{array}{l}
 \left\{ (S^a)^{\gl, k}_{\alpha},(S^b)^{\gl,m}_{\beta}
\right\}\,=\delta^{ab}\left\{
\begin{array}{ll}
\frac{1}{\sqrt{l}}\,\sum\limits_{s=0}^{l-1}\, \omega^{ms} \,
C_{\alpha,\, \lambda^s\beta}\,\,(S^a)^{\gl,
k+m}_{\alpha+\lambda^s\beta}\,,&
\alpha\neq \,-\lambda^{s} \beta\\ & \\
\frac{p_{\alpha}}{\sqrt{l}}\,\omega^{s\,m}\,(S^a)^{\gh,k+m}_{\alpha}\,,&\alpha=
\,-\lambda^{s} \beta
\end{array}\right.\\
\\
 \left\{(\bar{S}^a)^{\gh\,k}_{\,\alpha},(S^b)^{\gl, m}_{\,\beta}
\right\} =
\delta^{ab}\frac{1}{\sqrt{l}}\,\sum\limits_{s=0}^{l-1}\,\omega^{-
ks }\,({\hat \alpha}, \lambda^{s}\beta) \,(S^a)^{\gl, k+m}_{\,\beta}\,,\\
\\
\left\{\bar{v}^{\gh}_{\alpha}, u_{\beta}\right\}=
\frac{1}{\sqrt{l}}\,\sum\limits_{s=0}^{l-1}\,( \hat{\alpha},\lambda^{s}\beta )\,,\\
\\
\left\{v_{\alpha},
(S^a)^{\gl,k}_{\alpha}\right\}\,=\left\{v_{\alpha},
(S^a)^{\gh,k}_{\alpha}\right\}\,=\left\{v_{\alpha},
(S^a)^{\gh,k}_{\alpha}\right\}\,=0\,,\\
\\
\left\{u_{\alpha},
(S^a)^{\gl,k}_{\alpha}\right\}\,=\left\{u_{\alpha},
(S^a)^{\gh,k}_{\alpha}\right\}\,=\left\{u_{\alpha},
(S^a)^{\gh,k}_{\alpha}\right\}\,=0\,.
\end{array}
   \eqn
Notice that the Poisson-Lie brackets for each marked point are dual
to the commutation relation given in the Appendix.
The Poisson brackets (\ref{PB}) are generated by the dynamical
r-matrix structure with the r-matrix of the form:
%
%
   \beq{prm1}
r(\bfu,z,w)=r(\bfu,z-w)=r_{\gh}(\bfu,z-w)+  r_{R}(\bfu,z-w)\,,
  \eq
   \beq{prm102}
r_{R}(\bfu,z)=\frac{1}{2}\,\sum\limits_{k=\,0}^{l-1}\,
\sum\limits_{\alpha\,\in\,R}\,
|\alpha|^2\,\varphi^{\,k}_{\,\alpha}(\bfu,z) \,\gt^{\,k}_{\,\alpha}
\otimes \gt^{-k}_{-\alpha}\,,~
 r_{\gh}(\bfu,z)=
\sum\limits_{k=\,0}^{l-1}\,\sum_{\alpha\,\in\,\Pi}\,\varphi^{\,k}_{\,0}(\bfu,z)\,
\gh^{\,k}_{\,\alpha}\otimes\gh^{-k}_{\,\alpha}\,.
  \eq
More exactly, the following two statements were proved in
\cite{LOSZ1}:
\begin{predl}
The r-matrix  (\ref{prm1})-(\ref{prm102}) and the Lax operator
(\ref{LR123})-(\ref{r123}) described above define the Poisson
brackets (\ref{PB}) via $RLL$-equation:
  \beq{RLLe}
  \left\{ L(w)\otimes 1,\,1\otimes L(w)
  \right\} \,=\left[L(w)\otimes 1+1\otimes L(w),
  r(z,w)\right]-
  \eq
 $$
-\frac{\sqrt{l}}{2}\,\sum\limits_{k=0}^{l-1}\sum\limits_{\alpha\in\,R}\,
|\alpha|^2 \,\partial_{\bfu}\,\varphi^{k}_{\alpha}(\bfu,z-w)
\,\bar{S}^{\gh\,0}_{\,\alpha}\,t^{\,k}_{\,\alpha}\otimes
t^{-k}_{-\alpha}\,.
 $$
\end{predl}
The Jacobi identity for the brackets (\ref{RLLe}) is provided by the
 next one statement:
 \begin{predl}
The $r$-matrix (\ref{prm1})-(\ref{prm102}) satisfies the classical
dynamical Yang-Baxter equation:
  \beq{dfg5}
[r_{12}(z,w),r_{13}(z,x)]+[r_{12}(z,w),r_{23}(w,x)]+[r_{13}(z,x),r_{23}(w,x)]-
 \eq
 $$
\sqrt{l}\sum\limits_{k=0}^{l-1}\sum\limits_{\alpha\in\,R}\frac{|\al|^2}{2}
\gt^{\,k}_{\,\al}\otimes \gt^{\,-k}_{\,-\al}\otimes
\bar{\gh}_{\al}^{0}\,\p_{\bfu}\vf_\al^k(\bfu,z-w)-
\frac{|\al|^2}{2}\gt^{\,k}_{\,\al}\otimes \bar{\gh}_{\al}^{0}\otimes
\gt^{\,-k}_{\,-\al}\,\p_{\bfu}\vf_\al^k(\bfu,z-x)+
 $$
 $$
\frac{|\al|^2}{2}\bar{\gh}_{\al}^{0}\otimes
\gt^{\,k}_{\,\al}\otimes\gt^{\,-k}_{\,-\al}\,\p_{\bfu}\vf_\al^k(\bfu,w-x)=0
 $$
 \end{predl}
The last term in (\ref{RLLe}) prevent the system to be integrable on
$\bfP$. However the action by  the Cartan subgroup $\clH'_0$ of
the invariant subgroup $ G'_0\subset G$ generates the moment map
condition:
  \beq{mmp31}
\sum\limits_{a=1}^n(S^a)^{\gh,0}_{\alpha}=0\,,\ \ \forall
\alpha\,\in\,\ti\Pi\,.
  \eq
As it was told, after reduction with respect to $\ti\clH_0$
(\ref{predu}) we come to $\bfP^{red}$. On  $\bfP^{red}$ this term
vanishes. Then
  \beq{syb}
 \left\{ L^{red}(w)\otimes 1,\,1\otimes L^{red}(w)
\right\} \,=\left[L^{red}(w)\otimes 1+1\otimes L^{red}(w), \ti
r(z,w)\right]  \eq
 Here the
$r$-matrix is replaced on $\ti r$, because the Poisson structure on
$\bfP^{red}$ differs from the Poisson structure on   $\bfP$. The
difference is due to the Dirac terms coming from the reduction
$\bfP^{red}=\bfP//\ti\clH_0$. We don't need it explicit form.

The classical dynamical r-matrices corresponding to trivial bundles
were found in \cite{EV}. In this case the dynamical parameter $\bfu$
belongs to Cartan subalgebra $\gh\subset\gg$. The problem of
classifications of r-matrices if $\bfu\in\ti\gh\subset\gh$ was
formulated in  \cite{EV}. For trigonometric  r-matrices without the
spectral parameter it was done in \cite{Sh} in terms of symmetries
of the extended Dynkin graph. The corresponding elliptic version was
considered in \cite{EtSch,EtSch02} and \cite{LOSZ1}.


\subsection{Symmetries of phase space}

The explicit form of the Lax operator (\ref{r123}) allows us
describe the action of the Bernstein-Schwarzman  group on the
dynamical variables.  As it was explained above, this action is
generated by the gauge subgroup $\clG_{\ti H_0}$ (\ref{rgs}).
 In addition  we consider the action of the modular group
$\SLZ=\left\{ \left(
  \begin{array}{cc}
    a & b \\
    c & d \\
  \end{array}
\right)\right\}
$.

\bigskip
\begin{center}
\begin{tabular}{|c|c|c|c|}\hline
             &$\ti W_0=\{s\}$&$\ti R^{\vee}\oplus\tau R^{\vee}$
&$\SLZ$
\\
 \hline \hline
${\bf v}$ &$s{\bf v}$& ${\bf v}+\ka\ga $ & ${\bf
v}(c\tau+d)+2\pi\imath\ka \frac{c{\bf u}}{(c\tau+d)}-2\pi\imath
c\sum_a x_a(\bfS^a)^{\gh,0}$
\\
\hline
${\bf u}$ &$s{\bf u}$  & ${\bf u}+\ga_1+\ga_2\tau$
 &${\bf u}(c\tau+d)^{-1}$
\\
\hline
$(\bfS^a)^{\gh,-k}_{-\alpha}$ & $(\bfS^a)^{\gh,-k}_{-s\al}$   &   $(\bfS^a)^{\gh,-k}_{-\alpha}$
 & $(c\tau+d)(\bfS^a)^{\gh,-k'}_{-\alpha'}$
\\
\hline
$(\bfS^a)^{\gh,0}_{-\alpha}$ & $(\bfS^a)^{\gh,0}_{-s\al}$   &   $(\bfS^a)^{\gh,0}_{-\alpha}$
 & $(c\tau+d)(\bfS^a)^{\gh,0}_{-\alpha}$
\\   \hline
$(\bfS^a)^{\gl,k}_{-\al}$ &$(\bfS^a)^{\gl,k}_{-s\al}$&
$\chi_{\lan\ga,\al\ran}(x_a)(\bfS^a)^{\gl,k}_{-\al}$
   & $(c\tau+d)\exp\left(-2
\pi i\frac{ cx_a\langle
\bfu,\al\rangle}{c\tau+d}\right)(\bfS^a)^{\gl,k'}_{-\al'}$
\\   \hline
$\tau$   &$\tau$  &$\tau$                   &$\frac{a\tau+b}{c\tau+d}$
\\ \hline
$x_a$     &  $x_a$  &$x_a$        &    $x_a(c\tau+d)^{-1}$
\\ \hline
\end{tabular}
\vspace{0.5cm}

\textbf{Table 1}\\
Symmetry properties of the phase space variables
\end{center}

In order to calculate the modular transformations of the spin
variables we use the moment constraint
$\sum_a(\bfS^a)^{\gh,0}_{-\alpha}=0$ and relations
(\ref{A.15a})-(\ref{A.17a}). Notice that the Lax operators are not
invariant with respect to the modular group. Indeed, from
(\ref{A.15a})-(\ref{A.17a}) we have
  \beq{A.17d}
\varphi^k_\al(\frac{\bfu}{c\tau+d},\frac{z}{c\tau+d}|\frac{a\tau+b}{c\tau+d})\equiv\exp\left(2
\pi i z\langle\varrho,\al\rangle\right)\phi(\langle
\bfu,\al\rangle+\tau \langle \varrho,\al\rangle+\frac{k}{l},z|\tau)
  \eq
  $$
 =\exp\left(2
\pi i z\left(\frac{ c\langle \bfu,\al\rangle}{c\tau+d}+  \langle
\varrho,\al\rangle a+\frac{k}{l} c\right) \right)\phi\left(\langle
\bfu,\al\rangle + (\langle \varrho,\al\rangle\,,\,\frac{k}l)\mat{a}bcd
 \left(\begin{array}{l}\tau\\ 1\end{array}\right),z|\tau\right)
   $$
   $$
=\exp\left(2 \pi i z\frac{ c\langle
\bfu,\al\rangle}{c\tau+d}\right)\varphi^{k'}_{\al'}(\bfu,z|\tau)\,,
   $$
where the last equality should be considered as definition of $k'$
and $\al'$, i.e.
  $$
\langle \bfu,\al'\rangle+\tau \langle
\varrho,\al'\rangle+\frac{k'}{l}=\langle \bfu,\al\rangle + (\langle
\varrho,\al\rangle\,,\,\frac{k}l)\mat{a}bcd
 \left(\begin{array}{l}\tau\\ 1\end{array}\right)\,.
  $$
However the Lax operators become invariant if the modular group acts
together with some gauge transformation. The results presented in
the Table are listed with regard to this gauge action. First, we can
gauge away the factor $\exp\left(2 \pi i z\frac{ c\langle
\bfu,\al\rangle}{c\tau+d}\right)$ by some transformation
$h=h(\bfu,z)$ from the Cartan subgroup. Second, we can act by some
$g$ such that $\hbox{Ad}_{g} \gt^{\,k}_{\,\al}=\gt^{\,k'}_{\,\al'}$.
Finally, we have:
 \beq{mmp3177}
\kappa\p_z+L(\bfS,\bfv,\frac{\bfu}{c\tau+d},\frac{z}{c\tau+d}|\frac{a\tau+b}{c\tau+d})=\hbox{Ad}_{gh}\left(\kappa\p_z+
L(\bfS',\bfv',\bfu',z|\tau)\right)\,,
  \eq
where the primed variables are given in the last (right) column of
the Table.

It follows from the last column that we can consider the independent variables taking
values in the moduli space of elliptic curves with marked points
 $$
\gM_{1,n}=\gT_{1,n}/\SLZ\,.
 $$
The Poincar\'e Cartan bundle (\ref{pcu}) become non-trivial over
$\gM_{1,n}$:
 $$
 \begin{array}{cc}
{\cal P}(G)& \\
\downarrow& \widetilde{FBun}^0_\zeta(\Si_{\tau},\bfx,G)~(\ref{phe})\\
\gM_{1,n} &
 \end{array}
  $$



\subsection{Hamiltonians}

Consider the structure of the extended phase space $\clP$
(\ref{pcu}) for the bundles over elliptic curves. Replace the
Teichm\"{u}ller space $\clH^+=\{\tau\,|\,\Im m\tau>0\}$ by the
quotient - the moduli
 $\clH^+/{\rm SL}(2,\mZ)$. Let
  $$
 \gM_n=\{(\tau,\bfx(x_1,\ldots,x_n))~|~\tau\in\clH^+/{\rm SL}(2,\mZ)\,,~x_a\neq x_b\}\,.
  $$
Then the space $\clP$ has the structure  \beq{pcu1}
 \begin{array}{cc}
{\cal P}(G)& \\
\downarrow&\clR_\zeta(\Si_\tau,\bfx,G) \\
 \gM_n &
 \end{array}
  \eq
 The local coordinate on the fiber $\clR_\zeta(\Si_\tau,\bfx,G)$ is
 $(\bfv,\bfu,\vec\bfS)$ with symplectic form $\om^{\clR}$ (\ref{sr}).

Let us calculate the Hamiltonians (\ref{qha}) using the form of the
Beltrami differential (\ref{bel}) (\ref{bel1}). In the coordinates
$(w,\ti w)$ the density of measure on $\Si_\tau$
 $$
d\si=\frac{dzd\bz}{\rho_0}=\frac{dwd\ti
w}{\rho}\,,~~\rho_0=\tau_0-\bar\tau_0\,,~~ \rho=\tau-\bar\tau_0
 $$
It follows from the structure of $L$ that under the condition
(\ref{mmp31})
 $\tr (L^2(w))$ is some periodic function on $\Si_\tau$ with second order poles
 at the marked points (\ref{res1}). Therefore,
the Hamiltonians can be computed from the decomposition:
   \beq{lse}
\oh\tr(L^2(w))=H_\tau+\sum_{a=1}^n(H_aE_1(w-x_a)+C_{2}^aE_2(w-x_a))\,.
  \eq
This expansion corresponds to the integrals
  $$
H_\tau=\oh\int_{\Si_\tau}\tr (L^2)\mu_\tau^0\,,~~
H_a=\oh\int_{\Si_\tau}\tr (L^2)\mu_a^0\,.
  $$
Then the quadratic Casimir functions (corresponding to orbits
attached at $x_a$) are
   \beq{res}
C_2^a=\frac{1}{2}\,\sum\limits_{k=\,0}^{l-1}\,
\sum\limits_{\alpha\,\in\,R}\, |\alpha|^2\,(S^a)^{\gl,k}_{\alpha}
(S^a)^{\gl,-k}_{-\alpha}+
\sum\limits_{k=\,0}^{l-1}\,\sum_{\alpha\,\in\,\Pi}\,
(S^a)^{\,\gh,k}_{\,\alpha}(S^a)^{\,\gh,-k}_{\,\alpha}\,,\ \
a=1,...,n\,.
  \eq
 The "Gaudin" (or "Schlesinger") Hamiltonians take the form
 \begin{small}
  \beq{hh32}
 \begin{array}{c}
H_a=l\sum\limits_{\alpha\in \Pi}v^{\gh}_\al(S^a)^{\,\gh,0}_\al+
\\ \ \\
+\sum\limits_{c\neq a}^n\sum\limits_{k=0}^{l-1}\left(\frac{1}{2}
\sum\limits_{\alpha \in R}|\alpha|^2
 \varphi^{k}_{\alpha}(\bfu,z_a-z_c) \,(S^a)^{\gl,k}_{\alpha}
(S^c)^{\gl,-k}_{-\alpha} +
 \sum\limits_{\alpha\in \Pi} \varphi_{0}^{k}(\bfu,z_a-z_c)
 \,
(S^a)^{\,\gh,k}_{\,\alpha}(S^c)^{\,\gh,-k}_{\,\alpha}\right)\,.
\end{array}
  \eq
 \end{small}
Finally, the zero Hamiltonian corresponding to the particles
interaction is
 \begin{small}
  \beq{hh323}
 \begin{array}{c}
H_\tau=\frac{l}{2}\sum\limits_{\alpha\in \Pi}
  \sum\limits_{s=0}^{l-1}v^{\gh}_\al
v^{\gh}_{\lambda^s\hat{ \alpha}}+
\\ \ \\
+\sum\limits_{b\neq d}^n\sum\limits_{k=0}^{l-1}\left(\frac{1}{2}
\sum\limits_{\alpha \in R}|\alpha|^2
 f^{k}_{\alpha}(\bfu,z_b-z_d) \,(S^b)^{\gl,k}_{\alpha}
(S^d)^{\gl,-k}_{-\alpha} +
 \sum\limits_{\alpha\in \Pi} f_{0}^{k}(\bfu,z_b-z_d)
 \,
(S^b)^{\,\gh,k}_{\,\alpha}(S^d)^{\,\gh,-k}_{\,\alpha}\right)\,.
\end{array}
  \eq
  \end{small}
The functions $f^{k}_{\alpha}(\bfu,z)$ are defined in
(\ref{t3036})-(\ref{t3064}).


 \subsection{M-operators}
 The $M_s$-operators can be calculated from the Lax equation (\ref{lmp}), since the $L$-operator
 is already known. However, following \cite{BV} we construct the $M$-operators from the r-matrix
structure (\ref{RLLe}) defined for the r-matrix (\ref{prm1})-(\ref{prm102}).

 Consider first the autonomous case (\ref{lae}). Define  the Lax
 pairs for the integrable flows described by the Hamiltonians
 (\ref{hh32}), (\ref{hh323}).
Let us compute the  (off-shell)
brackets
   \beq{lse0309}
\oh\tr_2\{1\otimes L^2(z),L(z)\otimes
1\}=[L(z),M_w(z)]+\Delta(z,w)\,,
  \eq
where $\tr_2$ denotes the trance in the second component of the
tensor product
   \beq{lse03}
M_w(z)=-\tr_2\left(r(z,w)L_2(w)\right)
  \eq
while $\Delta(z,w)$ comes from the last term in (\ref{RLLe}):
    \beq{lse04}
\Delta(z,w)=\tr_2\Big(\frac{\sqrt{l}}{2}\,\sum\limits_{k=0}^{l-1}\sum\limits_{\alpha\in\,R}\,
|\alpha|^2 \,\partial_{\bfu}\,\varphi^{k}_{\alpha}(\bfu,z-w)
\,\bar{S}^{\gh\,0}_{\,\alpha}\,t^{\,k}_{\,\alpha}\otimes
t^{-k}_{-\alpha}L_2(z)\Big)\,.
  \eq
This term vanishes on the first class constraint (\ref{mmp31}).
Therefore, on-shell  (\ref{mmp31})  we have
   \beq{lse06}
\oh\tr_2\{1\otimes L^2(z),L(z)\otimes 1\}\left.\right|_{\hbox{on
shell}\ (\ref{mmp31})}=[L(z),M_w(z)]\,.
  \eq
Taking into account (\ref{lse}) we conclude that the M-operators of
the Hamiltonians (\ref{hh32}), (\ref{hh323}) can be evaluated by
expansion of $M_w(w)$ (\ref{lse03}) similarly to (\ref{lse}). In
this way we obtain a set of Lax equations
   \beq{lse11}
\p_{t_a}L(z)=[L(z),M_a(z)]\,,\ \ a=1,...,n
  \eq
 and
   \beq{lse112}
\p_{t_\tau}L(z)=[L(z),M_{\tau}(z)]
  \eq
on shell (\ref{mmp31}). The later M-operators are of the form:
  \beq{LR1234}
 \begin{array}{l}
  M_a(z)=(M^a)_{R}(z)+(M^a)_{\gh}(z)+(M^a)_{\gh}^{0}(z)\,,
  \end{array}
  \eq
with
  \beqn{r1234}
  \begin{array}{l}
   (M^a)_{R}(z)=-\frac{1}{2}\sum\limits_{k=\,0}^{l-1} \,
   \sum\limits_{\beta\,\in\, R}\,| \beta |^2\, \varphi^{\,k}_{\,\beta} (\bfu,z-x_a) \,
   (S^a)^{\gl,-k}_{-\beta} \, \gt^{\,k}_{\,\beta}\,,\\
   \\
   (M^a)_{\gh}(z)=-\sum\limits_{k=\,1}^{l-1}
   \sum\limits_{\alpha\,\in\,\Pi} \, \phi(\frac{k}l,z-x_a) \, (S^a)^{\gh,-k}_{\alpha}\, \gh^{\,k}_{\,\alpha}\,,\\
   \\ (M^a)_{\gh}^{0}(z)=-
   \sum\limits_{\alpha\,\in\,\Pi}E_{1}(z-x_a)\,(S^a)^{\gh,0}_{\alpha}
   \gh^{\,0}_{\,\alpha}\,.
  \end{array}
   \eqn
and
  \beq{LR1237}
 \begin{array}{l}
  M^\tau(z)=(M^\tau)_{R}(z)+(M^\tau)_{\gh}(z)+(M^\tau)_{\gh}^{0}(z)\,,
  \end{array}
  \eq
with
  \beqn{r1237}
  \begin{array}{l}
   (M^\tau)_{R}(z)=\frac{1}{2}\sum\limits_{a=1}^n\,\sum\limits_{k=\,0}^{l-1} \,
   \sum\limits_{\beta\,\in\, R}\,| \beta |^2\, f^{\,k}_{\,\beta} (\bfu,z-x_a) \,
   (S^a)^{\gl,-k}_{-\beta} \, \gt^{\,k}_{\,\beta}\,,\\
   \\
   (M^\tau)_{\gh}(z)=\sum\limits_{a=1}^n\,\sum\limits_{k=\,1}^{l-1}
   \sum\limits_{\alpha\,\in\,\Pi} \, f(\frac{k}l,z-x_a) \, (S^a)^{\gh,-k}_{\alpha}\, \gh^{\,k}_{\,\alpha}\,,\\
   \\ (M^\tau)_{\gh}^{0}(z)=\frac{1}{2}\sum\limits_{a=1}^n\,
   \sum\limits_{\alpha\,\in\,\Pi}\left(E_{1}^2(z-x_a)-\wp(z-x_a)\right)\,(S^a)^{\gh,0}_{\alpha}
   \gh^{\,0}_{\,\alpha}\,.
  \end{array}
   \eqn
As announced the constraints  (\ref{mmp31}) should be supplied by
some gauge fixation (of action of the Cartan subgroup $\clH_0$ of
the invariant subgroup). The entire reduction
$\bfP^{red}=\bfP//\ti\clH_0$ changes (in principle) M-operators and
r-matrices (see example of the spinless Calogero model
\cite{Babelon2}). We do not consider these type of reductions here
because there is no any "good" recipe for the choice of the gauge
fixation.


 \subsection{Painlev\'e-Schlesinger equations}
In the beginning of the Section we mentioned that the monodromy
preserving equations arise in our approach as non-autonomous version
of equations in Hitchin systems \cite{LO11}. At the level of Lax or
zero curvature equation that the same pairs of matrices as in
(\ref{lse11}) or (\ref{lse112}) satisfy the monodromy preserving
equations
   \beq{lse115}
\p_{x_a}L(z)-\p_z M_a(z)=[L(z),M_a(z)]\,,\ \ a=1,...,n
  \eq
 and
   \beq{lse1125}
\p_{\tau}L(z)-\p_z M_\tau=[L(z),M_{\tau}(z)]
  \eq
as well. The phenomenon is known as the  Painlev\'e-Calogero
Correspondence \cite{LO97} (see also \cite{Takasaki01} for developed
version and \cite{Sul,ZZ,ZZ02,ZZ03} for the quantum versions).

Notice that the time variables $t_a$ in (\ref{lse11}) are replaced
by the positions of marked points $x_a$ in (\ref{lse115}) while
$M_a(z)$ are the same in both sets of equations. In the same manner
the time variable $t_\tau$ in (\ref{lse112}) is replaced by the
moduli $\tau$  in (\ref{lse1125}) while $M_{\tau}$ is the same for
both equations.
Technically, the {\em Painlev\'e-Calogero Correspondence} follows
from
  \beq{pc21}
d_{x_a}L(z)=\p_z M_a(z)\,,
  \eq
where $d_{x_a}$ is the derivative by the explicit dependence on
$x_a$ only and
  \beq{pc22}
d_{\tau}L(z)=\p_z M_\tau\,,
  \eq
where $d_{\tau}$ is the derivative by the explicit dependence on
$\tau$ only. Equations (\ref{pc21}) are trivially follows from the
type of dependence $L(z,x_a)=L(z-x_a)$, $M_a(z,x_a)=M_a(z-x_a)$
while (\ref{pc22}) follows from the heat equation (\ref{ap212}).

In this way we get the Painlev\'e-Schlesinger equations
(\ref{lse115}) and (\ref{lse1125}). Let us also mention that the
M-operator can be derived in terms of the modification $\Xi(z)$ from
the assumption that the Painlev\'e-Calogero Correspondence holds
true \cite{AASZ}.

\subsection{KZB equations}


\subsubsection{KZB equations on elliptic curves}

The Knizhnik-Zamolodchikov-Bernard (KZB) equations appear as
horizontality condition of some  sections (conformal blocks) for the
so-called KZB connection. The KZB equations (see (\ref{t1}) below)
can be considered as the {\em quantization} of the monodromy
preserving equations. The KZB connection can be defined on elliptic
curve \cite{Felder3}. In the case of arbitrary characteristic class
it is given by the following differential operators \cite{LOSZ3}:
   \beq{t2}
 \nabla_a=\p_{z_a}+\hat{\p}^a+\sum\limits_{c\neq a}r^{ac}\,,
   \eq
   \beq{t202}
\nabla_\tau=2 \pi i
\p_\tau+\Delta+\frac{1}{2}\sum\limits_{b,d}f^{bd}\,,
  \eq
with
   \beq{t203}
r^{ac}=\frac{1}{2}\sum\limits_{k=0}^{l-1} \sum\limits_{\alpha \in
R}|\alpha|^2
 \varphi^{k}_{\alpha}(\bfu,z_a-z_c) \gt^{k,a}_{\alpha} {\gt^{-k, c}_{-\alpha}} +
 \sum\limits_{k=0}^{l-1}\sum\limits_{\alpha\in \Pi} \varphi_{0}^{k}(\bfu,z_a-z_c)
 \gH_{\alpha}^{k,a}\gh_{\alpha}^{k,c}\,,
  \eq
  \beq{t3}
 f^{ac}=\frac{1}{2}\sum\limits_{k=0}^{l-1} \sum\limits_{\alpha \in R}
|\alpha|^2 f^{k}_{\alpha}(\bfu,z_a-z_c)
 \gt^{k,a}_{\alpha} {\gt^{-k, c}_{-\alpha}} +
 \sum\limits_{k=0}^{l-1}\sum\limits_{\alpha\in \Pi} f_{0}^{k}(\bfu,z_a-z_c) \gH_{\alpha}^{k,a}
 \gh_{\alpha}^{k,c}\,,
    \eq
where $\gt^{k,a}_{\alpha}=1\otimes...1\otimes
\gt^{k}_{\alpha}\otimes 1...\otimes 1$ (with $\gt^{k}_{\alpha}$ on
the a-th place) and similarly
 for the generators
 $\gH^{k,a}_{\alpha}$ and $\gh^{k,a}_{\alpha}$
 \footnote{For brevity we write $\gt^{k,a}_{\alpha}\,, \gh^{k,a}_{\alpha}$
 instead of representations of these generators
  in the spaces $V_{\mu_a}$.}.
The following short notations are used here:
  $$
 \hat{\p}^a=l \sum\limits_{\alpha \in \Pi }  \gh^{0,a}_{\alpha} \partial_{\hat{\alpha}}\,,~~
 \Delta=\frac{l}{2}\sum\limits_{\alpha\in \Pi}
  \sum\limits_{s=0}^{l-1} \p_{u_\alpha}\p_{u_{\lambda^s\hat{
  \alpha}}}\,.
   $$
From the definition it follows that  $r^{ac}=-r^{ca}$ and
$f^{ac}=f^{ca}$.
 Notice that
   \beq{t307}
f^{cc}=-\sum\limits_{k=0}^{l-1}\sum\limits_{\alpha\in R} |\alpha|^2
 \wp^{k}_{\alpha} \gt^{k,c}_{\alpha} \gt^{-k,c}_{-\alpha} -\sum\limits_{k=0}^{l-1}
 \sum\limits_{\alpha\in \Pi}
  \gH^{k,c}_{\alpha} \gh^{-k,c}_{\alpha}-2l\eta_{1} C_{2}^{c}\,,
  \eq
 where $C_{2}^{c}$ is the Casimir operator acting on the $c$-th component.
Recall that we study the following system of differential equations
(KZB):
   \beq{t1}
 \left\{
 \begin{array}{l} \nabla_aF=0,\ \ a=1...n\,,\\
\nabla_\tau F=0\,.
\end{array}
\right.
   \eq
There are two types of the compatibility conditions of KZB equations
(\ref{t1}):
  \beq{t5}
 \left\{
\begin{array}{l}
[\nabla_a,\nabla_b]F=0,\ \ a,b=1...n\,,
\\
\left[\nabla_a,\nabla_{\tau}\right]F=0,\ \ a=1...n\,.
\end{array}\right.
  \eq
It is important to mention that the solutions of (\ref{t1}) $F$ are
assumed to satisfy the following condition (compare with
(\ref{mmp31})):
  \beq{t4}
 \left(\sum\limits_{c=1}^n \gh^{0,c}_{\alpha}\right)F=0\,,\
\hbox{for}\ \hbox{any}\ \alpha\in \ti\Pi\,.
   \eq
\begin{predl}\label{predl1}
The upper equations in (\ref{t5}) $[\nabla_a,\nabla_b]=0$ are valid
for the r-matrix (\ref{prm1}) on the space of solutions of
(\ref{t1}) satisfying (\ref{t4}). They follow from the classical
dynamical Yang-Baxter equations:
   \beq{t501}
[r^{ab},r^{ac}]+[r^{ab},r^{bc}]+[r^{ac},r^{bc}]+
[\hat{\p}^a,r^{bc}]+[\hat{\p}^c,r^{ab}]+[\hat{\p}^b,r^{ca}]=0\,.
   \eq
\end{predl}

\begin{predl}\label{predl2}
The lower equations in (\ref{t5})
$\left[\nabla_a,\nabla_{\tau}\right]=0$ are valid for the r-matrix
(\ref{t303}) on the space of solutions of (\ref{t1}) satisfying
(\ref{t4}).
\end{predl}
The proofs of these statements are given in \cite{LOSZ3}.

\subsubsection{KZB equations and isomonodromy problem}

The KZB equations (\ref{t2}), (\ref{t202}) specified here for
elliptic curves can be written for curves of arbitrary genus
\cite{H2,Fe,I}. Their interrelations with the isomonodromy problem
on a sphere were investigated in \cite{Resh,Ha1}.
In general case as in the genus one the KZB equations have the form
 of the non-stationary  Schr\"{o}dinger equations
\beq{qkzb}
(\ka\p_s+\hat {H}_s)\Psi=0\,,
\eq
where $\hat {H}_s$ are the quantum quadratic Hitchin Hamiltonians.
To pass to the classical limit
we replace the wave functions (the conformal block) by its quasi-classical expression
 $$
\Psi=\exp \frac{\imath S}{\ka}\,,
 $$
where $S$ is the classical action ($S=\log\tau$ (\ref{tf})). The
classical limit $\ka\rar 0$ leads to the Hamilton-Jacobi equations
for ${S}$ \cite{LL}. They are equivalent to the equations of motion
1. in (\ref{2.19}). On the other hand, putting $\ka=0$ in
(\ref{qkzb}) and staying with the quantum Hamiltonians we come
 to the Schr\"{o}dinger equation $\hat {H}_s\Psi=0$.
At the classical level the limit $\ka\to 0$ implies the passage from
the isomonodromy flows to the isospectral flows, as it was  described
in Section \ref{idi}.


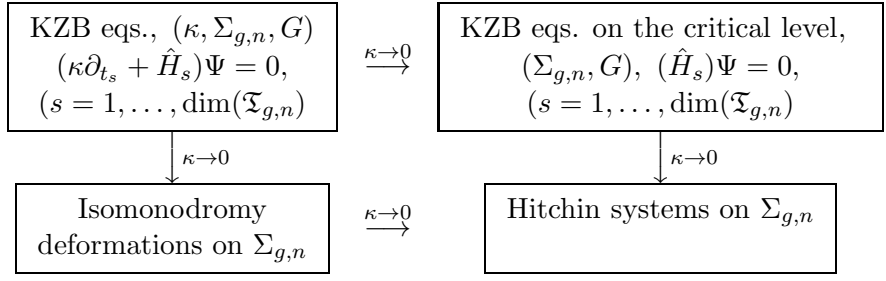
 $$
\def\normalbaselines{\baselineskip20pt
       \lineskip3pt    \lineskiplimit3pt}
\def\mapright#1{\smash{
        \mathop{\longrightarrow}\limits^{#1}}}
\def\mapdown#1{\Big\downarrow\rlap
       {$\vcenter{\hbox{$\scriptstyle#1$}}$}}
\begin{array}{ccc}
\boxed{
\begin{array}{c}
\mbox{KZB eqs.},~(\ka,\Si_{g,n},G)\\
(\ka\p_{t_s}+\hat{H}_s)\Psi=0,\\
(s=1,\ldots,\dim(\gT_{g,n})
\end{array}
}
  &\mapright{\ka\rar0} &
\boxed{
\begin{array}{c}
\mbox{KZB eqs. on the critical level},~\\
(\Si_{g,n},G),~(\hat{H}_s)\Psi=0,\\
(s=1,\ldots,\dim(\gT_{g,n})
\end{array}
}
\\
\mapdown{\ka\rar 0}&    &\mapdown{\ka\rar 0} \\
\boxed{
\begin{array}{c}
\mbox{ Isomonodromy} ~\\
\mbox{deformations on}~\Si_{g,n}
\end{array}
}
&\mapright{\ka\rar 0} &
\boxed{
\begin{array}{c}
\mbox{Hitchin systems on} ~\Si_{g,n}\\
       \\
 \end{array}
}
\end{array}
 $$


\subsection{Painlev\'e field theory}

Recall once again that in previous paragraphs we obtained the
monodromy preserving equations as non-autonomous versions of
mechanical integrable systems. More exactly, in the elliptic case we
replaced the linear problem and its compatibility condition (known
as Lax equations)
  \beq{eo01}
 \left\{
\begin{array}{l}
L(z)\Psi=\Lambda\Psi
\\
(\p_t+M(z))\Psi=0
\end{array}
\right.\hskip2cm \p_t L(z)=[L(z),M(z)]
  \eq
by another linear problem and the corresponding compatibility
condition (the monodromy preserving equations)
  \beq{eo02}
 \left\{
\begin{array}{l}
(\p_z+L(z))\Psi=0
\\
(\p_\tau+M(z))\Psi=0
\end{array}
\right.\hskip1cm \p_\tau L(z)-\p_zM(z)=[L(z),M(z)]
  \eq
At the same time it is well known that mechanical integrable systems
can be "extended" to (1+1) integrable field theories described by
the following linear problem and compatibility condition
(Zakharov-Shabat equations)
  \beq{eo03}
 \left\{
\begin{array}{l}
(\p_x+{\ti L}(z))\Psi=0
\\
(\p_t+\ti M(z))\Psi=0
\end{array}
\right.\hskip1cm \p_t \ti L(z)-\p_x \ti M(z)=[\ti L(z),\ti M(z)]\,,
  \eq
where the phase space consists of ${\mathbb S}^1$-valued fields
while $x$ is the coordinate on the unit circle ${\mathbb S}^1$. For
the Hitchin systems the transition
(\ref{eo01})$\rightarrow$(\ref{eo03}) was described in \cite{LOZ1}
(see also \cite{Z111,Z11102}).

In \cite{AALOZ} we suggested the {\em field-theoretical
generalization of the monodromy preserving equations}. It arises
from the linear problem
  \beq{eo04}
 \left\{
\begin{array}{l}
(\p_z +\p_x+L'(z))\Psi=0
\\
(\p_\tau+M'(z))\Psi=0
\end{array}
\right.
  \eq
with the compatibility condition
  \beq{eo05}
\p_\tau L'(z)-\p_xM'(z)-\p_zM'(z)=[L'(z),M'(z)]\,.
  \eq
For example the field generalization of the Painlev\'e VI equation
has the following form.
 Consider four three-vector fields
  $
 \bfS_b^\al(x)\,,\ \al\in (\mZ_2\times\mZ_2)\setminus(0,0)\,,\ \ b\in\mZ_2\times\mZ_2\,.
  $
 and let $\bfJ^I(\be,\p_x,\tau)\,,$ $\bfJ^{II}(\be,b,c,\p_x,\tau)$
 be the following pseudo-differential operators
  \beq{j1}
\bfJ^{I}(\al,b,c,\p_x,\tau)=E_2\left(\omega_\alpha-\frac{\bar{k}}{2\pi\imath}\partial_x,\tau\right)\,,~~
\bfJ^{II}(\al,b,c,\p_x,\tau)=f_{\alpha}\left(\frac{\bar{k}}{2\pi\imath}\partial_x,\omega_c-\omega_b\right)\,.
  \eq
 Then the field equation has the form of four
  interacting non-autonomous Euler-Arnold tops \cite{Ar1,Ar2}, related to the loop group $L(\SL)$
    \beq{Naft}
   \begin{array}{c}
  \frac{\p\,}{\p\tau}\bfS_b^\al(x)=
  \sum\limits_{\be\neq\al}\Bigl(\bfS_b^{\al-\be}(x)\bfJ^I(\al,\be,\p_x,\tau)
  \bfS_b^\be(x)
  +\sum\limits_{c\neq b}\bfS_b^{\al-\be}(x)\bfJ^{II}(\al,\be,b,c,\p_x,\tau)
  \bfS_c^\be(x)\Bigr)
  \end{array}
    \eq
and subjected  to the constraints
$\bfS^\al_b(x)=(-1)^{b\times\al}\bfS^\al_b(-x)$.
  The equation (\ref{Naft}) itself can be considered as the field
 generalization of the elliptic Schlesinger system \cite{CLOZ1}
 while in particular case of ${\rm sl}_2$-connection and four marked points
 at half-periods of the elliptic curve and on the constraints it generalizes PVI.
 Namely, it can be shown that for the zero modes (\ref{Naft}) along with the constraints  becomes the equation for the
 non-autonomous Zhukovsky-Volterra gyrostat (\ref{i401})
 for three variables $S^\al$ -- zero modes of $S^\al_0$. The later
 is one of possible forms of Painlev\'e VI equation as we will see
 below (in Section \ref{p6}).








\section{Symplectic Hecke Correspondence}

\setcounter{equation}{0}

We consider the Hecke transformation in the case
 $g=1$, $n=1$. The phase space has dimension of a coadjoint orbit
 $2\sum_{j=1}^{rank G}(d_j-1)$.
In this case $L$ satisfies the  conditions
 \beq{fm}
L(z+1)=\clQ L(z)\clQ^{-1}\,,~~~
L(z+\tau)=\La L(z)\La^{-1}\,,
 \eq
where $\clQ$ and $\La$ are solutions of (\ref{gce}), and
 \beq{repl}
\bp L(z)=\bfS\de(z,\bz)\,.
 \eq
In other words $Res|_{z=0}\,L(z)=\bfS$. These conditions fix $L$.
To make dependence on the characteristic class $\zeta(E_G)$ explicit  we will write
 $L(z)^{\varpi^\vee_j}$,  if the Lax matrix satisfies the quasi-periodicity conditions  with
 $\La=\La_{\varpi^\vee_j}$, $\clQ_{\varpi^\vee_j}$
 where $\La_{\varpi^\vee_j}$ $\clQ_{\varpi^\vee_j}$
 are solutions of (\ref{gce}) with
$\zeta=\bfe(-\varpi^\vee_j)$, $\,\varpi^\vee_j\in P^\vee$.

The modification $\Xi(\ga)$  of $E_G$ changes the characteristic
class. It acts on $L^{\varpi^\vee_j}$ as follows
 \beq{mhb}
L^{\varpi^\vee_j}\Xi(\ga) =\Xi(\ga)L^{\varpi^\vee_j+\ga} \,.
 \eq
It is the singular gauge transformation mentioned in the
Introduction.
%
%
The action (\ref{mhb}) allows one to write down condition on $\Xi(\ga)$.
Since $L^{\varpi^\vee_j}$ has a simple pole at $z=0$
the modified Lax matrix  $L^{\varpi^\vee_j+\ga}$ should have also a
 simple pole at $z=0$.
Decompose $L^{\varpi^\vee_j}$ and $L^{\varpi^\vee_j+\ga}$ in the
Chevalley basis (\ref{CD1}),
 (\ref{CBA})
 $$
L^{\varpi^\vee_j}=L_{\gh}(z)+\sum_{\al\in R}L_\al(z)E_\al\,,~~~
L^{\varpi^\vee_j+\ga}=\ti L_{\gh}(z)+\sum_{\al\in R}\ti L_\al(z)E_\al\,.
 $$
Expand $\al$ in the basis of simple roots (\ref{ale1})
 $\al=\sum_{j=1}^lf_j^\al\al_j$ and $\ga$
in the basis of fundamental coweights $\ga=\sum_{j=1}^lm_j\varpi_j^\vee$.
Assume  that
$\lan\ga,\al_j\ran\geq 0$ for simple $\al_j$.
In other words $\ga$ is a dominant coweight.
  Then
$\lan\ga,\al\ran=\sum_{j=1}^lm_jn_j^\al$ is an integer number,
 positive for $\al\in R^+$ and negative for $\al\in R^-$.
From (\ref{mhb})  we find
 \beq{shc}
 L^{\varpi^\vee_j+\ga}_{\gh}(z)=L^{\varpi^\vee_j}_{\gh}(z)\,,~~~
 L^{\varpi^\vee_j+\ga}_\al(z)=z^{\lan\ga,\al\ran}L^{\varpi^\vee_j}_\al(z)\,.
 \eq
 In the neighborhood of $z=0$ $L_\al(z)$ should have the form
 \beq{cbl}
L^{\varpi^\vee_j}_\al(z)=a_{\lan\ga,\al\ran}z^{-\lan\ga,\al\ran}+
a_{\lan\ga,\al\ran+1}z^{-\lan\ga,\al\ran+1}+
\ldots\,,~~~(\al\in R^-)\,,
 \eq
otherwise the transformed Lax operator becomes singular.
 It means that the type of the modification $\ga$
is not arbitrary, but depends on the local behavior of the Lax
operator. It allows (in principle) one to find the dimension of the
space of the Hecke transformation.

Now consider  a global behavior of $L(z)$ (\ref{fm}). Then we find that
 $\Xi(\ga)$ should intertwine the quasi-periodicity conditions
 $$
\Xi(\ga,z+1)\clQ_{\varpi^\vee_j}=\clQ_{\varpi^\vee_j+\ga}\Xi(\ga,z)\,,~~
\Xi(\ga,z+\tau)\La_{\varpi^\vee_j}=\La_{\varpi^\vee_j+\ga}\Xi(\ga,z)\,.
 $$
%
 For $G=\SLN$, $\ga=\varpi^\vee_1$ and special residue of $L(z)$,
 the
 solutions of these equations were found in \cite{LOZ1}.
%
%
%
 On the reduced space $\bfP^{red}$ the equations of motion corresponding to integrals
 $I_{jk}$ acquire the Lax form
  $$
 \ka\p_{t_s}L-\ka\p_z M_{t_s}+[M_{t_s},L]=0\,.
  $$
 The operators $M_{t_s}$ are reconstructed via $L$ and $r$-matrix.

\subsection{Example}

Let us demonstrate the modification procedure with a example of
$\SLN$. The characteristic classes of underlying bundles are
elements of ${\mathbb Z}_N$. The values $+1$ and $\exp{\frac{2\pi
i}{N}}$ correspond to the most different bundles and dynamical
systems. These are the elliptic Calogero model and the elliptic top
\cite{LOZ1}. In the isomonodromic case both models become
non-autonomous \cite{CLOZ1,CLOZ102,KLO}. However, the L-A pairs are
the same as in autonomous mechanics due to the Painlev\'e-Calogero
Correspondence. The Lax matrices of the models are fixed by their
residues at the single marked point $z=0$ on the elliptic curve
$\Sigma_\tau$
  \beq{i010}
 \begin{array}{c}
\displaystyle{
 \mathop{\hbox{Res}}\limits_{z=0}L^{Cal}(z)
 = \nu\left(
\begin{array}{c}
0\ \ 1\ \ 1\ ...\ 1
\\
1\ \ 0\ \ 1\ ...\ 1
\\
.......
\\
1\ \ 1\ ...\ 1\ \ 0

\end{array}
\right),\ \ \ \ \
 \begin{array}{l}
\mathop{\hbox{Res}}\limits_{z=0}
L^{top}(z)=S=\sum\limits_{m,n}T_{mn}S_{mn}
\\
\hbox{Spec}(S)=\hbox{diag}(\nu,...,\nu,-(N-1)\nu)
 \end{array}
 }
 \end{array}
  \eq
and the boundary conditions
  \beq{i011}
 \begin{array}{c}
 \displaystyle{ \ \ \ \ L^{Cal}(z+1)=L^{Cal}(z),\ \ \ \ \ \ \ \ \ \ \ \
\ \ \ L^{top}(z+1)=QL^{top}(z)Q^{-1}}
\\
 \displaystyle{L^{Cal}(z+\tau)=\bfe(-\bfu)L^{Cal}(z)\bfe(\bfu),\ \ \ \ \ \
 L^{top}(z+\tau)=\Lambda L^{top}(z)\Lambda^{-1}\,,
 }
 \end{array}
  \eq
where
  \beq{i012}
 \begin{array}{c}
 \displaystyle{
 \bfu_{ij}=\delta_{ij}u_i\,,\ \
\Lambda_{ij}=\delta(mod_N(i+1),j)\,,\ \
Q_{ij}=\delta_{ij}\bfe(\frac{i}{N})
  }
 \end{array}
  \eq
 and
  \beq{i013}
 \begin{array}{c}
 \displaystyle{
\{T_{mn}=\bfe(\frac{mn}{2N})Q^m\Lambda^n\}\,,\ \ m,n\in{\mathbb Z}_N
 }
 \end{array}
  \eq
is the sin-algebra basis. They are of the following form
  \beq{i0121}
 \begin{array}{c}
 \displaystyle{L^{cal}(z)=\sum\limits_{i,j=1}^N
 E_{ij}\Big(\delta_{ij}v_i+\nu(1-\delta_{ij})\Phi(z,u_i-u_j)\Big)\,,\
 \ \left(E_{ij}\right)_{ab}=\delta_{ia}\delta_{jb}\,,
 }
 \end{array}
  \eq
  \beq{i01297}
 \begin{array}{c}
 \displaystyle{L^{top}(z)=\sum\limits_{n^2+m^2\neq
 0}S_{mn}T_{mn}\vf_{mn}(z)\,.
 }
 \end{array}
  \eq
provide the Hamiltonian functions
  \beq{i01213}
 \begin{array}{c}
 \displaystyle{H^{cal}=\frac{1}{2}\sum\limits_{k=1}^N
 v_k^2-\sum\limits_{i<j}\nu^2\wp(u_i-u_j)\,,
 }
 \end{array}
  \eq
  \beq{i0123}
 \begin{array}{c}
 \displaystyle{H^{top}=\frac{1}{2}\sum\limits_{n^2+m^2\neq
 0}S_{mn}S_{-m,-n}\wp(\frac{m+n\tau}{N})\,.
 }
 \end{array}
  \eq
The modification is given by the matrix \cite{LOZ1}
  \beq{i014}
 \begin{array}{c}
 \displaystyle{
 {\Xi}_{il}(z,U,\tau)=\theta \left[
\begin{array}{c}
\frac{i}{N}-\frac{1}{2}
\\
\frac{N}{2}
\end{array} \right](z-Nu_l,N\tau)\, D_l\,,\ \ \ D_l=(-1)^l \prod_{j<k;j,k\ne l}
\vartheta^{-1}(u_k-u_j,\tau)\,.
 }
 \end{array}
  \eq
It acts as gauge transformation and, therefore, establishes the
following relation between the models:
  \beq{i015}
 \begin{array}{c}
 \displaystyle{
L^{top}(z)=\Xi(z) L^{Cal}(z)\Xi^{-1}(z)-\kappa\p_z\Xi(z)
\Xi^{-1}(z)+\frac{\kappa}{N}E_1(z)\,.
 }
 \end{array}
  \eq
The last term can be removed into definition of $L^{top}$ or
$L^{Cal}$ since it is proportional to the identity matrix.



\section{Painlev\'e VI}\label{p6}
\setcounter{equation}{0}

\subsection{Three forms of Painlev\'e VI}



In this Section we describe interrelations between three forms of
the Painlev\'e VI equation and corresponding linear problems. These
three forms are:

\vspace*{3mm}

\noindent {\bf 1.} {\em Rational form} is the original one. It was
found by Gambier \cite{Gambier,Painleve1,book1,IN} and proved to be
the last and the most general in the list of the second order
ordinary differential equations satisfying the Painlev\'e property:
  \beq{i2}
 \begin{array}{c}
\displaystyle{\frac{d^2X}{dT^2}=\frac{1}{2}\left(\frac{1}{X}+\frac{1}{X-1}+
\frac{1}{X-T}\right)\left(\frac{dX}{dT}\right)^2-
\left(\frac{1}{T}+\frac{1}{T-1}+\frac{1}{X-T}\right)\frac{dX}{dT}+}
 \\ \ \\
\displaystyle{+\frac{X(X-1)(X-T)}{T^2(T-1)^2}\left(\al+\be\frac{T}{X^2}+
\gamma\frac{T-1}{(X-1)^2}+\delta\frac{T(T-1)}{(X-T)^2}\right)\,.}
 \end{array}
  \eq
It contains four free complex constants $\al\,, \be\,, \ga\,,
\delta$.

\vspace*{3mm}

\noindent {\bf 2.} {\em Elliptic form} was found by P. Painlev\'e
\cite{Painleve1906,Manin}:
  \beq{i1}
\frac{d^2u}{d\tau^2}=\sum\limits_{a=0}^{3}\nu_a^2\wp'(u+\om_a)\,,\ \
\{\om_a\,,\ k=0\,,...\,,3\}=\{0\,, \frac{1}{2}\,,
\frac{1+\tau}{2}\,, \frac{\tau}{2}\}\,.
  \eq
The equation is defined on the elliptic curve $\Sigma_\tau$ with
half-periods $\om_a$ and moduli $\tau$ which is the time variable
here. The relation between (\ref{i2}) and (\ref{i1}) is given by the
following change of variables:
  \beq{i3}
 \begin{array}{c}
\displaystyle{(u\,,\tau)\rightarrow\left(X(u,\tau)=\frac{\wp(u)-e_1}{e_2-e_1}\,,\
\ T(\tau)=\frac{e_3-e_1}{e_2-e_1}\right)\,,\ e_\al
\equiv\wp(\om_\al)\,,}
\\ \ \\
(\nu_0^2\,, \nu_1^2\,, \nu_2^2\,, \nu_3^2)=
-4\pi^2(\al\,,-\be\,,\gamma\,,-\delta+\frac{1}{2})\,.
 \end{array}
  \eq
In such a form the equation is obviously Hamiltonian. The mechanical
model is given by the Hamiltonian function
  \beq{i303}
H^{PVI}=\frac{1}{2}v^2-\sum\limits_{a=0}^{3}\nu_a^2\wp(u+\om_a)
  \eq
and the canonical Poisson bracket
  \beq{i304}
\{v,u\}=1\,.
  \eq
The mechanics is non-autonomous since two of the half-periods and
the Weierstrass $\wp$-function depend on $\tau$.

\vspace*{3mm}

\noindent {\bf 3.} {\em Non-autonomous version of the
Zhukovsky-Volterra gyrostat} was introduced and proved to be
equivalent to the Painlev\'e VI in \cite{LOZ2}:
  \beq{i401}
 \p_\tau S=[S,J(S)]+[S,\nu']\,,
  \eq
where $S$ is ${\rm sl}_2^*$-valued dynamical variable. In the basis
of Pauli matrices
  \beq{i402}
 \begin{array}{c}
\displaystyle{S=\sum\limits_{\al=1}^3S_\al\sigma_\al\,,\ \
J(S)=\sum\limits_{\al=1}^3 J_\al(\tau)S_\al\sigma_\al\,,\ \ J_\al(\tau)=\wp(\om_\al)=e_\al\,,}\ \al=1\,,2\,,3\,,\\ \ \\
\displaystyle{\nu'=\sum\limits_{\al=1}^3\nu'_\al\sigma_\al\,,\ \
\nu'_a=-\tilde{\nu}_a \exp(-2\pi
i\,\om_a\p_\tau\om_a)\left(\frac{\vth'(0)}{\vth(\om_a)}\right)^2\,,\
a=0\,,1\,,2\,,3\,,}
 \end{array}
  \eq
 where $\tilde{\nu}_a$ are four free complex constants. The
 autonomous version (when the time variable is not related to the moduli $\tau$) of (\ref{i401})
 is known as the Zhukovsky-Volterra
 gyrostat \cite{Zhuk,Zhuk02}. Vector $(\nu'_1\,,\nu'_2\,,\nu'_3)$ plays role of
 the gyrostatic momentum while $J_\al(\tau)$ are the inverse components of
 the inertia tensor in the principal axes of inertia.

 The equation (\ref{i401}) is Hamiltonian with the Hamiltonian
 function
  \beq{i403}
 \begin{array}{c}
\displaystyle{H^{ZVG}=\frac{1}{2}\sum\limits_{\al=1}^3J_\al
S_\al^2+S_\al\nu'_\al }
 \end{array}
  \eq
and the Poisson-Lie brackets on ${\rm sl}_2^*$ are
  \beq{i404}
 \begin{array}{c}
\displaystyle{  \{S_\al,S_\be\}=\epsilon_{\al\be\ga}S_\ga  \,.}
 \end{array}
  \eq
The relation between (\ref{i401}) and (\ref{i1}) is given by the
following change of variables
  \beq{i405}
\begin{array}{l}
S_{1}=-v\frac{\vtb(0)}{\vth'(0)}\frac{\vtb(2u)}{\vth(2u)}
-\frac{\ka}{2}\frac{\vtb(0)}{\vth'(0)}\frac{\vtb'(2u)}{\vth(2u)}\,+
\\
\tilde{\nu}_0\frac{\vtb^2(0)}{\vtc(0)\vtd(0)}
\frac{{\vtc(2u)\vtd(2u)}}{\vth^2(2u)}
+\tilde{\nu}_1\frac{\vtb^2(2u)}{\vth^2(2u)}+
\tilde{\nu}_2\frac{\vtb(0)}{\vtd(0)}
\frac{\vtb(2u)\vtd(2u)}{\vth^2(2u)}+\tilde{\nu}_3\frac{\vtb(0)}{\vtc(0)}
\frac{\vtb(2u)\vtc(2u)}{\vth^2(2u)}\,,
\\ \ \\
iS_{2}=v\frac{\vtc(0)}{\vth'(0)}\frac{\vtc(2u)}{\vth(2u)}+
\frac{\ka}{2}\frac{\vtc(0)}{\vth'(0)}\frac{\vtc'(2u)}{\vth(2u)}\,-
\\
\tilde{\nu}_0\frac{\vtc^2(0)}{\vtb(0)\vtd(0)}
\frac{{\vtb(2u)\vtd(2u)}}{\vth^2(2u)}
-\tilde{\nu}_1\frac{\vtc(0)}{\vtb(0)}
\frac{\vtc(2u)\vtb(2u)}{\vth^2(2u)}-
\tilde{\nu}_2\frac{\vtc(0)}{\vtd(0)}
\frac{\vtc(2u)\vtd(2u)}{\vth^2(2u)}-
\tilde{\nu}_3\frac{\vtc^2(2u)}{\vth^2(2u)}\,,
\\ \ \\
S_{3}=-v\frac{\vtd(0)}{\vth'(0)}\frac{\vtd(2u)}{\vth(2u)}
-\frac{\ka}{2}\frac{\vtd(0)}{\vth'(0)}\frac{\vtd'(2u)}{\vth(2u)}\,+
\\
\tilde{\nu}_0\frac{\vtd^2(0)}{\vtb(0)\vtc(0)}
\frac{{\vtb(2u)\vtc(2u)}}{\vth^2(2u)}+
\tilde{\nu}_1\frac{\vtd(0)}{\vtb(0)}
\frac{\vtb(2u)\vtd(2u)}{\vth^2(2u)}+\tilde{\nu}_2
\frac{\vtd^2(2u)}{\vth^2(2u)}+ \tilde{\nu}_3\frac{\vtd(0)}{\vtc(0)}
\frac{\vtd(2u)\vtc(2u)}{\vth^2(2u)}\,.
\end{array}
  \eq
and the following identification of constants:
  \beq{i406}
\begin{array}{l}
{\tilde\nu}_0=\frac{1}{2}\left( \nu_0+\nu_1+\nu_2+\nu_3 \right)\,,
\\
{\tilde\nu}_1=\frac{1}{2}\left( \nu_0+\nu_1-\nu_2-\nu_3 \right)\,,
\\
{\tilde\nu}_2=\frac{1}{2}\left( \nu_0-\nu_1+\nu_2-\nu_3 \right)\,,
\\
{\tilde\nu}_3=\frac{1}{2}\left( \nu_0-\nu_1-\nu_2+\nu_3 \right)\,.
\end{array}
  \eq
Notice that while three constants
$(\tilde\nu_1\,,\tilde\nu_2\,,\tilde\nu_3)$ enter the equation
(\ref{i401}) explicitly while the last one $\nu'_0$ appears to be
related to the value of the Casimir function of the brackets
(\ref{i404}):
  \beq{i407}
 \begin{array}{c}
\displaystyle{\frac{1}{2}\sum\limits_{\al=1}^3
S_\al^2={\tilde\nu}_0^2={\nu'_0}^2\,, }
 \end{array}
  \eq
i.e. three linear combinations of the four Painlev\'e VI constants
from (\ref{i1}) are arranged into the gyrostatic momentum vector
while the last one linear combination is the length of the angular
momentum.

\subsection{Linear problems}

Let us describe the linear problems for the above given three forms
of Painlev\'e VI equation.

\vspace*{3mm}

\noindent {\bf 1.} {\em Rational form.} The linear problems for the
Painlev\'e equations in the rational form arise naturally from the
${\rm sl}_2$ Schlesinger systems \cite{Schlesinger},
\cite{JM1,JM2,JM3} on ${\mathbb{CP}}^1\setminus \{0\,, 1\,,
T\,,\infty\}$. The later describes behavior of the connection
  \beq{i501}
 \begin{array}{c}
 \displaystyle{
\left(\p_\zeta+A(\zeta)\right)d\zeta=\Big(\p_\zeta+\frac{A^0}{\zeta}+\frac{A^1}{\zeta-1}+\frac{A^T}{\zeta-T}\Big)d\zeta
 }
 \end{array}
  \eq
with logarithmic singularities at $\{0\,, 1\,, T\,,\infty\}$ and
${\rm sl}_2^*$-valued residues. The isomonodromy equation is the
compatibility condition for the linear problem
  \beq{i5011}
 \begin{array}{c}
 \left\{
 \begin{array}{l}
 \left(\p_\zeta+A(\zeta)\right)\Psi=0
 \\
 \left(\p_T+M(\zeta)\Psi\right)=0
 \end{array}
 \right.\,,\ \ \ \
 \displaystyle{
M(\zeta)=-\frac{A^T}{\zeta-T}\,.
 }
 \end{array}
  \eq
It has a form:
  \beq{i5012}
\p_TA(\zeta)-\p_\zeta M(\zeta)=[A(\zeta,M(\zeta))]\,.
  \eq
To get the Painlev\'e VI one should perform the reduction
  \beq{i502}
 \begin{array}{c}
 \displaystyle{
{\mathcal O}^0\times{\mathcal O}^1\times{\mathcal
O}^T\times{\mathcal O}^\infty\longrightarrow {\mathcal
O}^0\times{\mathcal O}^1\times{\mathcal O}^T\times{\mathcal
O}^\infty//{\rm SL}(2,{\mathbb C})
 }
 \end{array}
  \eq
where ${\mathcal O}^a$ is the coadjoint orbit, i.e. the space of
coalgebra $A^a$ with some fixed eigenvalues which are free
constants. The reduction (\ref{i502}) is generated by the global
($\zeta$-independent) coadjoint action $A^c\rightarrow
{\hbox{Ad}}^*_{{\rm SL}(2,{\mathbb C})} A^c$, $c=0\,, 1\,,
T\,,\infty$. It provides the moment map equation
  \beq{i503}
A^0+A^1+A^T+A^\infty=0\,.
  \eq
We should also choose some coordinates on the reduced space. For the
case of $2\!\times\! 2$ L-A pair the recipe for the canonical
variables is very well-known: let $A_{12}(X)=0$ and $P=A_{11}(X)$,
then $\{P,X\}=1$ (see e.g. \cite{SoV}). After quite a long
calculation one can reproduce (\ref{i2}).

\vspace*{3mm}

\noindent {\bf 2.} {\em Elliptic form.} $2\!\times\!2$ linear
problem was suggested in \cite{Zotov04}. It is formulated in terms
of the following connection $\left(\p_z+L^{PVI}(z)\right)dz$ in the
holomorphic bundle over elliptic curve defined by transition
functions $g_1=\mat{1}{0}{0}{1}$ and
$g_\tau=\mat{\bfe(u)}{0}{0}{\bfe(-u)}$:
  \beq{i601}
 \begin{array}{c}
 \displaystyle{
L^{P_{VI}}(z)=\mat{v}{0}{0}{-v}+\sum\limits_{c=0}^3 L^{P_{VI}}_c\,,\
L^{P_{VI}}_c\!=\!\tilde\nu_c\mat{0}{\vf_c(2u,z+\om_c)}{\vf_c(-2u,z+\om_c)}{0}
 }
 \end{array}
  \eq
with $\tilde\nu_c$ defined in (\ref{i406}). Together with M-operator
  \beq{i602}
 \begin{array}{c}
 \displaystyle{
M^{P_{VI}}(z)=\sum\limits_{c=0}^3 M^{P_{VI}}_c\,,\
M^{P_{VI}}_c\!=\!\tilde\nu_c\mat{0}{f_c(2u,z+\om_c)}{f_c(-2u,z+\om_c)}{0}
 }
 \end{array}
  \eq
the Lax matrix (\ref{i601}) provides the Painlev\'e VI equation
(\ref{i1}) via
  \beq{i603}
 \begin{array}{c}
 \displaystyle{
\p_\tau L^{P_{VI}}(z)-\p_z
M^{P_{VI}}(z)=\frac{1}{\kappa}[L^{P_{VI}}(z),M^{P_{VI}}(z)]\,.
 }
 \end{array}
  \eq
Notice that the Lax matrix (\ref{i601}) is a section of the bundle
over $\Sigma_\tau=\mathbb C/\mathbb Z+\mathbb Z\tau$.
Alternatively, one can obtain the Painlev\'e VI equation from the
following L-A pair:
  \beq{i604}
 \begin{array}{c}
 \displaystyle{
\tilde L^{P_{VI}}(z)=\mat{v}{0}{0}{-v}+\sum\limits_{c=0}^3 \tilde
L^{P_{VI}}_c\,,\ \tilde
L^{P_{VI}}_c\!=\!\nu_c\mat{0}{\vf_c(z,\om_c+u)}{\vf_c(z,\om_c-u)}{0}
 }
 \end{array}
  \eq
  \beq{i605}
 \begin{array}{c}
 \displaystyle{
\tilde M^{P_{VI}}(z)=\sum\limits_{c=0}^3 \tilde M^{P_{VI}}_c\,,\
\tilde
M^{P_{VI}}_c\!=\!\nu_c\mat{0}{f_c(z,\om_c+u)}{f_c(z,\om_c-u)}{0}
 }
 \end{array}
  \eq
with $\nu_c$ from (\ref{i1}). Here the Lax matrix (\ref{i604}) is a
section of holomorphic bundle over the doubled elliptic curve
$\Sigma_{2,2\tau}=\mathbb C/2\mathbb Z+2\mathbb Z\tau$.

\vspace*{3mm}

\noindent {\bf 3.} {\em Non-autonomous Zhukovsky-Volterra gyrostat.}
The linear problem with spectral parameter on elliptic curve was
suggested in \cite{LOZ2}. Similarly to the previous case it is
formulated in terms of the following connection
$\left(\p_z+L^{ZVG}(z)\right)dz$ in the holomorphic bundle over
elliptic curve $\Sigma_\tau=\mathbb C/\mathbb Z+\mathbb Z\tau$
defined by transition functions\footnote{The scalar multiplies are
not very important here since $\p_z+L(z)$ is the connection in the
adjoint bundle.} $g_1=-Q=\mat{1}{0}{0}{-1}$ and
$g_\tau=-\bfe(-\frac{\tau}{4}-\frac{z}{2})\Lambda=-\bfe(-\frac{\tau}{4}-\frac{z}{2})\mat{0}{1}{1}{0}$:
  \beq{i701}
L^{ZVG}(z)= -\frac{\ka}{2}\p_z\ln\vth(z;\tau)\si_0+ \sum_{\al=1}^3
(S_\al\vf_\al(z)+\nu_\al\vf_\al(z-\om_\al))\si_\al \,.
  \eq
  \beq{i702} M^{ZVG}(z)=-\frac{\ka}{2}\p_\tau\ln\vth(z;\tau)\si_0
-\sum\limits_{\al=1}^3
S_\al\frac{\varphi_1(z)\varphi_2(z)\varphi_3(z)}{\varphi_\al(z)}\sigma_\al
+E_1(z)L^{ZVG}(\ka=0)\,.
  \eq
Then the monodromy preserving condition
  \beq{i703}
 \begin{array}{c}
 \displaystyle{
\p_\tau L^{ZVG}(z)-\p_z
M^{ZVG}(z)=\frac{1}{\kappa}[L^{ZVG}(z),M^{ZVG}(z)]
 }
 \end{array}
  \eq
is equivalent to equation (\ref{i401}).

Trigonometric and rational degenerations of L-A pairs for
Calogero-Gaudin-Schlesinger type systems can be found in
\cite{AA3,AA302,AA303,AA304}. One more $2\times 2$ elliptic Lax pair
for Painlev\'e VI equation arises  in the quantum version of the
Painlev\'e-Calogero Correspondence \cite{ZZ02,ZZ03}. In this case
the matrix element $L_{12}(z)$ has two simple zeros at $\pm u$ and
no poles.

\subsection{Elliptic form of rational connection}

To relate the rational and elliptic connections one needs to lift
the bundle over ${\mathbb{CP}}^1$ to some bundle over elliptic
curve. The way to do it was suggested in \cite{LZ}. Let us perform
the analogue of the substitution (\ref{i3}) for (\ref{i501}), i.e.
let us make the following change of spectral parameter in the
rational connection
$A(\zeta)=\Big(\frac{A^0}{\zeta}+\frac{A^1}{\zeta-1}+\frac{A^T}{\zeta-T}\Big)d\zeta$:
  \beq{i801}
 \begin{array}{c}
 \displaystyle{
\zeta=\zeta(z,\tau)=\frac{\wp(\frac{z}{2})-e_1}{e_2-e_1}\,,\ \ \ \
T=T(\tau)=\frac{e_3-e_1}{e_2-e_1}\,.
 }
 \end{array}
  \eq
This gives
  \beq{i802}
 \begin{array}{c}
 \displaystyle{
A(\zeta)d\zeta=\sum\limits_{\ga=0,1,T}
\frac{A^\ga\wp'(\frac{z}{2})}{\wp(\frac{z}{2})-\wp(\om^\ga)}\frac{dz}{2}=
\sum\limits_{\ga=0,1,T}A^\ga\left(E_1(\frac{z}{2}+\om^\ga)-E_1(\frac{z}{2})-E_1(\om^\ga)\right)dz\,,
 }
 \end{array}
  \eq
where the half-periods  $\om^0,\,\om^1,\,\om^T$ of $\Sigma_\tau$ are
identified with the notation $\om_a$ from (\ref{i1}) as follows:
  $$
\om^0=\om_1=\frac{1}{2}\,,\ \ \ \om^1=\om_2=\frac{1+\tau}{2}\,,\ \ \
\om^T=\om_3=\frac{\tau}{2}\,.
  $$
Notice that
  \beq{i8037}
 \begin{array}{c}
 \displaystyle{
E_1(\frac{z}{2}+\om_\ga)-E_1(\frac{z}{2})-E_1(\om_\ga)=-\vf_\al(z)-\vf_\be(z)
 }
 \end{array}
  \eq
for any different indices $\al,\be,\ga$ running over $\{1,2,3\}$.
Indeed, both sides of (\ref{i8037}) are double-periodic functions on
the doubled elliptic curve $\Sigma_{2,2\tau}$ with simple poles at
its half-periods $(0,1,\tau+1,\tau)$. It only remains to compare the
residues. Therefore, we can rewrite (\ref{i802}) in the following
way:
  \beq{i804}
 \begin{array}{c}
 \displaystyle{
A(\zeta)d\zeta=\sum\limits_{\al=1}^3 B^\al\vf_\al(z)dz\,,
 }
 \end{array}
  \eq
where
  \beq{i805}
 \left\{\begin{array}{l}
 \displaystyle{
 B^1=-A^1-A^T
 }
 \\
 \displaystyle{
 B^2=-A^0-A^T
 }
  \\
 \displaystyle{
 B^3=-A^0-A^1
 }
 \end{array}\right.
 \hspace*{12mm}
\hbox{or}
 \hspace*{12mm}
 \left\{\begin{array}{l}
 {
 A^{\infty}\!=}\frac{1}{2}{\left( +B^1+B^2+B^3 \right)
 }
 \\
 {
 A^0=}\,\frac{1}{2}{\left( +B^1-B^2-B^3 \right)
 }
 \\
 {
 A^1=}\,\frac{1}{2}{\left( -B^1+B^2-B^3 \right)
 }
  \\
 {
 A^T=}\,\frac{1}{2}{\left( -B^1-B^2+B^3 \right)
 }
 \end{array}\right.
  \eq
The expression
   \beq{i803}
 \begin{array}{c}
 \displaystyle{
L^{ell}(z)=\sum\limits_{\al=1}^3 B^\al\vf_\al(z)
 }
 \end{array}
  \eq
  is the
double-periodic function on the doubled elliptic curve
$\Sigma_{2,2\tau}$ with simple poles at its half-periods
$(0,1,\tau+1,\tau)$. The residues equal $2A^\infty$, $2A^0$, $2A^1$,
$2A^T$ respectively.

\subsection{Symplectic Hecke Correspondence}

The modification of a bundle can be considered as the procedure
which relates the bundles with different characteristic classes and
established the Symplectic Hecke Correspondence \cite{LOZ1,LOSZ1}.
In the ${\rm SL}(2,\mathbb C)$ case the characteristic classes are
elements of ${\mathbb Z}_2$, i. e. $\pm 1$. The trivial one "$+1$"
corresponds to the system (\ref{i601})-(\ref{i603}) while "$-1$" --
to (\ref{i701})-(\ref{i703}).
The Symplectic Hecke Correspondence predicts relation between
elliptic L-A pairs (\ref{i601})-(\ref{i603}) and
(\ref{i701})-(\ref{i703}) generated by modification (\ref{i014}). In
the ${\rm SL}(2,\mathbb C)$ case it is of the form:
  \beq{i901}
 \begin{array}{c}
 \displaystyle{
\Xi(z)=\mat{\theta_3(z-2u,2\tau)}{-\theta_3(z+2u,2\tau)}
{-\theta_2(z-2u,2\tau)}{\theta_2(z+2u,2\tau)}\,.
 }
 \end{array}
  \eq
It maps the holomorphic bundle for $P_{VI}$ into the one for $ZVG$.
Therefore, the corresponding connections are related  as follows:
  \beq{i902}
 \begin{array}{c}
 \displaystyle{
L^{ZVG}(z)=\Xi(z) L^{P_{VI}}(z)\Xi^{-1}(z)-\kappa\p_z\Xi(z)
\Xi^{-1}(z)\,.
 }
 \end{array}
  \eq
This relation provides the change of variables (\ref{i405}) (see
\cite{LOZ2}).

Similarly, the modification (\ref{i901}) relates the connections
(\ref{i604}) and (\ref{i803}) on the curve $\Sigma_{2,2\tau}$. The
connection (\ref{i604}) has non-trivial automorphism: it is
conjugated by $\mat{0}{1}{1}{0}$ when $z\rightarrow -z$ (the later
corresponds to changing the branch of the elliptic curve realized as
the two fold covering of ${\mathbb{CP}}^1$ ramified at four points).
It can be verified that the modification (\ref{i901}) trivializes
the automorphism. Finally, we have
  \beq{i9025}
 \begin{array}{c}
 \displaystyle{
L^{ell}(z)-\frac{\kappa}{2}E_1(z)\sigma_0=\Xi(z) \tilde
L^{P_{VI}}(z)\Xi^{-1}(z)-\kappa\p_z\Xi(z)\Xi^{-1}(z)\,.
 }
 \end{array}
  \eq
This relation provides the non-trivial parametrization of the space
${\mathcal O}^0\times{\mathcal O}^1\times{\mathcal
O}^T\times{\mathcal O}^\infty//{\rm SL}(2,{\mathbb C})$ in terms of
canonical variables $v,u$. It means that $B^\al$ from (\ref{i803})
and, therefore,  $A^0$, $A^1$, $A^T$ and $A^\infty$ are obtained as
functions of $v,u$. Explicit expressions were found in \cite{LZ}:
 \beq{zzz5} B^1\left.\right|_{\ka=0}=
\mat{-v\frac{\vtb(0)\vtb(u)}{\vth'(0)\vth(u)}
-\nu_1\frac{\vtb^2(0)}{\vtc(0)\vtd(0)}\frac{\vtc(u)\vtd(u)}{\vth^2(u)}}
{\nu_3\frac{\vtb(0)\vtb(u)}{\vtc(0)\vtc(u)}-
\nu_4\frac{\vtb(0)\vtb(u)}{\vtd(0)\vtd(u)}}
{\nu_3\frac{\vtb(0)\vtb(u)}{\vtc(0)\vtc(u)}+
\nu_4\frac{\vtb(0)\vtb(u)}{\vtd(0)\vtd(u)}}
{v\frac{\vtb(0)\vtb(u)}{\vth'(0)\vth(u)}+
\nu_1\frac{\vtb^2(0)}{\vtc(0)\vtd(0)}\frac{\vtc(u)\vtd(u)}{\vth^2(u)}
}
 \end{equation}
 \beq{zzz7} B^2\left.\right|_{\ka=0}= \left(
\begin{array}{c}
{-\nu_3\frac{\vtd(0)\vtd(u)}{\vtc(0)\vtc(u)}}\ \ \ \ \ \ \ \ \ \
{-v\frac{\vtd(0)\vtd(u)}{\vth'(0)\vth(u)}-
\nu_1\frac{\vtd^2(0)}{\vtb(0)\vtc(0)}\frac{\vtb(u)\vtc(u)}{\vth^2(u)}+
\nu_2\frac{\vtd(0)\vtd(u)}{\vtb(0)\vtb(u)}}
\\
{-v\frac{\vtd(0)\vtd(u)}{\vth'(0)\vth(u)}-
\nu_1\frac{\vtd^2(0)}{\vtb(0)\vtc(0)}\frac{\vtb(u)\vtc(u)}{\vth^2(u)}+
\nu_2\frac{\vtd(0)\vtd(u)}{\vtb(0)\vtb(u)}}\ \ \ \ \ \ \ \ \ \
{\nu_3\frac{\vtd(0)\vtd(u)}{\vtc(0)\vtc(u)}}
\end{array}
\right)
 \end{equation}
 \beq{zzz6} B^3\left.\right|_{\ka=0}= \left(
\begin{array}{c}
{-\nu_4\frac{\vtc(0)\vtc(u)}{\vtd(0)\vtd(u)}}\ \ \ \ \ \ \ \ \ \
{-v\frac{\vtc(0)\vtc(u)}{\vth'(0)\vth(u)}-
\nu_1\frac{\vtc^2(0)}{\vtb(0)\vtd(0)}\frac{\vtb(u)\vtd(u)}{\vth^2(u)}
+\nu_2\frac{\vtc(0)\vtc(u)}{\vtb(0)\vtb(u)}}\\
{v\frac{\vtc(0)\vtc(u)}{\vth'(0)\vth(u)}+
\nu_1\frac{\vtc^2(0)}{\vtb(0)\vtd(0)}\frac{\vtb(u)\vtd(u)}{\vth^2(u)}
+\nu_2\frac{\vtc(0)\vtc(u)}{\vtb(0)\vtb(u)}}\ \ \ \ \ \ \ \ \ \
{\nu_4\frac{\vtc(0)\vtc(u)}{\vtd(0)\vtd(u)}}
\end{array}
\right)
 \end{equation}
 \beq{zzz8} \ka\p_z\Xi\Xi^{-1}=
\frac{\ka}{2}\mat{\frac{\vtb(0)\vtb'(u)}{\vth'(0\vth(u)}\vf_{10}(z)+E_1(z)}
{\frac{\vtc(0)\vtc'(u)}{\vth'(0\vth(u)}\vf_{01}(z)+
\frac{\vtd(0)\vtd'(u)}{\vth'(0\vth(u)}\vf_{11}(z)}
{-\frac{\vtc(0)\vtc'(u)}{\vth'(0\vth(u)}\vf_3(z)+
\frac{\vtd(0)\vtd'(u)}{\vth'(0\vth(u)}\vf_{11}(z)}
{-\frac{\vtb(0)\vtb'(u)}{\vth'(0\vth(u)}\vf_{10}(z)+E_1(z)}
 \end{equation}


\part{Appendices}

\section*{Appendix A: Simple Lie groups}
\setcounter{equation}{0}
\def\theequation{A.\arabic{equation}}

\addcontentsline{toc}{section}{Appendix A: Simple Lie groups}

Most of facts and notations are borrowed from \cite{Bou,OV}.

\emph{\textbf{Roots and weights.}}\\
 $V$ - a  vector space over $\mR$, $\dim\,V=n$;\\
  $V^*$ is its dual and  $\lan~,~\ran$ is a pairing between $V$ and  $V^*$.
 $R=\{\al\}$ is  a root system in $V^*$. \\
The dual  system $R^\vee=\{\al^\vee\}$ is the root system in $V$.\\
If $V$ and $V^*$ are identified by a scalar product $(~,~)$, then
$\al^\vee=\frac{2\al}{(\al,\al)}$.\\
The group of automorphisms of $V^*$ generated by reflections
  \beq{ref}
s_\al\,:~x\mapsto x-\lan x,\al\ran\al^\vee
  \eq
is \emph{the Weyl group} $W(R)$.\\
 \emph{Simple roots} $\Pi=(\al_1,\ldots,\al_l)$ form a basis in $R$
  \beq{ale1}
\al=\sum_{j=1}^nf^\al_j\al_j\,, ~~f^\al_j\in\mZ\,,
  \eq
and all $f^\al_j$ are positive (in this case $\al\in R^+$ is a
positive root), or negative ($\al$ is a negative root).  $R=R^+\cup
R^-$.
 \emph{The level } of $\al$ is the sum
  \beq{ale}
f_\al=\sum_{\al_j\in\Pi}f_j\,.
  \eq
{The Cartan matrix} is
  \beq{CMa}
a_{jk}=\lan
\al_j,\al_k^\vee\ran\,,~~\al_j\in\Pi\,,~\al_k^\vee\in\Pi^\vee\,.
  \eq
%
%
 The simple roots generate the root lattice
 $ Q=\sum_{j=1}^nn_j\al_j\,,~~(n_j\in\mZ\,,\,\al_j\in\Pi)$
 in $V^*$.
There exists a unique  \emph{a maximal root}  $-\al_0\in R^+$
  \beq{mro}
-\al_0 =\sum_{\al_j\in\Pi}n_j\al_j\,.
  \eq
Its level is equal to $h-1$, where
  \beq{co11}
h=1+\sum_{\al_j\in\Pi}n_j
  \eq
 is \emph{the Coxeter number}.
\emph{The positive Weyl chamber} is
  \beq{wca}
C^+=\{x\in V\,|\,\lan x,\al\ran>0\,,~\al\in R^+\}\,.
  \eq
 The Weyl group acts simply-transitively  on the set of the  Weyl chambers.
 The simple coroots $\Pi^\vee=(\al^\vee_1,\ldots,\al^\vee_l)$ form a basis in $V$ and
generate\emph{ the coroot lattice}
  \beq{cjrl}
Q^\vee=\sum_{j=1}^nn_j\al^\vee_j\subset V\,,~~~n_j\in\mZ\,.
  \eq
 \emph{The weight lattice}
$P=\sum_{j=1}^nm_j\varpi_j\subset V^*\,,~m_j\in\mZ$
 is dual to the coroot lattice (\ref{cjrl}).
 \emph{The fundamental coweights} are dual to simple roots
  \beq{bor}
\Upsilon^\vee=\{\varpi_j^\vee\in\gh\,,~j=1,\ldots,n\,|\, \lan\al_k,\varpi_j^\vee\ran=\de_{kj}
~~\al_j\in\Pi\}\,.
  \eq
They generate \emph{the coweight lattice}
  \beq{cwl}
 P^\vee=\sum_{j=1}^lm_j\varpi^\vee_j\,,~~m_j\in\mZ
  \eq
  dual to the root lattice $Q$.
%
The half-sum of positive roots is $\rho=\oh\sum_{\al\in
R^+}\al=\oh\sum_{j=1}^n\varpi_j$. For the dual vector in $V$ we have
  \beq{arho}
\rho^\vee=\oh\sum_{\al\in R^{\vee +}}\al^\vee=\sum_{j=1}^n\varpi^\vee_j\,.
  \eq
\emph{\textbf{Affine Weyl group.}}\vskip 3mm

\emph{The affine Weyl group} $W_a$ is  $Q^\vee\rtimes W$
  \beq{sh}
W_Q=Q^\vee\rtimes W\,,~~
x\to x-\lan\al,x\ran\al^\vee+k\be^\vee\,,~\al^\vee,\,\be^\vee\in R^\vee\,,~k\in\mZ\,.
  \eq
 \emph{The Weyl alcoves} are
connected components of the set $V\setminus\{\lan\al,x\ran\in\mZ\}$.
 Their closure are fundamental domains of the $W_a$-action.\\
An alcove belonging to $C^+$ (\ref{wca})
  \beq{gran}
C_{alc}=\{\,x\in V\,|\,\lan\al,x\ran> 0\,,~\al\in\Pi\,,~(\al_0,x)>- 1\,\}\,.
  \eq
 The shift operator $ x\to x+\ga\,,$ $\ga\in P^\vee$ generates
 a semidirect product
   \beq{q25}
 W_P=P^\vee\rtimes W\,.
   \eq
 The factor group is isomorphic to the center
 $  W_P /W_Q\sim P^\vee/Q^\vee\sim\clZ(\bG)$.

\bigskip
\emph{\textbf{Chevalley basis in $\gg$}.}\\
Let  $\gg$ be  a simple Lie algebra over $\mC$ of rank $n$ and $\gh$ is a Cartan subalgebra.
Let $\gh=V+iV$, where $V$ is the vector space defined above with the
root system $R$.
The algebra $\gg$ has the root decomposition
  \beq{CD1}
\gg=\gh+\gl\,,~~\gl=\sum_{\be\in R}\gR_\be\,,  ~~\dim_\mC\,\gR_\be=1\,.
  \eq
The Chevalley basis in $\gg$ is generated by
  \beq{CBA3}
\{E_{\be_j}\in\gR_{\be_j}\,,~\be_j\in R\,,~~H_{\al_k}\in\gh\,,~\al_k\in\Pi\}\,,
  \eq
where $H_{\al_k}$ are defined by the commutation relations
   $$
[E_{\al_k},E_{-\al_k}]=H_{\al_k}\,,~~[H_{\al_k},E_{\pm\al_j}]= a_{kj}E_{\pm\al_k}\,,~~\al_k\,,\al_j\in\Pi\,.
   $$
  \beq{cbcr} [H_{\al_j},E_{\al_k}]=a_{kj}E_{\al_k}\,,~~~[E_\al,E_{\be}]=C_{\al,\be}E_{\al+\be}\,,   \eq where $C_{\al,\be}$ are
structure constants of $\gg$.
They possess  the  properties
   \beqn{scp}
 \begin{array}{l}
C_{\alpha,\beta}=-C_{\beta,\alpha}\\
C_{\lambda\alpha,\beta }=C_{\alpha,\lambda^{-1}\beta}\,,~~~~\la\in W\,,\\
C_{\alpha+\beta,-\alpha}=\frac{|\beta|^2}{|\alpha+\beta|^2}\,C_{-\alpha,-\beta}
\end{array}
   \eqn

If $(~,~)$ is a scalar product in $\gh$ then $H_\al$ can be identified with coroots as
$H_\al=\al^\vee=\frac{2\al}{(\al,\al)}$ and
  \beq{kilh}
(H_\al,H_\be)=\frac{4(\al,\be)}{(\al,\al)(\be,\be)}=\frac{2}{(\al,\al)}a_{\al,\be}\,.
  \eq
The Killing form in the subspace $\gl$ is expressed in terms of $(\al,\al)$
  \beq{are6}
(E_\al,E_\be)=\de_{\al,-\be}\frac{2}{(\al,\al)}\,.
  \eq

\bigskip

\noindent



\bigskip

\noindent
\emph{\textbf{Centers of simple groups}}.\\
 A  simply-connected group $\bar G$
 in all cases apart $G_2$, $F_4$ and $E_8$ has a non-trivial center
 $\clZ(\bar G)\sim P^\vee/Q^\vee$.

\begin{center}

\begin{tabular}{|c|c|c| }
  \hline
   $\bG$ &Lie $(\bar{G})$ & $\clZ(\bar{G})$ \\
 \hline
SL$(n,\mC)$ &  $A_{n-1}$ & $\mu_n$  \\
Spin$_{2n+1}(\mC)$&  $B_n$ & $\mu_2$  \\
Sp$_n(\mC)$&  $C_n$ & $\mu_2$   \\
Spin$_{4n}(\mC) $&  $D_{2n}$& $\mu_2\oplus\mu_2$     \\
Spin$_{4n+2}(\mC) $&  $D_{2n+1}$ & $\mu_4$   \\
$E_6(\mC)$ &  $E_6$ & $\mu_3$   \\
$E_7(\mC)$ &  $E_7$ & $\mu_2$   \\
  \hline
\end{tabular}
\\
\vspace{3mm}
\textbf{Table 2}\\
Centers of universal covering groups
\\
($\mu_N=\mZ/N\mZ$)
\end{center}
\vspace{5mm}
  $\clZ(\bG)$ is a cyclic group except $\gg=D_{4l}$, and  $ord \,(\clZ(\bG))=\det\,(a_{kj})$,
  where $(a_{kj})$ is the Cartan matrix.
  \beq{adg}
G^{ad}=\bar G/\clZ(\bar G)\,.
  \eq
In the cases $A_{n-1}$ ($n$ is non-prime), and $D_n$ the center
$\clZ(\bar G)$ has non-trivial subgroups $\clZ_l\sim\mu_l=\mZ/l\mZ$.
Then there exists the factor groups
  \beq{fgl}
G_l=\bG/\clZ_l\,,~~~G_p=G_l/\clZ_p\,,~~~G^{ad}=G_l/\clZ(G_l)\,,
  \eq
where $\clZ(G_l)$ is the center of $G_l$ and $\clZ(G_l)\sim\mu_p=\clZ(\bar G)/\clZ_l$.\\
The group $\bar G=Spin_{4n}(\mC)$ has a non-trivial center
  $$
 \clZ(Spin_{4n})=(\mu^L_2\times\mu^R_2)\,,~~\mu_2=\mZ/2\mZ\,,
  $$
where three subgroups can be described in terms of their generators as
  $$
\mu^L_2=\{(1,1)\,,(-1,1)\}\,,~~\mu^R_2=\{(1,1)\,,(1,-1)\}\,,~~\mu^{diag}_2=\{(1,1)\,,(-1,-1)\}\,.
  $$
 Therefore there are three intermediate subgroups between $\bar G=Spin_{4n}(\mC)$
 and $G^{ad}$
  \beq{spin4}
\begin{array}{ccccc}
   &   & Spin_{4n} &  \\
   & \swarrow & \downarrow & \searrow &  \\
  Spin_{4n}^{R}= Spin_{4n}/\G^L &   & SO(4n)= Spin_{4n}/\G^{diag}&  &Spin_{4n}^{L}=Spin_{4n}/\G^R\\
   & \searrow & \downarrow & \swarrow &  \\
   &  & G^{ad}=Spin_{4n}/(\mu^L_2\times\mu^R_2) &  &
\end{array}
  \eq

\bigskip

\noindent
\emph{\textbf{Characters and cocharacters.}}\\
Let $\clH$ be a Cartan subgroup $\clH\subset G$.
 Define the group of characters
\footnote{
The holomorphic maps of $\clH$ to $\mC^*$ such that $\chi(xy)=\chi(x)\chi(y)$ for $x,y\in\clH$.}
  \beq{cha}
\G( G)=\{\chi\,:\,\clH\to\mC^*\}\,.
  \eq
This group can be identified with a lattice group in $\gh^*$ as follows.
Let $\bfx=(x_1,z_2,\ldots,x_n)$ be an element of $\gh$, and
$\exp\,2\pi i\bfx\in\clH$. Define $\ga\in V^*$ such that
$\chi_\ga=\exp 2\pi i \lan\ga,\bfx\ran\in\G( G)$. Then
  \beq{char}
\G(\bar G)=P\,,~~\G(G^{ad})=Q\,,
  \eq
and $\G(G^{ad})\subseteq\G(G_l)\subseteq\G(\bar G)$.
The fundamental weights $\varpi_k\,$ $(k=1,\ldots,n)\,$
(simple roots $\al_k$) form a basis in $\G(\bar G)$
($\G(G^{ad})$).
Let $\clZ(\bar G))$ be a cyclic group and $p$ is a divisor of $ord\,(\clZ(\bar G))$
such that $l=ord\,(\clZ(\bar G))/p$.
Then the lattice $\G(G_l)$ is defined as
  \beq{cg}
\G(G)=Q+\varpi\mZ\,,~~ p\varpi\in Q\,.
  \eq
Define the dual groups   of cocharacters  $t(G)=\G^*(G)$
as holomorphic maps
  \beq{coch}
t(G)=\{\mC^*\to\clH\}\,.
  \eq
In another way
  \beq{tb1}
t(G)=\{\bfx\in \gh\,|\,\chi(e^{2\pi i\bfx})=1\}\,.
  \eq
A generic element of $t(G)$ takes the form
  \beq{cocharc1}
z^\ga=\exp\,2\pi i \ga\ \ \ln z\in\clH_G\,,~~\ga\in
\G^*(G)\,,~~~z\in\mC^*\,.
  \eq
In particular, the groups $t(\bar G)$ and $t(G_{ad})$
 are identified with the coroot and the coweight lattices
  \beq{tb}
t(\bar G)=Q^\vee\,,~~t(G^{ad})=P^\vee\,,
  \eq
and $t(\bar G)\subseteq t(G_l)\subseteq t(G^{ad})$.
It follows from (\ref{cg}) that
  \beq{coch1}
t(G)=Q^\vee+\varpi^\vee\mZ\,,~~l\varpi^\vee\in Q^\vee\,.
  \eq
The sublattice $t(G_l)\subset P^\vee$ defines the affine Weyl group
  \beq{wgl}
W_{t(G)}=t(G)\rtimes W
  \eq
(see (\ref{sh}), (\ref{q25})).
The center $\clZ(G)$ of $G$ is isomorphic to the quotient
  \beq{cG}
\clZ(G)\sim P^\vee/t(G)\,,
  \eq
while $\pi_1(G)\sim t(G)/Q^\vee$.
In particular,
  \beq{center}
\clZ(\bar G)=P^\vee/t(\bar G)\sim P^\vee/Q^\vee\,.
  \eq
Similarly, the fundamental group of $G^{ad}$ is
$\pi_1(G^{ad})\sim t(G^{ad})/Q^\vee\sim P^\vee/Q^\vee$.
The triple $(R,\,t(G),\,\G(G))$ is called \emph{the root data}.
%
%

\bigskip

\noindent
\emph{\textbf{Parabolic subgroups and flag varieties}}\\
 Let $\Pi'$ be a subset of simple roots $\Pi'\subset\Pi(\gg)$ and $\gg'$ be a semi-simple
subalgebra of $\gg$ corresponding to $\Pi'$. Let $\gh'$ be the Cartan subalgebra $\gg'$.
Then the Cartan algebra $\gh\subset\gg$ has the decomposition
$\gh=\gh^{'}\oplus\ti\gh$. Similarly, we have for the coalgebras
   \beq{ccd}
\gh^*=\gh^{'*}\oplus\ti\gh^*\,.
   \eq
These data defines a parabolic subalgebra $\gp$ of $\gg$. Let
$R^{'}$ are roots generated by $\Pi'$ and $\ti R=R\setminus R^{'}$.
The parabolic subalgebra is the semi-direct sum of the reductive
subalgebra -- \emph{(the Levi subalgebra)} and the nilpotent ideal
  \beq{psa}
\gp=\gs(\Pi')\oplus\gn^+(\ti R^+)\,,~~
\gs=\gg'(\Pi')\oplus\ti\gh\,,~~\gn^+= \oplus_{\al\in\ti R^+ }c_\al E_\al\,.
  \eq
Consider the decomposition of $\gg$ into parabolic and nilpotent subalgebras
  \beq{pns8}
\gg=\gp\oplus \gn^{-}\,,~~\gn^{-}=\oplus_{\al\in \ti R^-}c_\al E_\al\,.
  \eq
Let $P\subset G$ be the parabolic subgroup defined by $\gp$, $N^+$ is
 the unipotent subgroup $\gn^{+}=Lie\,(N^+)$ and $L$ is a Levi subgroup
 with Lie algebra
   \beq{les}
 Lie(L)=\gs=\gg'\oplus\ti\gh\,,~~(\ref{psa})\,.
   \eq
  The parabolic subgroup
 $P$ is the semidirect product
   \beq{para}
 P=L\ltimes N^+\,,~~Lie\,(N^+)=\gn^{+}\,.
  \eq
If $\Pi'=\varnothing$, then $P=B$ is a\emph{ Borel subgroup}.
For $G$ we have \emph{the Bruhat decompositions}
   \beq{aal}
G=\bigcup_{ w\in W}N^- wP
 =\bigcup_{ w\in W}N^+wP\,,~~Lie\,( N^-)=\gn^-\,.
  \eq
The coset $G/P=Flag$ is called \emph{the $G$-flag variety}. The $N^+$-orbits of $w$ in $Flag$
are \emph{the Schubert cells}.
If  $P=B$ is a Borel subgroup  the flag $Flag=G/B$ is a full $G$-flag.
Let $G^{comp}$ be the compact form of the complex group $G$. There
is the Iwasawa decomposition
 \beq{iw}
 G=G^{comp}P\,, ~~G^{comp}\cap
P=L^{comp}\,,~~L^{comp}=L\cap G^{comp}\,.
  \eq
 Thereby, the flag
varieties are orbits of the compact groups
 \beq{como}
 Flag =G/P\sim G^{comp}/L^{comp}\,.
 \eq
The parabolic subalgebras can be defined by means of fundamental
coweights. Let $\ga\in \Upsilon^\vee$ (\ref{bor}). Its Lie algebra
has the form (see (\ref{psa}))
 \beq{pist}
 Lie(P_\ga)=\gp_\ga=\gs_\ga(\Pi')\oplus\gn_\ga^+(\ti R^+)\,, ~~
 \Pi'=\{\al\in\Pi\,|\,\lan\al,\ga\ran=0\}\,.
  \eq
  The
corresponding parabolic subgroup $P_\ga$ subgroup is maximal and
 the semi-simple  subalgebra $\gg'$ (\ref{les}) has rank $n-1$.
We need a special class of parabolic subalgebras. We call a parabolic subalgebras and
the corresponding Levi algebras  \emph{admissible} if
 \beq{adm}
\lan\ga,\al\ran=1~\forall ~\al\in\ti R^+
 \eq
 Define admissible fundamental coweights as
 \beq{afc}
 \hat{\Upsilon}^\vee=\{\ga\in \Upsilon^\vee\,|\,\ga\notin\Pi^\vee\,,~
 \lan\ga,\al\ran=1~\forall ~\al\in\ti R^+\}\,.
 \eq
Note, that for classical algebras any non-trivial fundamental
coweight $\ga\in \Upsilon^\vee$, $\ga\notin\Pi^\vee$ defines
admissible parabolic and Levi subalgebras. It is not the case for
the exceptional  algebras $e_6$ and $e_7$.

Using notations of roots and weights from \cite{Bou} we list
 admissible fundamental coweights and semisimple components of admissible Levi algebras
 of simple Lie groups with non-trivial centers.

\begin{center}

\begin{tabular}{|c|c|c|c| }
  \hline
   $\bG$ &$\hat{\Upsilon}^\vee$ & $\gg'$ & $G/P$ \\
 \hline
SL$(n,\mC)$ &  $\Upsilon^\vee$ & sl$(n-p)\oplus$sl$(p)$& $Gr(n,p)=$S(U$(n-p)\times$U$(p)$) \\
Spin$_{2n+1}(\mC)$&  $\varpi^\vee_1$ & so$(2n-1)$ & SO$(2n+1,\mR)/$SO$(2n-1,\mR)\times$SO$(2)$  \\
Sp$_n(\mC)$&  $\varpi^\vee_1$ & sp$(n-1)$ & Sp$(n,\mR)/$Sp$(n-1,\mR)\times$SO$(2)$   \\
Spin$_{2n}(\mC) $&  $\varpi^\vee_1$& so$(2n-2)$ & SO$(2n,\mR)/$SO$(2n-2,\mR)\times$SO$(2)$ \\
Spin$_{2n}(\mC) $&  $\varpi^\vee_{n-1.n}$ &  sl$(n)$ & SO$(2n,\mR)/$SU$(n-1)\times$SO$(2)$ \\
$E_6(\mC)$ &  $\varpi^\vee_1$ &so$(10)$& $E_6^{Comp}/$SO$(10)\times $SO$(2)$   \\
$E_7(\mC)$ &  $\varpi^\vee_7$ & $E_6$  & $E_7^{Comp}/E_6^{Comp} \times $SO$(2)$\\
  \hline
\end{tabular}
\\
\vspace{3mm}
\textbf{Table 3}\\
Admissible Levi subalgebras\\
and flag varieties
\end{center}

If we extend the condition (\ref{adm})
  \beq{adm1}
\lan\ga,\al\ran=1~{\rm or~} 0~\forall ~\al\in\ti R^+
  \eq
then any parabolic subalgebra of admissible parabolic algebra is admissible. In particular,
the Borel subalgebras are admissible.

\bigskip\vskip2mm

\noindent \emph{\textbf{Coadjoint orbits}}\\
 The cotangent bundle $T^*G$ to $G$ is equipped with the symplectic form
  \beq{sfcb}
\om=\de\lan a,g^{-1}\de g\ran\,,~~a\in\gg^*\,,~g\in G\,.
  \eq
The form is invariant under the parabolic subgroup actions $f_{out},f_{int}\in P$
  \beq{ag1}
Ad^*_{f_{int}}(a)=f_{int}^{-1}af_{int}\,,~~g\to gf_{int}\,,
  \eq
  \beq{gg1}
a\to a\,,~~g\to f_{out}g\,.
  \eq
The elements $\ep_{int},\ep_{out}\in\gp$ generate the vector fields
defined by the transformations (\ref{ag1}), (\ref{gg1}). Their  Hamiltonians
$F_{int}$, $F_{out}$ $\,(\de F_{int,out}=i_{\ep_{int,out}}\om)$ assumes the form
  \beq{hin}
F_{int}=\lan\ep_{int},a\ran\,,~~
F_{out}=\lan\ep_{out},gag^{-1}\ran\,.
  \eq
 The moments of these actions take values in
$\gp^*_a=\gs\oplus\gn_a^-$ (see (\ref{psa}) ).
The moment corresponding to the ${int}$ action is equal to
 $\mu_{int}=Pr|_{\gp^*}(a)$. Fix its value as
$\mu_{int}=\nu\in\gs$. It means that solution of the moment equation is
  \beq{sme}
a=\nu+\xi\,, ~~\xi\in\gn^+\,,
  \eq
where $\nu$ is fixed and $\xi$ is an arbitrary element of $\gn^+$.

The coadjoint action of $P$ preserves $\nu\in\gs$ and the reduced symplectic
manifold
  \beq{rcb}
T^*G//P=\mu_{int}^{-1}(\nu)/P\,.
  \eq
 It follows from (\ref{ag1}) and (\ref{sme}) that the symplectic quotient
 $T^*G//P$ is defined by the pairs $(g,\nu+\xi)$, $g\in G$, $\xi\in\gn^+$ with
 the equivalence relation
   \beq{coo}
 (gb,(Ad_b^*)(\nu+\xi))\sim(g,\nu+\xi)\,,~~b\in P\,.
   \eq
 Note, that the coadjoint action of $P$ on $\xi\in\gn^+$ is the affine action due to the
 $\nu$ term. The group $P$ acts free on $\gn^+$.
It means that $T^*G//P$  is the principal homogeneous space
$PH/T^*(G/P)$ over the cotangent bundle to the flag variety $G/P$.
The cotangent bundle corresponds to the choice $\nu=0$.

Let us fix a gauge of the $P$-action by the choice $\xi=0$. It follows from (\ref{psa}) and
(\ref{para}) that the Levi subgroup $L$ preserves $\xi=0$.
Then from (\ref{coo}) we find that
  \beq{dor0}
\clO_\nu=(Ad^*)^{-1}_G\nu=G/L\,.
  \eq
The form $\om$ on $T^*G$ (\ref{sfcb}) becomes the  Kirillov-Kostant
form
  \beq{kkf}
\om^{KK}=\lan\nu, g^{-1}\de g g^{-1}\de g\ran\,.
  \eq
The dimension of the orbit is
   \beq{dor}
 \dim\,(\clO_\gp)=\sharp\,(\ti R)\,.
     \eq

\bigskip

\noindent
\emph{\textbf{Affine Lie algebras} \cite{Ka}}

The affine root system is defined as
  $$
R^{aff}=\left\{\hat\al=\al+n\,|\,\begin{array}{cc}
                              \al\in R\cup 0 & n\in\mZ\setminus 0 \,,\\
                              \al\in R & n=0\,.
                            \end{array}
\right\}\,.
  $$
  \beq{nar}
R^{aff}_+=R^{aff}~{\rm for~}n>0\,,~{\rm or~} \al\in R^+\,,~n=0\,,
~~R^{aff}_-=R^{aff}\setminus R^{aff}_+\,.
  \eq
The affine Lie algebra can be represented as
$L(\gg)\oplus \mC{\bf c}\oplus\mC {\bf d}$, where
  \beq{la}
L(\gg)=\gg\otimes C[t^{-1},t]=\left\{\,\sum_kx_kt^k\,,~x_k\in\gg\,\right\}\,,
  \eq
${\bf c}$ is the generator of the center and ${\bf d}$ is derivation of $L(\gg)$.
The affine roots corresponds to the affine root subspaces of $L(\gg)$
  $$
E_{\hat\al}=E_\al t^n\,,~~~H_{\hat\al}=H_{\al}t^n\,~(n\neq 0)\,.
  $$
Let $\clH$ be a Cartan subgroup of $G$.
 Realize the Weyl group $W$ of $\gg$  as the quotient $\clN(\clH)/\clC(\clH)$,
 where $\clN(\clH)$ ($\clC(\clH)$) is the normalizer (centralizer) of $\clH$.
Define two types of the affine Weyl groups:
  \beq{awg}
W_P=\{\hat w=wt^\ga\,,~w\in W\,,~\ga\in P^\vee\}\,,~~W_Q=\{\hat w=wt^\ga\,,~w\in W\,,
~\ga\in Q^\vee\}\,.
   \eq
 They act on the root vectors as
 $ E_{\hat\al}= E_\al t^n\to E_{\hat w(\hat\al)}=E_{w(\al)}t^{n+\lan\ga,\al\ran}$.

\bigskip

\noindent \emph{\textbf{Loop groups  \cite{PS}}}

\vspace{2mm}

Let $L(G)$ be the loop group corresponding to the loop Lie algebra (\ref{la})
  \beq{logr}
L(G)=G\otimes C[t^{-1},t]]=\left\{\,\sum_kg_kt^k\,,~g_k\in G\,\right\}\,.
  \eq
Define  the loop subgroups
    \beq{bpl}
L^+(G)=\{g_0+g_1t+\ldots=g_0+tL[t]]\}\,,~~g_j\in G\,,~g_0\in P\,,
   \eq
    \beq{bne}
N^-(G)=\{n_-+g_1t^{-1}+\ldots=n_-+t^{-1}L[t^{-1}]\}\,,~~n_-\in N^-\,,
   \eq
   \beq{pns}
N^+(G)=\{n_++g_1t+\ldots=n_++tL[t]]\}\,,~~n_+\in N^+\,,
   \eq
 where $Lie\,( N^+)=\gn^+=\sum{\al\in\ti R^+}c_\al E_\al $ (\ref{psa}).
 It follows from (\ref{pns}) that
   \beq{laf}
 L^+(G)=L\ltimes N^+(G)\,,
   \eq
 where $L$ is the Levi subgroup.

There are the loop analogues of the Bruhat decomposition
(\ref{aal}). The decomposition of the Cartan subalgebra (\ref{ccd})
defines the coweight sublattice
  \beq{tco}
\ti P^\vee=\{\ga\in P^\vee\,|\,\lan\ga,\al\ran=0\,,~\al\in\gh'\}
  \eq
The subgroup $\ti W_P$
of the affine Weyl group $W_P$ generated by the Weyl group $W'=W'(\gg')$, where $\gg'$ is
defined in (\ref{ccd}), and the sublattice $P^{'\vee}\subset P^\vee$.
 Consider the quotient $\ti W_P=W_P/W'_P$.  Similarly define
 $\ti W_Q=W_Q /W'_Q$ and $\ti W_{t(G)}=W_{t(G)}/W'_{t(G)}$.
 Similarly to (\ref{aal}) \emph{the affine Bruhat decomposition} assumes the form
     \beq{PS}
 L(G^{ad}) =\bigcup_{\hat w\in\ti  W_P}N^-(G^{ad})\hat wL^+(G^{ad})
 =\bigcup_{\hat w\in \ti W_P}N^+(G^{ad})\hat wL^+(G^{ad})\,,
      \eq
   \beq{PS1}
   L(\bG) =\bigcup_{\hat w\in \ti W_Q}N^-(\bG)\hat wL^+(\bG)
    =\bigcup_{\hat w\in\ti W_Q}N^+(\bG)\hat wL^+(\bG) \,
      \eq
      \beq{PS2}
 L(G) =\bigcup_{\hat w\in \ti W_{t(G)}}N^-(G)\hat wL^+(G)
 =\bigcup_{\hat w\in\ti  W_{t(G)}}N^+(G)\hat wL^+(G) \,.
      \eq
An element $g(t)\in L(G)$ can have the monodromy
$g(te^{2\pi\imath})=g(t)\zeta$, $\zeta=\bfe(\xi)$, where $\xi$ is a representative of
the quotient $P^\vee/t(G)\sim\clZ(G)$ (\ref{cG}).
 Define the subset of loops $ L_\zeta(G)$ homotopic $\bfe(\xi)$.
Then we come to the decomposition
  \beq{hml}
L(G)=\cup_{\zeta\in\clZ(G)} L_\zeta(G)\,.
  \eq
%
Define \emph{the affine flag variety} as the quotient
  \beq{affl0}
Flag^{aff}=L(G)/L^+(G)\,.
  \eq
A $N^+(G)$-orbit of $\hat w$ in  $Flag^{aff}$
  \beq{spm}
C_{\hat w}=\{n(t)\hat wL^+(G)\,|\,n(t)\in N^+(G)\}\,.
  \eq
is called\emph{ the affine Schubert cell}. From (\ref{PS})-(\ref{PS2}) we find that
  $$
Flag^{aff}=\bigcup_{\hat w\in W_{P,Q}}C_{\hat w}\,.
  $$
 The dimension of the affine Schubert cell in $Flag^{aff}$ is
  \beq{dsm}
{\rm dim} (C_{\hat w}) =l(\hat w)\,,
  \eq
where $l(\hat w)$  is \emph{the length} of $\hat w$.
 It is the number of negative affine roots (\ref{nar}), which $\hat w$ transforms
 to positive ones.

\section*{Appendix B: Generalized Sin-basis in simple Lie algebras}
\setcounter{equation}{0}
\def\theequation{B.\arabic{equation}}

\addcontentsline{toc}{section}{Appendix B: Generalized Sin-basis in
simple Lie algebras}

In this Section we briefly describe the construction suggested in
\cite{LOSZ1}.
Let $\gg$ be a complex simple Lie algebra, $\gh$ is a Cartan
subalgebra and $R$ is the root system. Then we have  decomposition
  \beq{CD2}
\gg=\gh+\gl\,,~~\gl=\sum_{\be\in R}\gR_\be\,,  ~~\dim_\mC\,\gR_\be=1\,.
  \eq
The Chevalley basis in $\gg$ is generated by
  \beq{CBA}
\{E_{\be_j}\in\gR_{\be_j}\,,~\be_j\in R\,,~~H_{\al_k}\in\gh\,,~\al_k\in\Pi\}\,,
  \eq
where $H_{\al_k}$ are defined by the commutation relations
  $$
[E_{\al_k},E_{-\al_k}]=H_{\al_k}\,,~~[H_{\al_k},E_{\pm\al_j}]= a_{kj}E_{\pm\al_k}\,,~~\al_k\,,\al_j\in\Pi\,.
  $$
  \beq{cbcr2}
[H_{\al_j},E_{\al_k}]=a_{kj}E_{\al_k}\,,~~~[E_\al,E_{\be}]=C_{\al,\be}E_{\al+\be}\,,
  \eq
Let us pass from the Chevalley basis (\ref{CBA}) to a new basis that
is more convenient to define bundles corresponding to non-trivial
characteristic classes. We call it \emph{the generalized sin basis}
(GS-basis), because for $A_{n}$ case and degree one bundles it
coincides with the sin-algebra basis (see, for example, \cite{FFZ}).

Let us take an element $\zeta\in\clZ(\bar G)$ of order $l$ and the
corresponding $\La^0\in W$ from (\ref{gce}).  Then as in Section
\ref{decS} $\La^0$ generates a cyclic group
$\mu_l=(\La^0,(\La^0)^2,\ldots,(\La^0)^l=1)$ isomorphic to a
subgroup of $\clZ(\bar G)$. Since  $\La^0\in W$ it preserves the
root system $R$.
 Define the quotient set $\clT_l=R/\mu_l$.
Then $R$ is represented  as a union of $\mu_l$-orbits
$R=\cup_{\clT_l}\clO$.
We denote by  $\clO(\babe)$  an orbit starting from the root $\be$
  $$
\clO(\babe)=\{\be\,,\la(\be)\,,\ldots,\la^{l-1}(\be)\}\,, ~~\babe\in \clT_l\,.
  $$
The number of elements in an orbit $\clO$ (the length of $\clO$) is $l/p_\al=l_\al$,
where $p_\al$ is a divisors of $l$.
Let $\nu_\al$ be a number of orbits $\clO_{\baal}$ of the length $l_\al$.
Then $\sharp\, R=\sum\nu_\al l_\al$.
Note, that if $\clO(\babe)$ has length $l_\be$ $\,(l_\be\neq 1)$, then the elements
$\la^k\be$ and $\la^{k+l_\be}\be$ coincide.


\paragraph{Basis in $\gl$ (\ref{CD2})}

Transform first the root basis $\clE=\{E_\be\,,~\be\in R\}$ in $\gl$.
Define  an  orbit in $\clE$
  $$
\clE_{\babe}=\{E_\be\,,E_{\la(\be)}\,,\ldots\,,E_{\la^{l-1}(\be)}\}
  $$
 corresponding to $\clO(\babe)$. Again
$\clE=\cup_{\babe\in\clT_l}\clE_{\babe}$.

For  $\clO(\babe)$ define the set of integers
  \beq{dc4}
J_{p_\al}=\{a=mp_\al\,|\,m\in\mZ\,, ~~a~ {\rm is ~ defined~}mod\,l\,\}\,, ~~~(p_\al=l/l_{\al})\,.
  \eq
"The Fourier transform" of the root basis on the orbit $\clO(\babe)$ is defined as
  \beq{ft}
\gt^k_{\babe}=\f1{\sqrt{l}}\,\sum_{m=0}^{l-1}\om^{ma}E_{\la^m(\be)}\,,~~
\om=\exp\,\frac{2\pi i}{l}\,,~~a\in J_\be\,.
  \eq
This transformation is invertible
$E_{\la^k(\be)}=\f1{\sqrt{l}}\sum_{a\in J_l}\om^{-ka}\gt^k_{\babe}$,
and therefore there is the one-to-one  map
$\clE_{\be} \leftrightarrow \{\gt^k_{\babe}\,,~a\in J_\be\}$.
In this way we have defined the new basis
  \beq{fbl}
\{\gt^k_{\babe}\,,~(a\in J_l\,,~\babe\in\clT_l)\}\,.
  \eq
Since $\la (E_\al)=E_{\la(\al)}$  we have for $\La\bfe(\bfu)$ $\,(\bfu\in\ti\gh_0)$
  \beq{nlt}
Ad_{\La}(\gt^k_{\babe})=\bfe(\lan\bfu,\be\ran-\frac{a}l)\gt^k_{\babe}\,,~~~~\bfe(x)=\exp\,(2\pi ix)\,.
  \eq
It means that
 $\,\gt^k_{\babe}$ $\,(\babe\in\clT_l)$ is a part of basis in $\gg_{l-a}$ (\ref{gra}).
 Moreover,
    \beq{nqt}
Ad_{\clQ}(\gt^k_{\babe})=\bfe(\lan\ka,\be\ran)\gt^k_{\babe}\,.
  \eq
 These relations follows from  (\ref{gce2}). 
 We also  take into account that $\clQ$ and $\La$
commute in the adjoint representation and
$\bfe\,(x)E_\al\bfe\,(-x)=\bfe\,\lan x,\al\ran E_\al$ for $x\in\ti\gh_0$.

Picking another element $\La'$ generating a subgroup $\clZ_{l'_1}$ $(l'\neq l)$
we come to another set of orbits and to another basis. We have as many types of bases
as many of non-isomorphic subgroups in $\clZ(\bar G)$.


\subsubsection*{The Killing form}

 Consider two orbits  $\clO(\baal)$ and $\clO(\babe)$, passing through $E_\al$
 and $E_\be$. Assume that there exists such integer $r$
that $\al=-\la^r(\be)$.
It implies that elements of two orbits are related as
$\la^n(\al)=-\la^m(\be)$ if $m-n=r$.  In other words, $-\be\in\clO(\baal)$.
 In particular, it means that orbits have the same length.
It follows from (\ref{ft}) and (\ref{are6}) that
  \beq{kfl}
(\gt^{c_1}_{\baal},\gt^{c_2}_{\babe})= \de_{\al,-\la^r(\be)}
\de^{(c_1+c_2,0\,\,(mod\,l))}\om^{-rc_1}\frac{2p_\al}{(\al,\al)}\,,
  \eq
where $p_\al=l/l_\al$, and $l_\al$ is the length of $\clO(\baal)$.  In particular,
$(\gt_{\baal}^a,\gt_{-\baal}^{-a})=\frac{2p_\al}{(\al,\al)}$.
In what follows we need a dual basis $\gT_{\baal}^b$
  \beq{dbt}
(\gT_{\baal_1}^{b_1},\gt^{b_2}_{\baal_2})=
\de^{(b_1+b_2,0\,\,(mod\,l))}\de_{\baal_1,-\baal_2}\,,
~~~\gT_{\baal}^b=\gt^{-b}_{-\baal}\frac{(\al,\al)}{2p_\al}\,.
  \eq
The Killing form in this basis is inverse to (\ref{kfl})
  $$
(\gT_{\baal_1}^{a_1},\gT_{\baal_2}^{a_2})=
\de_{\al_1,-\la^r(\al_2)}
\de^{(a_1+a_2,0\,\,(mod\,l))}\om^{ra_1}\frac{(\al_1,\al_1)}{2p_{\al_1}}\,.
  $$
In particular,
  \beq{kf2}
(\gT_{\al}^{a},\gT_{-\al}^{-a})=
\frac{(\al,\al)}{2p_\al}\,.
  \eq



\subsection*{A basis in the Cartan subalgebra}

Almost the same construction exists in $\gh$.
Again let $\La^0$ generates the group $\mu_l$. Since  $\La^0$ preserves
the  extended Dynkin diagram, its action preserves  the extended coroot system
 $\Pi^{\vee ext}=\Pi^\vee\cup \al^\vee_0$ in $\gh$.
 Consider the quotient $\clK_l=\Pi^{\vee ext}/\mu_l$.
Define an orbit $\clH(\baal)$ of length $l_\al=l/p_\al$   in $\Pi^{\vee ext}$
passing through $H_\al\in\Pi^{\vee ext}$
  $$
\clH(\baal)=\{H_\al\,,H_{\la(\al)}\,,\ldots,H_{\la^{l-1}(\al)}\}\,,
~~~\baal\in\clK_l=\Pi^{\vee ext}/\mu_l\,.
  $$
The set $\Pi^{\vee ext}$ is a union of $\clH(\baal)$
  $$
(\Pi^{\vee})^{ext}=\cup_{\baal\in\clK_l}\clH(\baal)\,.
  $$
Define "the Fourier transform"
  \beq{fth}
\gh^k_{\baal}=\f1{\sqrt{l}}
\sum_{m=0}^{l-1}\om^{mk}H_{\la^m(\al)}\,,~~\om=\exp\,\frac{2\pi
i}{l}\,, ~~k\in J_\al ~(\ref{dc4})\,.
  \eq
The basis $\gh^c_{\baal}\,$, $(c\in J_\al,~\baal\in\clK_l)$
 is over-complete in $\gh$.
Namely, let $\clH(\baal_0)$ be an orbit passing through
the minimal coroot $\{H_{\al_0},H_{\la(\al_0)},\ldots,H_{\la^{l-1}(\al_0)}\}$.
Then the element $\gh^0_{\bar{\al}_0}$ is a linear combination of elements
$\gh^0_{-\baal}\,$, $(\al\in\Pi)$ and we should exclude it from the basis.
We replace the basis $\Pi^\vee$ in $\gh$ by
  \beq{cb}
\gh^c_{\baal}\,,~(c\in J_\al)\,,~~\,,~~
\left\{
\begin{array}{ll}
  \baal\in\ti\clK_l=\clK_l\setminus\clH(\baal_0)\,, & c=0 \\
  \baal\in \clK_l\,,  & c\neq 0\,.
\end{array}
\right.
  \eq
As before there is a one-to-one map $\Pi^\vee\leftrightarrow\{\gh^c_{\baal}\}$.
The elements $(\gh^a_{\baal},\gt^k_{\baal})$ form GS basis in $\gg_{l-a}$ (\ref{gra}).


\subsubsection*{The Killing form}

The Killing form in the basis (\ref{cb}) can be found from (\ref{kilh})
  \beq{clA}
(\gh^{k}_{\baal},\gh^{b}_{\babe})=\de^{(k+j,0\,(mod\,l))}
\clA^k_{\al,\be}\,,~~~
\clA^k_{\al,\be}=\frac{2}{(\be,\be)}\sum _{s=0}^{l-1}\om^{-sk}
a_{\be,\la^s(\al)}\,,
  \eq
where $a_{\al,\be}$ is the Cartan matrix (\ref{CMa}).
The dual basis is generated by elements $\gh^a_{\baal}$
  \beq{dbh}
(\gH^k_{\baal}, \gh^j_{\babe}) =\de^{(k+j,0\,(mod\,l))}\de_{\al,\be}\,,~~~
 \gH^k_{\baal}=\sum_{\be\in\Pi}(\clA^k_{\al,\be})^{-1}\gh^{-k}_{\babe}\,,
~~\gh^{k}_{\babe}=\sum_{\al\in\Pi}(\clA^{-k}_{\al,\be})\gH^{-k}_{\baal}
  \eq
The Killing form in the dual basis takes the form
  \beq{kfdbh}
(\gh^{k_1}_{\baal_1},\gh^{k_2}_{\baal_2})=\de^{(k_1+k_2,0\,(mod\,l))}
(\clA^{k_1}_{\baal_1,\baal_2})^{-1}\,.
  \eq
%
\bigskip
In summary, we have defined the GS-basis in $\gg$
  \beq{GSB}
\{\gt^{k}_{\babe},\gh^j_{\baal}\,,~~(k,\babe,j,\baal)~
{\rm are~ defined ~in ~(\ref{fbl}),~ (\ref{cb})}\}\,,
  \eq
and the dual basis
  \beq{DGSB}
\{\gT^{k}_{\babe},\gH^j_{\baal}\,,~~(k,\babe,j,\baal)~
{\rm are~ defined ~in ~(\ref{dbt}),~ (\ref{dbh})}\}\,,
  \eq
along with the Killing forms.


\subsection*{Commutation relations \label{commrell}}

The commutation relations in the GS basis can be found from the
commutation relations in the Chevalley basis (\ref{cbcr}). Taking
into account the invariance of the structure constants with
respect to the Weyl group action $C_{\la\al,\la\be}=C_{\al,\be}$
it is not difficult to derive the  commutation relations in the GS
basis using its definition in the Chevalley basis (\ref{ft}),
(\ref{fth}). In the case of root-root commutators we come to the
following relations
   \beqn{com1}
 [\gt^{a}_{\alpha},\gt^{b}_{\beta}]\,=\left\{
\begin{array}{ll}
\frac{1}{\sqrt{l}}\,\sum\limits_{s=0}^{l-1}\, \omega^{bs} \,
C_{\alpha,\, \lambda^s\beta}\,\gt^{a+b}_{\alpha+\lambda^s\beta},&
\alpha\neq \,-\lambda^{s} \beta\\ &
\\
\frac{p_{\alpha}}{\sqrt{l}}\,\omega^{s\,b}\,\gh^{a+b}_{\alpha}&\alpha=
\,-\lambda^{s} \beta
\end{array}\right.
    \eqn
 The Cartan-root commutators are:
  \beqn{com2}
\begin{array}{l}
\left[\gh^{\,k}_{\,\alpha}, \gt^{\,m}_{\,\beta}\right] =
\frac{1}{\sqrt{l}}\,\sum\limits_{s=0}^{l-1}\,\omega^{-ks}\,\frac{2(\alpha,
\lambda^{s}\beta) }{(\alpha,\alpha)}\,\gt^{k+m}_{\,\beta}\\
 \left[\gH^{\,k}_{\,\alpha},\gt^{\,m}_{\,\beta}
\right] = \frac{1}{\sqrt{l}}\,\sum\limits_{s=0}^{l-1}\,\omega^{-
ks }\,\frac{(\alpha,\alpha)}{2}\,({\hat \alpha}, \lambda^{s}\beta)
\,\gt^{k+m}_{\,\beta}
\end{array}
   \eqn
Here we denote by ${\hat \alpha}$ the dual to the simple roots
elements in the Cartan subalgebra:
  \beqn{dr}
 ({\hat \alpha_{i}},
\beta_{j})=\delta_{ij}\,.
   \eqn
  It is more
 convenient to use the following normalized basis for Cartan
subalgebra:
   \beqn{g2}
 \bar{\gh}^{\,k}_{\,\alpha}=\frac{(\alpha,\alpha)}{2}\,\gh^{\,k}_{\,\alpha},\
 \ \ \ \ \bar{\gH}^{\,k}_{\,\alpha}=\frac{2}{(\alpha,\alpha)}\,\gh^{\,k}_{\,\alpha}
   \eqn
This reparametrization leads to the following commutation
relations:
   \beqn{crel2}
\begin{array}{l}
 \left[\bar{\gh}^{\,k}_{\,\alpha},
\gt^{\,m}_{\,\beta}\right] =
\frac{1}{\sqrt{l}}\,\sum\limits_{s=0}^{l-1}\,\omega^{-ks}\,(\alpha,
\lambda^{s}\beta)\,\gt^{k+m}_{\,\beta}\\
 \left[\bar{\gH}^{\,k}_{\,\alpha},\gt^{\,m}_{\,\beta}
\right] = \frac{1}{\sqrt{l}}\,\sum\limits_{s=0}^{l-1}\,\omega^{-
ks }\,({\hat \alpha}, \lambda^{s}\beta) \,\gt^{k+m}_{\,\beta}
\end{array}
   \eqn
The following simple formula expresses the decomposition of Cartan
element in the basis of simple roots:
  \beqn{decsr}
\bar{\gh}_{\beta}^{k}=\sum\limits_{\alpha\in
\Pi}\,(\hat{\alpha},\beta)\,\bar{\gh}_{\alpha}^{k},\ \ \ \beta \in
R\,.
   \eqn
The connection of dual bases is clear from the following expression:
   \beqn{connbas}
  \sum\limits_{\beta\in \Pi}\,
(\hat{\alpha}, \beta) \, \bar{\gh}_{\beta}^{k}
=\sum\limits_{\beta\in \Pi}\, (\alpha, \beta) \,
\bar{\gH}_{\beta}^{k}
    \eqn
 The Cartan elements have the following
 symmetry property:
    \beqn{sprop}
\bar{\gh}^{k}_{-\alpha}=-\bar{\gh}^{k}_{\alpha},\ \
\bar{\gH}^{k}_{-\alpha}=-\bar{\gH}^{k}_{\alpha},\ \
   \eqn






\subsection*{Invariant subalgebra}

Consider the invariant subalgebra  $\gg_0$.
It is generated by the basis $(\gt^0_{\babe}\,,\,\gh^0_{\baal})$ (\ref{GSB}).
In particular,  $\{\gh^0_{\baal}\}$ (\ref{fth}),  (\ref{cb}) form a basis in the Cartan subalgebra
$\ti{\gh}_0\subset\gh$ ($\dim\,\ti{\gh}_0=p<n$).

We  pass from $\{\gh^0_{\baal}\}$ to  a special basis  in  $\ti{\gh}_0$
  \beq{tir}
\ti\Pi^\vee=\{\ti{\al_k}^\vee\,|\,k=1,\ldots,p\}\, .
  \eq
It is constructed in the following way.
Consider a subsystem of simple coroots
   \beq{dpi1}
 \Pi_1^\vee= \Pi^{ext\vee}\setminus\clO(\baal_0^\vee)
   \eq
 (see (\ref{cb})).
In other words, $\Pi_1^\vee$ is a subset of simple coroots that does not
contain simple coroots from the orbit passing through $\al_0$.
 For $A_{N-1}$, $B_n$, $E_6$ and $E_7$ the coroot basis $\ti\Pi^\vee$ (\ref{tir})
  is a result of
 an averaging along the $\la$ orbits in  $\Pi_1^\vee$
  \beq{invc}
\ti\al^\vee=\sum_{m=1}^{l-1}H_{\la^m(\al)}\,,~~~H_{\al}\in\Pi_1^\vee\,.
  \eq
In $C_n$ and $D_n$ cases this construction is valid for almost all
coroots except the last on the Dynkin diagram
 (see Remark 10.1 in  \cite{LOSZ2}).
Consider the dual vectors $\ti\Pi=
\{\ti{\al_k}\,|\,k=1,\ldots,p\,,~ \lan\ti{\al_k},\ti{\al_k}^\vee\ran=2\}$
in $\ti{\gh}_0^*$.
\begin{predl}
The set of vectors in $\ti\gh_0^*$
  \beq{tiPi}
\ti\Pi=\{\ti{\al_k}\,|\,k=1,\ldots,p\}\,,
  \eq
is a system of simple roots of a simple Lie subalgebra  $\ti\gg_0\subset\gg_0$
defined by the root system $\ti R=\ti R(\ti\Pi)$ and the Cartan matrix
 $\lan\ti{\al_k},\ti{\al_j}^\vee\ran$.
\end{predl}
The check of this statement is done in  \cite{LOSZ2}  case by case.

 Let  $ R_1= R_1(\Pi_1)$ be a subset of roots generated by simple roots $\Pi_1=\Pi^{ext}\setminus\clO(\al_0)$.
 It is invariant under $\la$ action.
 The root system $\ti R$ of  $\ti\gg_0$ corresponds to the $\la$ invariant set of $R_1$.
 Consider the complementary set of roots $R\setminus R_1$
 and  the set among orbits
   \beq{clt'}
 \clT'_l=(R\setminus R_1)/\mu_l\,.
    \eq
  It is a subset of all orbits  $\clT_l=R/\mu_l$.
 Therefore, $\clT_l=\ti R\cup\clT'_l$.
 The $\la$-invariant subalgebra $\gg_0$ contains the subspace
  \beq{gd27}
V=\{\sum_{\babe\in \clT'_l}a_{\babe}\gt^0_{\babe}\,,~~a_{\babe}\in\mC\}\,.
  \eq
 Then $\gg_0$ is a sum of $\ti{\gg}_0$ and $V$
  \beq{gd24}
\gg_0={\gg'}_0\oplus V\,.
  \eq
The components of this decomposition are orthogonal with respect to the Killing form (\ref{clA}),
 and $V$ is a representation of $\gg'_0$
We enumarate below the explicit forms of $\gg_0$ for all simple
algebras from our list.

Let $\gh'$ be a subalgebra of $\gh$ with the basis $\gh_{\baal}^j\,$ $j\neq 0$ (\ref{fth})
and $\ti\gh_0$ is a Cartan subalgebra of $\ti\gg_0$.
Then
  \beq{deco2}
\gh=\ti\gh_0\oplus\gh'\,.
  \eq
We summarize the information about invariant subalgebras in Table 4.

\bigskip
\begin{center}
\begin{tabular}{|c|c|c|c|c|c|c|}
  \hline
  &             &         &                  &          &   & \\
  $\Pi$  & $\clZ(\bG)$ &  $\varpi_j^\vee$  &$\Pi_1$ & $l\!=$ord$\,(\La)$ & $\ti\gg_0$ &  $\gg_0$ \\
    &             &         &                  &          &  &  \\
  \hline
  1& 2& 3& 4& 5& 6& 7 \\
   \hline
  $A_{N\!-\!1},(N=pl)$ &$\mu_N$&$\varpi_{N-1}^\vee$& $\cup_{1}^l A_{p-1}$ & $N/p$ &  ${\bf sl_p}$ &
   ${\bf sl_p}\oplus_{j=1}^{l-1}{\bf gl_p}$\\
  $B_n$ & $\mu_2$&$\varpi_{n}^\vee$ & ${\bf so_{2n-1}}$  & 2& ${\bf so(2n-1)}$ & $ {\bf so(2n)}$ \\
  $C_{2l}\,,~(l>1)$ &$\mu_2$ &$\varpi_{2l}^\vee$ &  $A_{2l-1}$ & 2 &  ${\bf so(2l)}$ & ${\bf gl_{2l}}$ \\
  $C_{2l+1}$ &$\mu_2$  & $\varpi_{2l+1}^\vee$ &$A_{2l}$ & 2 & ${\bf so(2l+1)}$ &  ${\bf gl_{2l+1}}$ \\
  $D_{2l+1}\,,~(l>1)$ &$\mu_4$  & $\varpi_{2l+1}^\vee$ &$A_{2l-2}$ & 4 & ${\bf so(2l-1)}$&
   ${\bf so(2l)}\oplus{\bf so(2l)}\oplus\underline{1}$  \\
  $D_{2l+1}\,,~(l>1)$ &$\mu_4$  &$\varpi_{1}^\vee$ & $D_{2l}$ & 2 & ${\bf so(4l-1)}$&
   ${\bf so(4l)}\oplus\underline{1}$  \\
  $D_{2l}\,,~(l>2)$ &$\mu_2\oplus\mu_2$ &$\varpi_{2l}^\vee$ & $A_{2l-1}$ & 2 &
   ${\bf so(2l)}$ &  ${\bf so(2l)}\oplus{\bf so(2l)}$ \\
  $D_{2l}\,,~(l>2)$ &$\mu_2\oplus\mu_2$ & $\varpi_{1}^\vee$ &$D_{2l-1}$ & 2 &
   ${\bf so(4l-3)}$&  ${\bf so(4l-2)}\oplus\underline{1}$ \\
  $E_6$ &$\mu_3$&$\varpi_{1}^\vee$ & $D_4$ & 3 & ${\bf g_2}$ & ${\bf so(8)}\oplus 2\cdot\underline{1}$\\
  $E_7$ &$\mu_2$&$\varpi_{7}^\vee$ & ${\bf e_6}$ & 2 & ${\bf f_4}$ &  ${\bf e_6}\oplus\underline{1}$\\
  \hline
\end{tabular}
\bigskip

\textbf{Table 4}\\
Invariant subalgebras $\ti\gg_0=\gg_{\ti\Pi}$ and $\gg_0$ of simple Lie algebras.\\
The coweights generating central elements are displaced in column 3.
\end{center}


\bigskip

In  the invariant simple algebra $\gg'_0$ instead of
 the basis $(\gh^0_{\baal},\gt^0_{\babe})$  we can use the Chevalley basis
and incorporate it in the GS-basis
  \beq{hal}
\{\gh^0_{\baal}\,, \gt^0_{\babe}\}\,\to\,\{\ti\gg_0=(H_{\ti\al}\,,\ti\al\in\ti\Pi\,,~E_{\ti\be}\,,\ti\be\in\ti R)\,,~
V=(\gt^0_{\babe}\,,\babe\in\clT')\}\,.
  \eq

\begin{rem}
For any $\xi\in Q^\vee$ a solution of (\ref{gce3}) is $\La=Id$. In
this case $\gg'_0=\gg$ and GS-basis is the Chevalley basis.
\end{rem}


\subsection*{The GS-basis from a canonical basis in $\gh$}

Let $(e_1,e_2,\ldots,e_n)$ be a canonical basis in $\gh$,
$((e_j,e_k)=\de_{jk})$.
\footnote{For $A_n$ and $E_6$ root systems it is convenient to choose canonical
bases in $\gh\oplus\mC$.}
Since $\La$ preserves $\gh$ we can
consider the action of $\mu_l$ on the canonical basis.
Define an orbit  of length $l_s=l/p_s$
passing through $e_s$
$\clO(s)=\{e_s,\la(e_{s}),\ldots,\la^{(l-1)}e_{s)}\}$.

The Fourier transform along $\clO(s)$  takes the form
  \beq{ahk2}
\gh_{s}^c=\f1{\sqrt{l}}\sum_{m=0}^{l-1}\om^{mc}\la^m(e_{s})\,,~~
c\in J_{p_s}\,, ~~\om=\exp\,(\frac{2\pi i}{l})\,,
  \eq
where $J_{p_s}=\{c=mp_s\, mod(l)\,|\,m\in\mZ\}$.
Consider the quotient $\clC_l=(e_1,e_2,\ldots,e_n)/\mu_l$.
Then we can pass from the canonical basis to the GS basis
  $$
(e_1,e_2,\ldots,e_n)\longleftrightarrow \{\gh_{s}^c\,,~s\in \clC_l\}\,.
  $$
%
%
%
The Killing form is read of from (\ref{ahk2})
  \beq{kfcb1}
(\gh_{s_1}^{j_1},\gh_{s_2}^{j_2})=\de_{(s_1,s_2)}\de^{(j_1,-j_2)}\,.
  \eq
Then the dual generators are
  \beq{kfcb}
\gH_{s}^{k}=\gh_{s}^{-k}\,.
  \eq
The commutation relations in $\gg$  take the form
  \beq{crh1}
[\gh^{k_1}_{s},\gt^{k_2}_{\babe}]=
\f1{\sqrt{l}}\sum_{r=0}^{l-1}\om^{-rk_1}\lan\la^r(\be),e_s\ran
\gt^{k_1+k_2}_{\babe}\,,
   \eq
  $$
[\gt^{k_1}_{\baal},\gt^{k_2}_{\babe}]=\f1{p_\al\sqrt{l}}\om^{rk_2}\sum_s
(\al^\vee,e_s)\gh^{k_1+k_2}_{s}\,,~~ {\rm if~}\al= -\la^r(\be){\rm ~for~ some~} r\,.
  $$
We obtain the last relation from (\ref{ft}) and from the expansion $\gh^{k}_\al=\sum_s(\al^\vee,e_s)\gh^k_s$. Alternatively,
the same relations can be written as given in (\ref{com1})-(\ref{com2}):
   $$
 [\gt^{k}_{\alpha},\gt^{j}_{\beta}]\,=\left\{
\begin{array}{ll}
\frac{1}{\sqrt{l}}\,\sum\limits_{s=0}^{l-1}\, \omega^{js} \, C_{\alpha,\,
\lambda^s\beta}\,\gt^{k+j}_{\alpha+\lambda^s\beta},& \alpha\neq \,-\lambda^{s} \beta\\ &
\\
\frac{p_{\alpha}}{\sqrt{l}}\,\omega^{s\,j}\,\gh^{k+j}_{\alpha}&\alpha= \,-\lambda^{s} \beta
\end{array}\right.
   $$
   $$
 \begin{array}{l}
\left[\gh^{\,k}_{\,\alpha}, \gt^{\,m}_{\,\beta}\right] =
\frac{1}{\sqrt{l}}\,\sum\limits_{s=0}^{l-1}\,\omega^{-ks}\,\frac{2(\alpha,
\lambda^{s}\beta) }{(\alpha,\alpha)}\,\gt^{k+m}_{\,\beta}\\
 \left[\gH^{\,k}_{\,\alpha},\gt^{\,m}_{\,\beta}
\right] = \frac{1}{\sqrt{l}}\,\sum\limits_{s=0}^{l-1}\,\omega^{- ks }\,\frac{(\alpha,\alpha)}{2}\,({\hat \alpha},
\lambda^{s}\beta) \,\gt^{k+m}_{\,\beta}.
\end{array}
   $$



\section*{Appendix C: Elliptic functions}
\setcounter{equation}{0}
\def\theequation{C.\arabic{equation}}

\addcontentsline{toc}{section}{Appendix C: Elliptic functions}

Most of facts and notations are borrowed from \cite{Mum,We}. The
identities can be proved directly by comparing residues and
quasi-periodicities on the lattice ${{\mathbb Z}+\tau{\mathbb Z}}$.

\emph{Notations.}
   $$
(\bfe(x)=\exp 2\pi\imath(x)\,,~~q=\bfe(\oh\tau))\,.
   $$
   $$
\om_1\,,~\om_2-{\rm~fundamental~half-periods}\,,~\tau=\om_2/\om_1\,.
   $$
\bigskip
\noindent {\it The theta  function}:
   \beq{A.1a}
\vth(z|\tau)=\sum_{n\in {\bf Z}}(-1)^{n-\oh}\bfe\oh \Bigl(
(n+\oh)^2\tau+(2n+1)z\Bigr)=
   \eq
   $$
=2q^{\f1{4}} \sum_{n=0}^\infty(-1)^{n}q^{n(n+1)}\sin\pi z
   $$
\bigskip
\noindent {\it The  Eisenstein functions}
   \beq{A.1}
E_1(z|\tau)=\p_z\log\vth(z|\tau),
~~E_1(z|\tau)|_{z\to0}\sim\f1{z}-2\eta_1z\,,
   \eq
   \beq{A.2}
E_2(z|\tau)=-\p_zE_1(z|\tau)= \p_z^2\log\vth(z|\tau)\,,
~~E_2(z|\tau)|_{z\to0}\sim\f1{z^2}+2\eta_1\,.
   \eq
Here
   \beq{A.6}
\eta_1(\tau)=\frac{3}{\pi^2}
\sum_{m=-\infty}^{\infty}\sum_{n=-\infty}^{\infty '}
\frac{1}{(m\tau+n)^2}=\frac{24}{2\pi
i}\frac{\eta'(\tau)}{\eta(\tau)}\,,
   \eq
where
   $$
\eta(\tau)=q^{\frac{1}{24}}\prod_{n>0}(1-q^n)\,.
   $$
is the Dedekind function.

\bigskip
\noindent {\it Relation to the Weierstrass functions}
   \beq{a100}
\zeta(z,\tau)=E_1(z,\tau)+2\eta_1(\tau)z\,,
   \eq
   \beq{a101}
\wp(z,\tau)=E_2(z,\tau)-2\eta_1(\tau)\,.
   \eq

\bigskip
\noindent \emph{Functions $\phi$:}
   \beq{A.3}
\phi(u,z)= \frac {\vth(u+z)\vth'(0)} {\vth(u)\vth(z)}\,.
   \eq
It has a pole at $z=0$ and
   \beq{A.3a}
\phi(u,z)=\frac{1}{z}+E_1(u)+\frac{z}{2}(E_1^2(u)-\wp(u))+\ldots\,.
   \eq
Let
   \beq{A3c}
f(u,z)=\p_u\phi(u,z) =\phi(u,z) (E_1(u+z)-E_1(u))\,.
   \eq

\bigskip
\noindent {\it Addition formulae.}
   \beq{d2}
\phi(u,z)f(v,z)-\phi(v,z)f(u,z)=(E_2(v)-E_2(z))\phi(u+v,z)\,.
   \eq
   \beq{d3}
\phi(u,z)\phi(-u,z)=(E_2(z)-E_2(u))\,.
   \eq
   \beq{ft5}
\phi(u,z)\phi(-u,w)=\phi(u,z-w)[E_1(u+z-w)-E_1(u)+E_1(w)-E_1(z)]\,.
   \eq
 Let $\sum_{a=1}^nc_a=0$. Then
   \beq{e1q}
 \left(\sum_{a=1}^nc_aE_1(w-x_a)\right)^2=\sum_{a=1}^n
 \left(c^2_a\wp(w-x_a)+\sum_{b\neq a}c_ac_bE_1(x_a-x_b)E_1(w-x_a)\right)\,.
   \eq

\bigskip
\noindent {\it Heat equation}
   \beq{A.4b}
\p_\tau\phi(u,w)-\f1{2\pi i}\p_u\p_w\phi(u,w)=0\,.
   \eq

\bigskip
\noindent {\it Parity}
   \beq{A.300}
\phi(u,z)=\phi(z,u)\,,~~\phi(-u,-z)=-\phi(u,z)\,.
   \eq
   \beq{pei}
E_1(-z)=-E_1(z)\,,~~E_2(-z)=E_2(z)\,.
   \eq
   \beq{pfu}
f(-u,-z)=f(u,z)\,.
   \eq

\bigskip
\noindent {\it Quasi-periodicity}
   \beq{A.11}
\vth(z+1)=-\vth(z)\,,~~~\vth(z+\tau)=-q^{-\oh}e^{-2\pi iz}\vth(z)\,,
   \eq
   \beq{A.12}
E_1(z+1)=E_1(z)\,,~~~E_1(z+\tau)=E_1(z)-2\pi i\,,
   \eq
   \beq{A.13}
E_2(z+1)=E_2(z)\,,~~~E_2(z+\tau)=E_2(z)\,,
   \eq
   \beq{A.14}
\phi(u,z+1)=\phi(u,z)\,,~~~\phi(u,z+\tau)=e^{-2\pi \imath
u}\phi(u,z)\,.
   \eq
   \beq{A.15}
f(u,z+1)=f(u,z)\,,~~~f(u,z+\tau)=e^{-2\pi \imath
u}f(u,z)-2\pi\imath\phi(u,z)\,.
   \eq
 The most important object for
construction of Lax operators and $r$-matrices is the function
defined as follows:
   \beq{t303}
 \varphi^{k}_{\alpha}(\bfu,z)=e^{2\pi i\lan\varrho,\alpha\ran z} \phi
 \Big( \lan\bfu+\varrho \tau, \alpha\ran+\frac{k}{l},z \Big)\,,~~(\varrho=\rho/h)\,,
    \eq
     \beq{t3032}
 \varphi^{k}_{0}(\bfu,z)= \phi
(\frac{k}{l},z)\,,
    \eq
   \beq{t305}
 \varphi^{0}_{0}(\bfu,z)=E_{1}(z)
   \eq
and
   \beq{t3036}
f^{k}_{\alpha}(\bfu,z)=e^{2\pi i\lan\varrho,\alpha\ran z} f\Big(
\lan\bfu+\varrho \tau, \alpha\ran+\frac{k}{l},z \Big)\,,
    \eq
   \beq{t3037}
   f^{k}_{\alpha}(\bfu,0)=-E_{2}\Big(<u+\varrho \tau,
\alpha>+\frac{k}{l}\Big) =-\wp\Big(<u+\varrho \tau,
\alpha>+\frac{k}{l}\Big)-2\eta_1\,,
    \eq
   \beq{t3064}
 f^{0}_{0}(\bfu,z)=f^0_0(z)=\frac{1}{2}\Big( E_{1}^{2}(z)-\wp(z)
\Big)\,.
    \eq
%
%
The heat equation takes the form:
   \beq{ap212}
2\pi i \p_{\tau} \varphi^{k}_{\alpha}(\bfu,z)=\p_{z}
f^{k}_{\alpha}(\bfu,z)\,.
   \eq
To save space we sometimes omit the $u$-dependence of the function
in the formulae below.

\bigskip

\noindent {\em Fay identity:}
   \beq{ap3}
\phi(u_1,z_1)\phi(u_2,z_2)-\phi(u_1+u_2,z_1)\phi(u_2,z_2-z_1)-\phi(u_1+u_2,z_2)\phi(u_1,z_1-z_2)=0\,.
   \eq
Differentiating over $u_2$ we find:
   \beq{ap31}
\phi(u_1,z_1) f(u_2,z_2)-\phi(u_1+u_2,z_1)f(u_2,z_2-z_1)=
   \eq
   $$
\phi(u_2,z_2-z_1)f(u_1+u_2,z_1)+\phi(u_1,z_1-z_2)f(u_1+u_2,z_2)\,.
   $$
Substituting here
   \beq{ap32}
\begin{array}{l}
u_1=<u+\rho\tau,\alpha+\beta>+\frac{k+m}{l}\,,\\
u_2=-<u+\rho\tau,\beta>-\frac{m}{l}\,,\\
z_1=z_a-z_c=z_{ac}\,,\\
z_2=z_b-z_c=z_{bc}
 \end{array}
   \eq
and multiplying by appropriate exponential factor we can rewrite it
in the form:
   \beq{ap33}
\varphi^{k}_{\alpha}(z_{ac})
f^{m}_{\beta}(z_{ab})-\varphi^{m}_{\beta}
(z_{ab})f^{k}_{\alpha}(z_{ac})
+\varphi^{k+m}_{\alpha+\beta}(z_{ab})f^{k}_{\alpha}(z_{cb})-
\varphi^{k+m}_{\alpha+\beta}(z_{ac}) f^{-m}_{-\beta}(z_{bc})=0\,.
   \eq
Taking the limit $m=0, \beta=0$ and using the expansion
   \beq{ap214}
\phi(u,z)\sim\frac{1}{u}+E_{1}(z)+u f^0_0(z)+...
   \eq
we find:
   \beq{ap215}
\varphi^{k}_{\alpha}(z_{ac}) f^0_0(z_{ab})-E_{1}(z_{ab})
f^{k}_{\alpha}(z_{ac})+ \varphi^{k}_{\alpha}(z_{ab})
f^{k}_{\alpha}(z_{bc})-\varphi^{k}_{\alpha}(z_{ac})f^0_0(z_{cb})
=\frac{1}{2}\p_{u}f^{k}_{\alpha}(z_{ac})\,.
   \eq
More Fay identities:
   \beq{ap29}
\varphi^{k}_{\alpha}(z_{ac})f^{m}_{\beta}(z_{ac})-\varphi^{m}_{\beta}(z_{ac})
f^{k}_{\alpha}(z_{ac})= \varphi^{k+m}_{\alpha+\beta}(z_{ac})(
\wp^{k}_{\alpha}-\wp^{m}_{\beta} )\,,
   \eq
   \beq{ap210}
\varphi^{m}_{\beta}(z_{ac})f^{-m}_{-\beta}(z_{ac})-
\varphi^{-m}_{-\beta}(z_{ac})f^{m}_{\beta}(z_{ac})={E_{2}^{\prime}}^{m}_{\beta}\,,
   \eq
   \beq{ap211}
\varphi^{k}_{\beta}(z_{ac})\wp^{k}_{\beta}-\varphi^{k}_{\beta}(z_{ac})f^0_0(z_{ac})+
E_{1}(z_{ac})f^{k}_{\beta}(z_{ac})= \frac{1}{2}
\p_{u}f^{k}_{\beta}(z_{ac})\,.
   \eq
The last one follows from
   \beq{ap212a}
\p_{u}\phi(u,z)=\phi(u,z)( E_{1}(z+u)-E_{1}(u))\,,
   \eq
and
   \beq{ap213}
(E_{1}(z+u)-E_{1}(u)-E_{1}(z))^2=\wp(z)+\wp(u)+\wp(z+u)\,.
   \eq

\bigskip
\noindent {\it Modular properties.}\\ Let $\mat{a}bcd\in{\rm
SL}(2,{\mathbb Z})$, i.e. $a,b,c,d\in{\mathbb Z}$ and $ad-bc=1$.
Then
  \beq{A.156}
\te(\frac{z}{c\tau+d}|\frac{a\tau+b}{c\tau+d})= \ep e^{\frac{\pi
i}{4}}(c\tau+d)^\oh \exp\left(\frac{i\pi
cz^2}{c\tau+d}\right)\te(z|\tau)\,,~~(\ep^8=1)\,,
  \eq
  \beq{A.15a}
E_1(\frac{z}{c\tau+d}|\frac{a\tau+b}{c\tau+d})=(c\tau+d)E_1(z|\tau)
+2\pi i c z\,,
  \eq
  \beq{A.16}
E_2(\frac{z}{c\tau+d}|\frac{a\tau+b}{c\tau+d})=(c\tau+d)^2E_2(z|\tau)-
2\pi i c(c\tau+d)\,,
  \eq
 \beq{A.17a}
\phi(\frac{u}{c\tau+d},\frac{z}{c\tau+d}|\frac{a\tau+b}{c\tau+d})=({c\tau+d})\exp\left(2
\pi i\frac{ czu}{c\tau+d}\right)\phi(u,z|\tau)\,.
  \eq


\section*{Appendix D: Characteristic classes and conformal groups}
\setcounter{equation}{0}
\def\theequation{D.\arabic{equation}}

\addcontentsline{toc}{section}{Appendix D: Characteristic classes
and conformal groups}

In this Section we briefly describe the construction of conformal
groups suggested in \cite{LOSZ1}.

\subsection*{Conformal groups}

Here we introduce an analog of the group $\GLN$ for other simple groups apart from $\SLN$.
Let
  \beq{phi1}
\phi\,:\, \clZ(\bG)\hookrightarrow \left(\mC^*\right)^r
  \eq
 be an embedding of the
 center $\clZ(\bG)$ into algebraic torus $\left(\mC^*\right)^r$ of
  minimal dimension ($r=1$ for a cyclic center and $r=2$ for
  $\mu_2\times\mu_2$).
 Note that any two embeddings for $\clZ(\bG)$, $(\bG\neq
 \SLN)$ are
 conjugated from the left: $\phi_1=A\phi_2$ for some automorphism
 $A$ of the torus $\left(\mC^*\right)^r$.
For these groups we deal with
 $\mu_2,\mu_3,\mu_4$ or $\mu_2\times\mu_2$.
In these cases the non-trivial  roots of unity coincide or they are
 inverse to each other. In the latter case $A\,:\,x\to x^{-1}$.

 Consider the "anti-diagonal" embedding
 $ \clZ(\bG)\to \bG\times \left(\mC^*\right)^r\,,$
$\,\zeta\mapsto (\zeta,\phi(\zeta)^{-1})\,,$ $\,\zeta\in \clZ(\bG)$.
 The image of this map is a normal
subgroup since $\clZ$ is the center of $\bG$.
\begin{defi}
The quotient
  $$
C\bG=\left(\bG\times \left(\mC^*\right)^r\right)/\clZ(\bG)
  $$
is called  the conformal version of $\bG$.
\end{defi}
In the similar way the conformal version can be defined for any $G$
with a non-trivial center. If the center of $G$ is trivial as for
$G^{ad}$ then $CG=G\times\mC^*$.

The group $C\bG$ does not depend on embedding  in $\mC^r$
   due to above remark about conjugacy of $\phi$'s.
 We have a natural inclusion $\bG\subset C\bG$.
 Consider the quotient  torus
 $Z^{\vee}=\left(\mC^*\right)^r/\clZ(\bG)\sim\left(\mC^*\right)^r$.
  The last isomorphism is defined by $\lambda\to \lambda^N$ for cyclic center
 and $(\lambda_1,\lambda_2)\to (\lambda_1^2,\lambda_2^2)$ for $D_{even }$.
  The sequence
   \beq{bgcg}
 1\to \bG\to C\bG\to Z^{\vee}\to 1
   \eq
 is the analogue of
  $$
1\to \SLN\to \GLN\to \mC^* \to 1\,.
  $$
 On the other hand, we have embedding $\left(\mC^*\right)^r\to C\bG$
  with the quotient $C\bG/\left(\mC^*\right)^r=G^{ad}$.
 Then the sequence
   \beq{cgad}
 1\to\left(\mC^*\right)^r\to C\bG\to G^{ad}\to 1
   \eq
 is similar  to the sequence
   $$
 1\to{\mathbb C}^*\to \GLN\to \PGLN\to 1\,.
   $$
Let $\pi$ be an irreducible  representation of $\bG$ and  $\chi$ is a character
 of the torus $\left(\mC^*\right)^r$.
It follows from (\ref{bgcg})
that an irreducible  representation $\ti\pi$ of $C\bG$ is defined as
   \beq{tipi}
\ti\pi=\pi\boxtimes\chi((\mC^*)^r)\,,~{~\rm such~that}~\pi|_{\clZ(\bG)}=\chi\phi\,,~~~(\phi~(\ref{phi1}))\,.
   \eq
 Assume for the simplicity that  $\pi$ is a fundamental representation.
  It means that  the highest weight $\nu$ of $\pi$ is a fundamental weight.
Let $\varpi^\vee$ be
 a fundamental coweight generating $\clZ(\bG)$ for $r=1$. In other words,
 $\zeta=\bfe(\varpi^\vee)$ is a generator of $\clZ(\bG)$ ($\zeta^N=1$, $\,N=$ord$(\clZ(\bG))$.
 Then $\pi|_{\clZ(\bG)}$ acts as a scalar
$\bfe\lan\varpi^\vee,\nu\ran$. The highest weight can be expanded in the basis of
simple roots $\nu=\sum_{\al\in\Pi}c^\nu_\al\al$. Then the coefficients $c^\nu_\al$
are rows of the inverse Cartan matrix.
They have the form $k/N$, where $k$ is an integer. Therefore the scalar
  \beq{scal}
\bfe\lan\varpi^\vee,\nu\ran=\bfe\,
\Bigl(\sum_{\al\in\Pi}c^\nu_\al\de_{\lan\varpi^\vee,\al\ran}\Bigr)
  \eq
is a root of unity. On the other hand, let $\chi_m(\mC^*)=w^m$ ($w\in\mC^*$)
 be a character of $\mC^*$,  and
$\phi(\zeta)=\bfe\,(l/N)$.
In terms of weights the definition of $\ti\pi$ (\ref{tipi}) takes the form
$\bfe\lan\varpi^\vee,\nu\ran=\bfe\,\Bigl(\frac{ml}N\Bigr)$.
It follows from this construction that characters of $C\bG$ are defined by
the weight lattice $P$ and the integer lattice $\mZ$ with an additional restriction
  $$
\chi_{(\ga,m)}(\bfx,w)=\exp\,2\pi i\lan\ga,\bfx\ran w^m\,,~~
\lan\ga,\varpi^\vee\ran=\frac{ml}N+j\,,~~\ga\in P\,,~~m,j\in\mZ\,,~~\bfx\in\gh\,.
  $$
The case $D_{even}$ $(r=2)$ can be considered in the similar way.

\begin{rem}
Simple groups can be defined as subgroups of $GL(V)$ preserving
some multi-linear forms in $V$. For examples, in the defining representations
these forms are bilinear  symmetric forms for $SO$,  bilinear
 antisymmetric forms for $Sp$, a trilinear form for $E_6$ and a form of
 fourth order for $E_7$.
 In a generic situation $G$ is defined as a subgroup of $GL(V)$ preserving a
 three tensor in $V^*\otimes V^*\otimes V$ \cite{GP}.
 The conformal versions of these groups can be
 alternatively defined as transformations preserving the forms up to dilatations.
 We prefer to use here the algebraic construction, but this approach justifies
  the name "conformal version".
 \end{rem}

The conformal versions can be also defined in terms of exact representations of $\bG$.
 Let $V$ be such a representation and assume that $\clZ(\bG)$
 is a cyclic group.
  Then $C\bG$ is a subgroup of $GL(V)$ generated by $G$ and dilatations
 $\mC^*$. The character $\det\,V$ is equal to $\lambda^{\dim\,(V)}$, where $\la$ is equal to (\ref{scal})
 for fundamental representations.

 For $D_{even }$ we use two representations, for example
 the left and right spinors $Spin^{L,R}$. The conformal group
 $CSpin_{4k}$ is a subgroup
 of $GL(Spin^L\oplus Spin^R)$ generated by $Spin_{4k}$ and $\mC^*\times\mC^*$,
 where the first factor $\mC^*$ acts by dilatations on $Spin^L$ and
the second factor acts on $Spin^R$. The character
$\det\,Spin^L$ ($\det\,Spin^R$) is equal to $\lambda_1^{\dim\,(Spin_{4k}^L)}$
($\lambda_2^{\dim\,(Spin_{4k}^R)}$), $\,(\dim\,(Spin_{4k}^{L,R})=2^{2k-1})$.


\subsection*{Characteristic classes and degrees of vector bundles}

From the exact sequence (\ref{cgad}) and vanishing of the second cohomology of
 a curve $H^2(\Si,\clO^*)=0$ with coefficients in analytic sheaf
 we get that
 any $G^{ad}({\cal O})$-bundle (even topologically non-trivial one with
 $\zeta(G^{ad}({\cal O})\neq 0$)  can be lifted to a  $C\bG ({\cal O})$-bundle.

 Let $V$ be an exact representation either irreducible or the sum
$Spin^L\oplus Spin^R$ for $D_{2k}$. Then from (\ref{phi1})
one has an embedding of $\clZ(\bG)$ to the automorphisms of $V$
  \beq{phv}
\phi_V\,:\,\clZ(\bG)\hookrightarrow
\left(\mC^*\right)^r={\rm Aut}_{\bG}(V)\,.
  \eq
In particular case, when $V$ is a  fundamental defining representation
 the center acts by multiplication  by (\ref{scal}).

 Let ${\cal P}_{C\bG}$ be a principal $C\bG({\cal O})$-bundle.
Denote by $E( V)={\cal P}_{C\bG}\otimes_{C\bG}V$ (or $E(Spin^{L,R})$) a vector bundle
 induced by a representation $V$ ($Spin^{L,R}$ for $D_{even}$).

\begin{theor}
\footnote{For $G=\GLN$ this theorem was proved in \cite{NS}}
Let $E_{ad}=E(Ad)$ be the adjoint bundle with the  characteristic class
 ${\zeta}(E_{ad})$. The image of $\zeta(E_{ad})$
 under $\phi_V$ (\ref{phv}) is
   $$
 \phi_V(\zeta(E_{ad}))=\left\{
 \begin{array}{ll}
  \exp\bigl(-2\pi i\,{\rm deg}\,(E_{\bG}(V))/{ \dim V}\bigr)& \clZ(\bG)-{\rm is~ cyclic}\,, \\
   \exp\bigl(-2\pi i{\rm deg}\,(E_{Spin_{4k}^{L,R}})/2^{2k-1}\bigr)\,.
 \end{array}
 \right.
   $$
\end{theor}

\emph{ Proof.} Consider the commutative diagram
  $$
\begin{CD}
  @.1@. 1@. @.\\
@. @AAA@AAA@.@.\\
  1 @>>>Z^{\vee}({\cal O}_\Si)@>\sim>> Z^{\vee}({\cal O}_\Si)@>>> 1@.\\
@. @AA[N]A@AA A@AAA@.\\
  1 @>>>\left({\cal O}_\Si^*\right)^r@>>>C\bG({\cal O}_\Si)@>>>G^{ad}({\cal O}_\Si)@>>>1\\
@. @AAA@AAA@AAA@.\\
 1@>>>\clZ(\bG)@>>>\bG({\cal O}_\Si)@>>>G^{ad}({\cal O}_\Si)@>>>1\\
@. @AAA@AAA@AAA@.\\
@.1@. 1@.1 @.\\
\end{CD}
  $$
and corresponding diagram of \^{C}ech cochains. Let $\psi$ be a
1-cocycle with values in $G^{ad}({\cal O}_\Si)$. Consider its
preimage as a cocycle with values in $C\bG ({\cal O}_\Si)$. Due to
definition of $C\bG$ this cocycle is a pair of cochains $(\Psi,\nu)$
with values in $\bG({\cal O}_\Si)$ and $\left({\cal O}_{\Si}^*
\right)^r$ such that $\phi_V(d\Psi)d\nu=1\in \left({\cal O}^*
\right)^r$, where $d$ is the \^{C}ech coboundary operator. The
cohomology class of $d\Psi$ by definition is the characteristic
class ${\bf c}$, so $\phi_V$ of it is opposite to the class of
$d\nu$: $\,\phi_V(\zeta(E_{ad}))=(d\nu)^{-1}$. Since $\nu$ acts in
$V$ as a scalar $\nu^{{\rm dim}V}$, it is a one-cocycle as a
determinant of this action. It represents the determinant of the
bundle $E(V)$. In this way $\nu$ is a preimage of the cocycle
$\nu^{{\rm dim}V}$ under the taking $N=\dim\,(V)$-th power ${\cal
O}^*\stackrel{[N]}{\rightarrow}{\cal O}^*\,,$ $\,\nu\to \nu^{N}\,,$
$\,N=\dim\,(V)$.

Consider the long exact sequence
  $$
1\to\mu_{N} \to{\cal O}_\Si^*\stackrel{[N]}{\rightarrow}{\cal O}_\Si^*\to 1\,,
~~~(\mu_{N}=\mZ/N\mZ)\,.
   $$
It  induces the map $H^1(\Si,{\cal O}_\Si^*)\ {\rightarrow}H^2(\Si,\mu_N)$.
 The cocycle $d\nu$ lies in the cohomology class which is an image
of the class of $\det\,E(V)=\nu^N$ under the
coboundary map $H^1(\Sigma, {\cal O}^*)\to H^2(\Sigma, \mu_N)$.  Denote it by
 ${\rm Inv}_N=$Image$(\det\,E(V))$. Thus, by the definition, the class
  of $d\nu$ equals to ${\rm Inv}_{N}(\det\,E(V))=
{\rm Inv}_{N}(\zeta_1(E(V)))$.

The statement of the theorem follows from  the following
proposition
\begin{predl}
 Let $\gamma$ be a 1-cocycle with values in
${\cal O}^*$. Then ${\rm Inv}_N(\gamma)=\exp\left(\frac1N2\pi
i\,{\rm deg}(\gamma)\right)$.
\end{predl}
\emph{Proof}\\
 Consider the diagram
  $$
\begin{CD}
0@>>>\mu_N@>>>{\cal O}_\Si^*@>[N]>>{\cal O}_\Si^*@>>>0\\
@. @.@AA\exp A@AA\exp A @.\\
 @.0 @>>>{\cal O}_{\Si}@>\times N >>{\cal O}_{\Si}@>>>0 \\
 @. @.@AAA @AAA  @.\\
 @. @.2\pi i {\mathbb Z}@>\times N >>2\pi i {\mathbb Z}@. \\
\end{CD}
  $$

Let $\gamma$ be a 1-cocycle of ${\cal O}_{\Si}^*$. By definition its
image in $H^2(X,\mu_N)$ is equal to the coboundary of 1-cochain
$\gamma^{1/N}$ of ${\cal O}_{\Si}^*$, $(\gamma^{1/N})^N=\gamma $.
Let $\log(\gamma)$ be a preimage of the cycle $\gamma$ under
exponential map; $\log(\gamma)$ is a 1-cochain of ${\cal O}_{\Si}$
and its coboundary equals to degree of $\gamma$ times $2\pi i$. As
the multiplication by $N$ is invertible on ${\cal O}_{\Si}$, the
cochain $\frac1N\log(\gamma)$ is well-defined. Due to commutativity
of the diagram we can choose $\exp\left(\frac1N\log(\gamma)\right)$
as $\gamma^{1/N}$. Hence, the image of $\gamma$ in $H^2(X,\mu_N)$
equals to coboundary of $\exp\left(\frac1N\log(\gamma)\right)$.

The case $r=2$,   can be analyzed in the same way. The theorem is
proved.
$\Box$

\bigskip

Let as above $\varpi^\vee$ be a fundamental coweight generating a center $\clZ(\bG)$
and $\nu$ is weight of representation of $\bG$ in $V$.
Then it follows from Theorem 4.1 and (\ref{scal} ) that
  \beq{cde}
\hbox{deg}\,(E(V))=\dim\,(V)(\lan\varpi^\vee,\nu\ran+k)\,,~~~k\in\mZ\,.
  \eq
  It follows from our considerations that replacing the transition matrix
  $$
\La\to\ti\La(z)=\bfe\,(\lan\varpi^\vee,\nu\ran( z+\frac{\tau}2))\La
  $$
defines the bundle of conformal group $CG$ of degree (\ref{cde}).
For the fundamental representations of $\bG$ we have the following
realization of  (\ref{cde}): \vspace{3mm}
\begin{center}
\begin{tabular}{|c|l|l|l| }
  \hline
   $\bG$ &$\nu$, &$V$ & deg$\,(E(V))$ \\
 \hline
SL$(n,\mC)$ &  $\varpi^\vee_1$ & $\underline{n}$ &$-1+kn$ \\
Spin$_{2n+1}(\mC)$&  $\varpi^\vee_n$ & $\underline{2^n}$ & $2^{n-1}(1+2k)$ \\
Mp$_n(\mC)$& $\varpi^\vee_1$  & $\underline{2n}$ &  $n(1+2k)$ \\
Spin$^{L,R}_{4n}(\mC)$& $\varpi^\vee_{n,n-1}$ & $\underline{2^{2n-1}}$ &$2^{2n-2}(1+2k)$ \\
Spin$_{4n+2}(\mC) $& $\varpi^\vee_n$  & $\underline{2^n}$ & $2^{n-2}(1+4k)$   \\
$E_6(\mC)$ &  $\varpi^\vee_1  $ & $\underline{27}$ & $9(1+3k)$  \\
$E_7(\mC)$ &  $\varpi^\vee_1$ & $\underline{56}$ & $28(1+2k)$  \\
  \hline
\end{tabular}
  $$
(k\in\mZ)
  $$
\vspace{5mm}
\textbf{Table 5.}\\
 Degrees of
 bundles for conformal groups.\\
 Mp$_n(\mC)$ is the universal covering of Sp$_n(\mC)$.
\end{center}



\begin{small}

\end{small}

\end{document}